\documentclass[aps,prd,preprint,nofootinbib ,superscriptaddress,showpacs,ctexart]{revtex4-1} 
\pdfoutput=1 % if your are submitting a pdflatex (i.e. if you have
\usepackage{float}
\usepackage[bottom]{footmisc}
\usepackage{caption}
\usepackage{textcomp}
\usepackage{amsmath}
\usepackage{cases}
\usepackage{amssymb}
\usepackage{graphicx}
\usepackage{subfig}
\usepackage{diagbox}
\usepackage{color}
\usepackage{ulem}
\usepackage{enumerate}
\usepackage{setspace}
\usepackage[unicode=true, bookmarks=false, breaklinks=false,pdfborder={0 0 1},colorlinks=false] {hyperref}

\makeatletter

\newcommand{\bmat}{\left(\begin{array}}
\newcommand{\emat}{\end{array}\right)}
\newcommand{\beq}{\begin{equation}}
\newcommand{\eeq}{\end{equation}}

\makeatletter % `@' now normal "letter"
\@addtoreset{equation}{section}
\makeatother  % `@' is restored as "non-letter"

\def\alt{\mathrel{\mathpalette\gl@align<}}
\def\agt{\mathrel{\mathpalette\gl@align>}}
\def\gl@align#1#2{\lower.6ex\vbox{\baselineskip\z@skip\lineskip\z@
\ialign{$\m@th#1\hfil##\hfil$\crcr#2\crcr\sim\crcr}}}
\def\su5u1{SU(5) \times U(1)}
\def\fsu5u1{SU(5) \times U(1)'}
\def\so10{SO(10)}
\def\sq20{SO(10) \times SO(10)}

\def\bwt{\begin{widetext}}
\def\ewt{\end{widetext}}
\def\be{\begin{equation}}
\def\ee{\end{equation}}
\def\bea{\begin{eqnarray}}
\def\eea{\end{eqnarray}}
\def\bean{\begin{eqnarray*}}
\def\eean{\end{eqnarray*}}
\def\bary{\begin{array}}
\def\eary{\end{array}}
\def\bit{\begin{itemize}}
\def\eit{\end{itemize}}

\makeatother

\begin{document}

\begin{center}

{\Large \bf Quantum Spectrum of BPS Instanton States in $ 5d $ Guage Theories   \\}
%\vspace{3mm}

\end{center}

\vspace{7 mm}

\begin{center}

{ Shan Hu }

\vspace{6mm}
{\small \sl Department of Physics, Hubei University,} \\
\vspace{3mm}
{\small  \sl Wuhan 430062, P. R. China} \\

\vspace{6mm}

{\small \tt hushan@hubu.edu.cn} \\

\end{center}

\vspace{8 mm}

\begin{abstract}\vspace{1cm}

We construct $ 1/2 $ BPS 1-instanton states in the Hilbert space of $ 5d $ gauge theories. The states are local with the scale size $ \rho=0 $. With the degeneracy from the gauge orientation taken into account, the BPS spectrum consists of an infinite number of states in the definite representations of the gauge group and the $ (J,0) $ representation of the $ SO(4)\cong SU(2)_{L} \times SU(2)_{R} $ rotation group. In the brane webs realization of the $ 5d $ $ \mathcal{N}=1 $ gauge theory, BPS states are string webs inside the 5-brane webs. For a theory with the gauge group $ SU(N) $ and the Chern-Simons level $ \kappa $, we give a classification of string webs in the Coulomb branch and show that for each web, one can always find an instanton state carrying the same spin and the same electric charge. The action of supercharges on instanton states gives the instanton supermultiplets. For the $ \mathcal{N} =1$ $ SU(2) $ gauge theory with $ N_{f} \leq 4$, we construct the instanton vector and hyper multiplets, which, together with the original multiplets of the theory, furnish the complete representation of the $ E_{N_{f}+1} $ group at UV.

\end{abstract}

\maketitle

\begin{spacing}{1.8}
\tableofcontents
\end{spacing}

\section{Introduction}

Five dimensional gauge theories contain a topologically conserved current whose integration gives the instanton number
\begin{equation}
 Q_{I}   =\frac{1}{32\pi^{2}}\int d^{4}y\; \epsilon^{mnkl}tr(F_{mn}F_{kl})\;.
\end{equation}
For the arbitrary finite energy gauge field configuration $ A $, the corresponding gauge potential eigenstate $  \vert A \rangle $ will satisfy $  Q_{I}   \vert A \rangle = q \vert A \rangle$ for an integer $ q$. So the total Hilbert space $ \mathcal{H} $ can be decomposed as 
\begin{equation}
\mathcal{H}= \oplus^{+\infty}_{q=-\infty}   \mathcal{H} _{q}
\end{equation}
with $ \mathcal{H} _{q} $ the eigenspace of $ Q_{I}    $ with the eigenvalue $ q $. When the $ 5d $ gauge theory has a UV completion as a $ 6d $ theory, the instanton number is interpreted as the KK momentum along the extra dimension \cite{5,6}. When the UV fixed point of the $ 5d $ gauge theory exhibits the enhanced flavor symmetry, the instanton number also enters into the enhanced flavor charge \cite{L1,tw10a,GNO,tw10aa,tw10ac,1bb}.

We want to find the lowest energy states in each sub-Hilbert space $ \mathcal{H} _{q}  $. For supersymmetric theories, it is also desirable to look for the BPS states in $\mathcal{H} _{q}  $. The classical configurations with the minimum energy and preserving $ 1/2 $ of supersymmetries are instantons labeled by $ 4qN $ collective coordinates when the gauge group is $ SU(N) $ \cite{tasi1}. Quantum mechanically, one should solve for the ground state of a $ 1d $ sigma model in the instanton moduli space. However, unlike the magnetic monopoles, the instanton moduli space suffers from a singularity at $ \rho=0 $ corresponding to the small instanton with the vanishing size \cite{GNOgl,GNOgll}. The singularity may be a reflection of the non-renormalizability of the $ 5d $ gauge theory, and could be resolved by turning on the non-commutivity \cite{GNOglll,GNOgllll}. An instanton is also conjectured to be the bound state of $ N $ partons, with $ \rho $ a parameter characterizing the distances between the constituents \cite{tw12,kik}. The partonic nature of instantons is more explicitly demonstrated in \cite{kik1,GNOg,L3}, where the $ 5d $ $\mathcal{N}=2$ theory is compactified on $ S^{1} $, and a $ D0 $ brane becomes $ N $ D-string segments completing a singly-wound D-string around $ S^{1} $. $ \rho^{2} $ is proportional to the sum of the distances between the D-string segments. In \cite{L3}, with the non-commutativity turned on, when $ N=2 $, two ground states of this multi-monopole system are obtained, representing two distinct ways the D-string segments gluing together. In the limit of the vanishing non-commutativity, only one survives, which is a lump localized in the moduli space with $ \rho=0 $. The classical picture is that $ N $ D-string segments tend to glue into a single one due to the interaction. With and without the non-commutativity, there are $ N $ and $ 1 $ ways to glue. Finally, in $ 5d $ supersymmetric gauge theories, when the scalar field gets the vacuum expectation value, $ \rho $ can be stabilized even in the classical solution. Such configurations are called dyonic instantons \cite{L31,6,L311}.

In this paper, instead of the quantization in moduli space, we will construct the ground state in $ \mathcal{H} _{q} $ directly, which is obtained from the action of the instanton operator $ I(x) $ on the vacuum $ \vert \Omega\rangle $. When the theory is supersymmetric, $  I(x) \vert \Omega\rangle $ is $ 1/2 $ BPS with $Q_{\dot{\alpha}}^{A} I(x) \vert\Omega \rangle=0 $ and saturates the bound $ (P_{0} +P_{5}) I(x) \vert\Omega \rangle=0$\footnote{Usually, a state with $ P_{0} =-P_{5} $ cannot be local but must be the momentum eigenstate with the eigenvalue $ 0 $. However, in $ 5d $, we have the selfdual tensor $ B_{\alpha\beta} $ for which, $P_{0} B_{\alpha\beta}=-P_{5} B_{\alpha\beta}    $ is the self-duality condition.}. $  I(x) \vert \Omega\rangle $ is related to the small instanton localized at $ x $ with $ \rho=0 $. Since $ x $ and $ \rho $ are fixed, the rest degeneracy all comes from the action of the gauge transformation $ U $. When the gauge group is $ SU(N) $ and $ q=1 $, the generic states are $  \{UI(x)\vert \Omega \rangle|\;\forall \; U \}  $, which could be organized into a series of states $\mathcal{I}_{\alpha_{1}\cdots \alpha_{2J}}(x)\vert \Omega \rangle  $ in the $ (J,0) $ representation of the rotation group $ SO(4)\cong SU(2)_{L}\times SU(2)_{R}  $ and some gauge representation of $ SU(N) $, where $ \alpha_{i} =1,2$ is the $ SU(2)_{L} $ spinor index. When $ N\geq 3 $, the allowed $ SU(N) $ representation of $\mathcal{I}_{\alpha_{1}\cdots \alpha_{2J}}(x)\vert \Omega \rangle  $ is restricted by (\ref{mnn}), depending on $ N $, the Chern-Simons level $ \kappa $ and the number of fundamental flavors $ N_{f} $. The situation is quite simple for $ N=2 $, where the gauge representation is uniquely determined by the spin. With the $ SU(2) $ gauge index $ b_{i} =1,2$ added, the BPS spectra are $ (\mathcal{I}_{\alpha_{1}\cdots \alpha_{2J}} )^{b_{1}\cdots b_{2J}}(x) \vert \Omega \rangle$ in the $ (J,J) $ representation of $ SU(2)_{L}  \times SU(2) $.

In string theory, the $ 5d $ $ \mathcal{N} =1$ gauge theory could be realized as the low energy effective theory on 5-brane webs \cite{Hoo,Hoo2,2,2k2,2k2k,2k2k2,2k2k2k,2k2k2k2}. $ 1/2 $ BPS states are string webs inside the brane webs \cite{tw15}. Quantization of a string web with $ n_{X} $ external legs gives a supermultiplet with the highest spin $ \frac{n_{X}}{2} $. The electric charge and the instanton number carried by the string web can be deduced from the mass in Coulomb branch. In addition to F-strings related to W bosons, D-strings as well as the string webs with more external legs and more electric charges also exist giving rise to the instanton particles with the arbitrarily high spins and large electric charges. When the gauge group is $ SU(2) $, the general string web with $q= 1 $ contains $ n_{X} $ external legs and carries $ \frac{n_{X}}{2} $ electric charges. After the quantization, the obtained highest spin states are in one-to-one correspondence with the instanton states $ (\mathcal{I}_{\alpha_{1}\cdots \alpha_{ n_{X} }} )^{b_{1}\cdots b_{ n_{X} }} (x)\vert \Omega \rangle$. When $ N\geq 3 $, for the $ SU(N)  $ gauge theory with the Chern-Simons level $ \kappa $, we will also show that for each string web, one can always find a related instanton state $ \mathcal{I}_{\alpha_{1}\cdots \alpha_{ n_{X}}}(x)\vert \Omega \rangle   $ carrying the same spin and the same electric charges.

We construct the instanton operator in the canonical formalism. The instanton operator is also introduced in the radial quantization formalism by turning on an instanton background on $ S^{4} \times \mathbb{R}_{+}$ \cite{4d}. It was found that the instanton configuration in $ S^{4} \times \mathbb{R}_{+}$ breaks all supersymmetrie unless $ \rho= 0 $ (or $ \rho = \infty $), where $ 1/2$ of supersymmetries are perserved \cite{tw10}. This is also consistent with the $ 5d $ superconformal index computation, where only the point-like instantons/anti-instantons localized at the south ($ \rho = \infty $)/north ($ \rho= 0 $) pole of $ S^{4} $ have the contribution \cite{5a}. $ 5d $ $ \mathcal{N}=1 $ gauge theories may undergo a global symmetry enhancement at the UV fixed point. In \cite{Hoo3, 3d, 5d}, by analysing the fermionic zero modes around the instanton operator on $ S^{4} $, the symmetry enhancement conditions are obtained. The criterion is the existence of a gauge invariant scalar in the triplet of the $ SU(2) $ R-symmetry, which is the lowest component of the current supermultiplet of the enhanced symmetry. In our formalism, the triplet scalar is built from a gauge invariant scalar operator $\mathcal{I}(x)  $. When $ N\geq 3 $, in a $ 5d $ $ \mathcal{N}=1 $ $ SU(N)  $ gauge theory with $ N_{f} $ hypermultiplets and the Chern-Simons level $ \kappa $, according to (\ref{mnn}), $\mathcal{I}(x)   $ exists only when $ \pm \kappa =N-\frac{1}{2} N_{f} $, which is also the symmetry enhancement condition found in \cite{Hoo3}. Moreover, when $ \pm \kappa =N-\frac{1}{2} N_{f}  $, a spin $ 1 $ state $(\mathcal{I}_{\alpha\beta})_{a}^{b} (x) \vert\Omega \rangle$ in the adjoint representation of $ SU(N) $ and $ N_{f} $ spin $\frac{1}{2}$ states $ (\mathcal{I}_{\alpha;s})^{b} (x) \vert\Omega \rangle $ with $  s=1,2,\cdots,N_{f} $ in the fundamental representation of $ SU(N) $ exist, which could be mapped into the original vector and hypers by the enhanced $ SU(2) $ symmetry.

For a $ 5d $ $ \mathcal{N} =1$ $ SU(2) $ gauge theory with $ N_{f} $ hypermultiplets, 
the symmetry enhancement pattern is $  U(1) \times SO(2N_{f}) \rightarrow E_{N_{f}+1}  $, where the $ U(1) $ charge is the instanton number and $ N_{f} \leq 7$. In the UV SCFT, the action of $ E_{N_{f}+1} $ flavor charges on the vector multiplet (hypermultiplets) could generate the instanton vector multiplet (hypermultiplets), which follow the same supersymmetry transformation rule and the same equations of motion as the original ones. When the coupling constant is finite, from the 5-brane webs construction, such instanton supermultiplets still exist but get the mass, breaking the $ E_{N_{f}+1} $ symmetry to $U(1) \times SO(2N_{f})  $. The instanton supermultiplets and the original supermultiplets live in the same $ E_{N_{f}+1} $ representation, and thus should be constructed in the same framework, even when the coupling constant is finite. This is our motivation to construct the instanton states in $ \mathcal{H}_{q} $ directly without relying on the moduli space approximation.

$ 5d $ $ \mathcal{N} =1$ gauge theories can also be engineered through a compactification of M-theory on a Calabi-Yau threefold \cite{22b1,22b2,22b3,22b4,22b5,22b6,22b7,22b8}. BPS particles come from M2-branes wrapping the holomorphic 2-cycles. Quantization of the M2-brane gives the supermultiplets in representations of the enhanced flavor group. In \cite{Hoo1}, a classification for the enhanced flavor symmetry representation of the vector and hyper multiplets is given. In section \ref{Nf}, when the gauge group is $ SU(2) $ and $ N_{f}\leq 4 $, we will construct the instanton vector and hyper multiplets, which, together with the original multiplets of the theory, compose the complete $   E_{N_{f}+1} $ representations in \cite{Hoo1}.

The rest of the paper is organized as follows. In section \ref{sec}, we construct the instanton operator in the canonical formalism. In section \ref{sup1}, we study the supersymmetry transformation of the instanton operator. Section \ref{gau1} and \ref{rot1} give the gauge transformation and the space rotation of the instanton states. In section \ref{spec1}, we present the spectrum of instanton states with the definite spin and the electric charges. In section \ref{web1}, we compare the BPS spectrum of instanton states and the string webs in 5-brane webs. In section \ref{mul1}, we construct the $ 1/2 $ BPS instanton supermultiplets. In section \ref{Nf}, for the $ \mathcal{N} =1$ $ SU(2) $ gauge theory with $ N_{f} \leq 4$, we give the instanton vector and hyper multiplets which are required to complete the $ E_{N_{f}+1} $ representations in UV. The conclusion and discussion are in section \ref{con1}.

 \section{Instanton operator in canonical formalism}\label{sec}

In a $ 5d $ gauge theory with the gauge group $G=U(N)  $ or $ SU(N) $ for $ N\geq 2 $, the $ 4d $ gauge field configuration $  A_{m} $ with the finite energy must approach the pure gauge at the infinity, i.e. $  A_{m}\rightarrow iu\partial_{m}u^{-1}$ when $ r \rightarrow \infty $, $ u \in G $, $ m=1,2,3,4 $. Such $  A_{m} $ can be classified by the the homotopy group $  \Pi_{3}(G) \cong \mathbb{Z} $. For the gauge potential eigenstate $\vert A_{m} \rangle  $ with the eigenvalue $ A_{m} $, with the instanton number operator $  Q_{I}$ given by
\begin{eqnarray}
\nonumber   Q_{I}   &=&\frac{1}{32\pi^{2}}\int d^{4}y\; \epsilon^{mnkl}tr\left( F_{mn}F_{kl}\right) =\frac{1}{8\pi^{2}}\epsilon^{mnkl}\int_{S_{\infty}^{3}} d^{3}S_{m}\;tr\left[  A_{n}(\partial_{k}A_{l}-\frac{2i}{3} A_{k}A_{l})\right] \\ &=& \frac{1}{24\pi^{2}}\epsilon^{mnkl}\int_{S_{\infty}^{3}} d^{3}S_{m}\;
 tr \left[ (u\partial_{n}u^{-1})(u\partial_{k}u^{-1})(u\partial_{l}u^{-1})\right]  \;,
\end{eqnarray}
we have $ Q_{I}   \vert A_{m} \rangle=q \vert A_{m} \rangle $, $ q \in  \mathbb{Z}$.

Consider a gauge transformation operator $ U $ with $ U^{-1}A_{m}U= gA_{m}g^{-1}+ig\partial_{m}g^{-1} $, $g\in G  $. If 
\begin{equation}
 \frac{1}{24\pi^{2}}\epsilon^{mnkl}\int_{S_{\infty}^{3}} d^{3}S_{m}\;
tr\left[  (g\partial_{n}g^{-1})(g\partial_{k}g^{-1})(g\partial_{l}g^{-1})\right]  =k\;,
\end{equation}
then 
\begin{equation}\label{23}
U^{-1}Q_{I}U=Q_{I}+ \frac{1}{24\pi^{2}}\epsilon^{mnkl}\int_{S_{\infty}^{3}} d^{3}S_{m}\;
tr\left[  (g\partial_{n}g^{-1})(g\partial_{k}g^{-1})(g\partial_{l}g^{-1})\right]  =Q_{I}+k\;.
\end{equation}
$[Q_{I},U]=kU  $. $ U  $ is an operator with the instanton number $ k $. For $ \vert A_{m}  \rangle $ in a homotopy class labeled by $ q $, $  U \vert A_{m}  \rangle $ is in a homotopy class labeled by $ q +k$. The ordinary gauge transformations have $ k=0 $. When $ k\neq 0 $, the action of $ U $ must be singular in at least one point.

The instanton operator $ I(x) $ can be defined as such $ U $ singular at the point $ x $. This is a direct extension of 't Hooft's seminal work \cite{Hooft} where topological operators, like monopole operators and the 't Hooft loops, are all realized as the singular gauge transformations. Consider a $ 5d $ theory with the  gauge field $ A_{m} $ and the matter field $ \Phi $, both in the adjoint representation of $ G$. The action of $ I(x) $ on $ \vert A_{m},\Phi\rangle $ is given by 
    \begin{equation}
I\vert A_{m},\Phi\rangle=\vert gA_{m}g^{-1} +ig\partial_{m}g^{-1} , g\Phi g^{-1} \rangle\;.
 \end{equation}
When $ G=SU(2) $, for $ I(x) $ carrying the instanton number $ 1 $, the related gauge transformation matrix $ g $ can be selected as 
  \begin{equation}\label{2}
 g(x,y)=\frac{ \sigma_{n}(y^{n}-x^{n})}{|y-x|}\;,
 \end{equation}
where $\sigma_{n}= ( i\vec{\tau}, 1_{2}) $ with $ \tau_{a} $, $ a=1,2,3 $ three Pauli matrices, $ n=1,2,3,4$, $ \bar{\sigma}_{n} \equiv \sigma^{\dagger}_{n}= (- i\vec{\tau}, 1_{2}) $. (See Appendix \ref{A} for the gamma matrix conventions.) Similarly, for $ I^{\dagger}(x) $ carrying the instanton number $- 1 $, the related gauge transformation matrix is
  \begin{equation}
 g^{\dagger}(x,y)=\frac{  \bar{\sigma}_{n}(y^{n}-x^{n})}{|y-x|}\;.
 \end{equation}

  Under the action of $ I $, $\Phi $, $A_{n}$, $   D_{n} \Phi$ and $ F_{mn} $ transform as 
 \begin{eqnarray}
\nonumber && I^{-1}\Phi I=g\Phi g^{-1} \;,\;\;\;\;\;\;\;\;  I^{-1} A_{n}I=gA_{n}g^{-1}+a_{n}    \;,\\  \label{1}&& I^{-1}  D_{n}\Phi I=gD_{n}\Phi g^{-1} 
\;,\;\;\;\;\;\;\;\;
   I^{-1}  F_{mn}I= gF_{mn}g^{-1}+f_{mn}\;,
\end{eqnarray} 
where 
\begin{eqnarray}
\nonumber && a_{n}=ig\partial_{n} g^{-1}=\lim_{\rho\rightarrow 0}  \frac{i\sigma_{mn}(y^{m}-x^{m})}{|y-x|^{2}+\rho^{2}}\;,\\  \label{mn}&& f_{mn}=\partial_{m}a_{n}-\partial_{n}a_{m}-i [a_{m},a_{n}]=\lim_{\rho\rightarrow 0}  \frac{2i\rho^{2}\sigma_{mn}}{(|y-x|^{2}+\rho^{2})^{2}}\;.
\end{eqnarray} 
$ a_{n} $ and $ f_{mn} $ are just the gauge potential and the field strength of a small instanton localized at $ x $ with $ \rho=0 $. $ f_{mn} $ vanishes everywhere except for $ x $. $\sigma_{mn}=\frac{1}{2}\epsilon_{mnkl}\sigma^{kl}  $, so
    \begin{equation}\label{cd}
f_{mn}=\frac{1}{2}\epsilon_{mnkl}   f^{kl}\;.
 \end{equation}
 $ f_{mn} $ can be decomposed into the self-dual part $  f_{\alpha\beta}=\frac{1}{2}f_{mn}\sigma_{\alpha\beta}^{mn} $ and the anti-self-dual part $f_{\dot{\alpha}\dot{\beta}}=\frac{1}{2}f_{mn}\bar{\sigma}_{\dot{\alpha}\dot{\beta}}^{mn}  $ with $ f_{\dot{\alpha}\dot{\beta}}=0 $. When $ G=SU(N) $, we  can make the replacement  
 \begin{equation}\label{10}
 g \rightarrow g \oplus 1_{N-2}\;,\;\;\;\;\;\;a_{m}\rightarrow a_{m}\oplus 0_{N-2}\;,\;\;\;\;\;\;f_{mn}\rightarrow f_{mn}\oplus 0_{N-2}
 \end{equation}
in (\ref{1}).

 When the $ 5d $ theory is viewed as the compactification of a $ 6d $ theory on $ x^{5} $, the momentum $ P_{5} $ along the $ x^{5} $ dimension is related with the instanton number $ Q_{I} $ via
 \begin{equation}\label{511}
P_{5}=-\frac{Q_{I}}{R_{5}}\;,
 \end{equation}
where $ R_{5} $ is the compactification radius along $ x^{5} $. The Hamiltonian is given by  
 \begin{equation}
P_{0}=\dfrac{1}{4\pi^{2}R_{5}}\int d^{4}y \;tr \left( \frac{1}{4} F_{mn}F^{mn} + \ldots   \right)  \;,
 \end{equation}
where ``$ \ldots $" are terms without involving $F_{mn}  $. From (\ref{1}), (\ref{mn}), (\ref{10}) and $ \int d^{4}y \;f_{mn}f^{mn}=8\pi^{2}1_{2} $, we have
\begin{eqnarray}
\nonumber I^{-1}(x)P_{0}I(x) &=& P_{0}+\dfrac{1}{4\pi^{2}R_{5}}\int d^{4}y \;tr \left( \frac{1}{4} f_{mn}f^{mn} + \frac{1}{2} f_{mn} gF^{mn}g^{-1}  \right)   \\ \nonumber&=& P_{0}+\dfrac{1}{4\pi^{2}R_{5}}\int d^{4}y \;tr \left( -\frac{1}{4} f_{mn}f^{mn} + \frac{1}{2} f_{mn} I^{-1}(x)F^{mn} I(x)  \right)  \\ &=& P_{0}- \frac{1}{R_{5}} +\frac{i}{4}I^{-1}(x)(F^{\alpha}_{\beta})_{\alpha\;0}^{\beta}(x)I(x)\;,
\end{eqnarray}
and thus
\begin{equation}\label{k2}
[P_{0},I(x)] =-[P_{5},I(x)]=-\frac{1}{R_{5}} I(x)+\frac{i}{4}(F^{\alpha}_{\beta})_{\alpha\;0}^{\beta}(x)I(x)\;,
\end{equation}
where 
\begin{equation}
(F^{\alpha}_{\beta})_{\alpha\;0}^{\beta}(x)=\lim_{\rho\rightarrow 0}  \int d^{4}y \;\frac{2\rho^{2}}{\pi^{2}R_{5}(|y-x|^{2}+\rho^{2})^{2}}(F^{\alpha}_{\beta})_{\alpha}^{\beta}(y) \propto (F^{\alpha}_{\beta})_{\alpha}^{\beta}(x)\;.
\end{equation}
$(F^{\alpha}_{\beta})_{\alpha}^{\beta} =\frac{1}{2} (\sigma_{mn})^{\alpha}_{\beta}(F^{mn})_{\alpha}^{\beta}$. $ (F^{\alpha}_{\beta})_{\alpha\;0}^{\beta}(x)=0 $ if $ (F^{\alpha}_{\beta})_{\alpha}^{\beta}  $ is finite at $ x $. $ I(x) $ saturates the BPS bound with
\begin{equation}\label{2.19}
 [P_{0}+P_{5},I(x)]=0\;. 
\end{equation}
When acting on the regular state $\vert \Psi \rangle$, $ I(x) $ creates a small instanton singularity at $ x $ so that 
\begin{equation}\label{po}
[P_{0},I(x)]\vert \Psi \rangle =-[P_{5},I(x)]\vert \Psi \rangle=-\frac{1}{R_{5}} I(x)\vert \Psi \rangle+\frac{i}{4}(F^{\alpha}_{\beta})_{\alpha\;0}^{\beta}(x)I(x)\vert \Psi \rangle=\frac{1}{R_{5}} I(x)\vert \Psi \rangle\;,
\end{equation}
in accordance with (\ref{23}). For the vacuum state $ \vert\Omega\rangle $,
\begin{equation}\label{218}
P_{0}I(x)\vert\Omega\rangle =-P_{5}I(x)\vert\Omega\rangle =\frac{1}{R_{5}} I(x)\vert\Omega\rangle\;.
\end{equation}

\section{Supersymmetry transformation of the instanton operator}\label{sup1}

Now consider the $ 5d $ $ SU(N) $ gauge theory with the $2 \mathcal{N} $-extended supersymmetry for $\mathcal{N} =1, 2$. The theory has a vector multiplet consisting of a gauge field $ A $, real scalars $ \phi^{AB} $, and the symplectic-Majorana fermions $ \lambda^{A} $ in the adjoint representation of $ SU(N) $. $ A,B=1,\cdots,2\mathcal{N} $. $ \Omega_{AB} $ is the symplectic invariant of the R-symmetry group $ Sp(\mathcal{N})_{R} $. 
\begin{equation}
\xi_{A}=\xi^{B}\Omega_{BA}\;,\;\;\;\;\;\;\;\;\xi^{A}=\xi_{B}\Omega^{BA}\;,\;\;\;\;\;\;\;\;\Omega_{AB}\Omega^{BC}=\delta_{A}^{C}\;.
\end{equation}
When $\mathcal{N}=2  $, $ \Omega_{AB} \phi^{AB} =0  $. The symplectic-Majorana condition is
\begin{equation}
\bar{\lambda}_{A}=(\lambda^{T})^{B}\Omega_{BA}C \;,
\end{equation}
where $ C $ is the charge conjugation matrix given by (\ref{C}).

The supercharges are also symplectic-Majorana spinors satisfying 
\begin{equation}
\bar{Q}_{A}=(Q^{T})^{B}\Omega_{BA}C\;.
\end{equation}
For 
\begin{equation}       
Q^{A}=\left[                  
  \begin{array}{c}   
Q_{\alpha}^{A} \\  
Q^{A \dot{\alpha}} \\  
  \end{array}
\right]  \;,
\end{equation}
where $ \alpha,\dot{\alpha}=1,2 $, 
\begin{equation}       
\bar{Q}_{A}=\left[                  
Q_{\alpha}^{B}\Omega_{BA}\epsilon^{\alpha\beta},
Q^{B \dot{\alpha}} \Omega_{BA}\epsilon_{\dot{\alpha}\dot{\beta}}
\right]  \;.
\end{equation}

Let $ P_{\alpha\dot{\alpha}}= \sigma_{\alpha\dot{\alpha}}^{m}P_{m}$ represent the $ 4d $ momentum, then the $ 5d $ superalgebra is given by 
\begin{eqnarray}
     &&\label{kk4k}\{Q_{\alpha}^{A},Q_{\dot{\alpha}}^{B}\}=-2i\Omega^{AB}P_{\alpha\dot{\alpha}}-2iZ^{AB}_{\alpha\dot{\alpha}}\;,   \\ && \label{kk4} \{Q_{\alpha}^{A},Q_{\beta}^{B}\}=2i\Omega^{AB}\epsilon_{\alpha\beta}(P_{0}-P_{5})+2iZ^{AB}\epsilon_{\alpha\beta}+2Z_{\alpha\beta}^{AB}\;, \\ &&  \{Q_{\dot{\alpha}}^{A},Q_{\dot{\beta}}^{B}\}=2i\Omega^{AB}\epsilon_{\dot{\alpha}\dot{\beta}}(P_{0}+P_{5})-2iZ^{AB}\epsilon_{\dot{\alpha}\dot{\beta}}+2Z_{\dot{\alpha}\dot{\beta}}^{AB}\;.\label{k4}
\end{eqnarray}
The electric and the magnetic central charges are \cite{6}
\begin{eqnarray}
    Z^{AB} &=& \int d^{4}y \; tr \left( \Lambda \phi^{AB}\right)\;,   \\   Z_{m}^{AB}&=&\frac{1}{8\pi^{2}R_{5}}\int d^{4}y \;\partial^{l} tr\left( \epsilon_{mnkl} F^{nk}\phi^{AB}\right) =\frac{1}{8\pi^{2}R_{5}}\int d^{4}y \; tr\left( \epsilon_{mnkl} F^{nk}D^{l}\phi^{AB}\right) \;, \\  \nonumber Z_{mn}^{AB}&=&\frac{i}{8\pi^{2}R_{5}}\int d^{4}y \; \partial_{[m}tr\left( \phi_{N}^{A}D_{n]}\phi^{NB}\right) \\ &=&\frac{i}{8\pi^{2}R_{5}}\int d^{4}y \;  tr\left( D_{[m}\phi_{N}^{A}D_{n]}\phi^{NB}\right) -\frac{1}{8\pi^{2}R_{5}}\int d^{4}y \; tr\left( [\phi_{N}^{A},\phi^{NB}]F_{mn}\right) 
\end{eqnarray}
with $ \Lambda = \frac{\delta \mathcal{L}}{\delta A^{0}} $ the generator of the $ SU(N) $ transformation. 
\begin{equation}
  Z^{AB} =-Z^{BA}\;,\;\;\;  Z_{m}^{AB} =-Z_{m}^{BA}\;,\;\;\;   Z_{mn}^{AB} =Z_{mn}^{BA} \;,\;\;\;   \Omega_{AB} Z^{AB} =\Omega_{AB} Z_{m}^{AB} =0\;.
\end{equation}
We may define $Z^{AB}_{\alpha\dot{\alpha}}=\sigma^{m}_{\alpha\dot{\alpha}}Z^{AB}_{m}  $, $ Z_{\alpha\beta}^{AB}=\frac{1}{2}\sigma^{mn}_{\alpha\beta}Z^{AB}_{mn} $ and $ Z_{\dot{\alpha}\dot{\beta}}^{AB}=\frac{1}{2}\sigma^{mn}_{\dot{\alpha}\dot{\beta}}Z^{AB}_{mn} $ for later use. When acting on the fundamental fields, $ Z^{AB} $ generates a gauge transformation with the transformation parameter $ \phi^{AB} $. 
\begin{equation}\label{zab}
[Z^{AB},A_{\alpha\dot{\alpha}}]= i D_{\alpha\dot{\alpha}}  \phi^{AB}\;,\;\;\;\;\;\;[Z^{AB},\Phi]=-[\phi^{AB},\Phi]
\end{equation}
for the matter field $ \Phi $.

The supercharges are given by
\begin{equation}
Q^{A}=\frac{1}{4\pi^{2}R_{5}}\int d^{4}x \;tr\left( \frac{1}{2}F_{\mu\nu}\gamma^{\mu\nu} \gamma^{0} \lambda^{A}+\cdots\right) \;,
\end{equation}
where ``$ \ldots $" are terms without involving $F_{mn}  $. Under the action of $ I(x) $, from (\ref{1}), (\ref{cd}) and (\ref{10}), 
\begin{equation}
 I^{-1}Q^{A}I  = Q^{A}+\frac{1}{4\pi^{2}R_{5}}\int d^{4}y \;tr \left[ \frac{1}{4}f_{mn}\gamma^{mn}(1+\gamma^{5})  \gamma^{0} g\lambda^{A}g^{-1}\right] \;.
\end{equation}
For 
\begin{equation}       
( \lambda^{A})_{b}^{a}=\left[                
  \begin{array}{c}   
(\lambda_{\alpha}^{A})_{b}^{a} \\  
(\lambda^{A \dot{\alpha}})_{b}^{a}\\  
  \end{array}
\right]\;,
\end{equation}
where $ a,b=1,\cdots,N $ are $ SU(N) $ indices, with (\ref{mn}) plugged in, we have
\begin{eqnarray}
\nonumber   I^{-1}(x)Q_{\alpha}^{A}I(x)   &=&Q_{\alpha}^{A}-\frac{i}{8\pi^{2}R_{5}}\int d^{4}y \; (\sigma^{mn})^{\beta}_{\alpha}tr \left(f_{mn} g\lambda_{\beta}^{A}g^{-1}\right) \nonumber  \\ &=& Q_{\alpha}^{A}-I^{-1}(x)\left[ \lim_{\rho\rightarrow 0} \int d^{4}y \; \frac{2\rho^{2}}{\pi^{2}R_{5}(|y-x|^{2}+\rho^{2})^{2}}(\lambda_{\beta}^{A})^{\beta}_{\alpha}(y)
\right]I(x)\;,\\I^{-1}(x)Q^{A \dot{\alpha}} I(x)
&=&Q^{A \dot{\alpha}}\;.
\end{eqnarray}
So
\begin{equation}\label{k1}
[Q_{\alpha}^{A},I(x)]
=\frac{4}{R_{5}}(\lambda_{\beta}^{A})^{\beta}_{\alpha \;0}(x)I(x)
\;,\;\;\;\;\;\;\;\;[Q^{A \dot{\alpha}},I(x)]=0\;,
\end{equation}
where 
\begin{equation}\label{90909}
(\lambda_{\beta}^{A})^{\beta}_{\alpha \;0}(x)=-\lim_{\rho\rightarrow 0} \int d^{4}y \; \frac{\rho^{2}}{2\pi^{2}(|y-x|^{2}+\rho^{2})^{2}}(\lambda_{\beta}^{A})^{\beta}_{\alpha}(y)\propto (\lambda_{\beta}^{A})^{\beta}_{\alpha }(x)
\end{equation}
is the zero mode of $ \lambda^{A} $ on the small instanton background. The action of $  Q_{\alpha}^{A}$ on $ I(x) $ generates the zero mode $ (\lambda_{\beta}^{A})^{\beta}_{\alpha \;0} $, while the action of $Q^{A\dot{\alpha}}  $ makes $ I(x) $ invariant.

Actions of central charges and the momentum $P_{\alpha\dot{\alpha}} $ on $ I(x) $ are given by
\begin{eqnarray}
     &&[Z^{AB},I(x)]=0\;, \label{k3}  \\ &&\label{k331}  [Z^{AB}_{\alpha\dot{\alpha}},I(x)]
=-\frac{i}{2}(D_{\beta\dot{\alpha}}\phi^{AB})^{\beta}_{\alpha\;0}(x)I(x)\;, \\ && \label{k33a} [Z_{\alpha\beta}^{AB},I(x)]=\frac{i}{2}([\phi_{N}^{A},\phi^{NB}])_{\alpha\beta\;0}(x)I(x)\;,\;\;\;\;\;\;[Z_{\dot{\alpha}\dot{\beta}}^{AB},I]=0\;, \label{k33}\\ &&  [P_{\alpha\dot{\alpha}},I(x)]
=-\frac{i}{2}(F_{\beta\dot{\alpha}})^{\beta}_{\alpha\;0}(x)I(x)\;,
\end{eqnarray}
where
\begin{eqnarray}
 && (D_{\beta\dot{\alpha}}\phi^{AB})^{\beta}_{\alpha\;0}(x)  =\lim_{\rho\rightarrow 0} \int d^{4}y \;  \frac{2\rho^{2}}{\pi^{2}R_{5}(|y-x|^{2}+\rho^{2})^{2}}(D_{\beta\dot{\alpha}}\phi^{AB})^{\beta}_{\alpha}(y)\;,  \\&&([\phi_{N}^{A},\phi^{NB}])_{\alpha\beta\;0}(x) =  \lim_{\rho\rightarrow 0} \int d^{4}y \;  \frac{2\rho^{2}}{\pi^{2}R_{5}(|y-x|^{2}+\rho^{2})^{2}}([\phi_{N}^{A}(y),\phi^{NB}(y)])_{\alpha\beta}\;,\\&&(F_{\beta\dot{\alpha}})^{\beta}_{\alpha\;0}(x)=\lim_{\rho\rightarrow 0} \int d^{4}y \; \frac{2\rho^{2}}{\pi^{2}R_{5}(|y-x|^{2}+\rho^{2})^{2}}(F_{\beta\dot{\alpha}})^{\beta}_{\alpha}(y)\;,
\end{eqnarray}
$D_{\beta\dot{\alpha}}=\sigma^{m}_{\beta\dot{\alpha}} D_{m} $, $  F_{\beta\dot{\alpha}}=\sigma^{m}_{\beta\dot{\alpha}}F_{0m} $. $ I(x) $ commutes with $ Z^{AB} $ but carries the magnetic charge. From (\ref{k1}), (\ref{k4}), (\ref{k3}) and (\ref{k33}), (\ref{2.19}) is recovered in supersymmetric case.

When $ \mathcal{N} =1$, $ A,B=1,2 $, aside from the vector multiplet $ (\phi,\lambda^{A} ,A ) $, $ N_{f} $ hypermultiplets $ (q_{s}^{A},\psi_{s}) $ in the fundamental representation of $ SU(N) $ can also be added, where $ q_{s}^{A}$ and $ \psi_{s}$ are complex scalars and complex fermions. $ s=1,\cdots,N_{f} $. For the $ \mathcal{N}=1 $ supersymmetry, the superalgebra (\ref{kk4k})-(\ref{k4}) reduces to 
\begin{eqnarray}
     &&\label{qwabb}\{Q_{\alpha}^{A},Q_{\dot{\alpha}}^{B}\}=-2i\epsilon^{AB}(P_{\alpha\dot{\alpha}}+Z_{\alpha\dot{\alpha}})\equiv-2i\epsilon^{AB}K_{\alpha\dot{\alpha}}\;,   \\ &&\label{qwab}  \{Q_{\alpha}^{A},Q_{\beta}^{B}\}=2i\epsilon^{AB}\epsilon_{\alpha\beta}(P_{0}-P_{5}+Z)\equiv 2i\epsilon^{AB}\epsilon_{\alpha\beta}H\;, \\ &&\label{qwa}  \{Q_{\dot{\alpha}}^{A},Q_{\dot{\beta}}^{B}\}=2i\epsilon^{AB}\epsilon_{\dot{\alpha}\dot{\beta}}(P_{0}+P_{5}-Z)\equiv 2i\epsilon^{AB}\epsilon_{\dot{\alpha}\dot{\beta}}h\;.
\end{eqnarray}
The Jacobi identities give
\begin{equation}\label{ja}
[Q_{\alpha}^{A},H]=0\;,\;\;\;\;[Q_{\dot{\alpha}}^{A},h]=0\;,\;\;\;\;[Q_{\beta}^{A},K_{\alpha\dot{\alpha}}]=\epsilon_{\alpha\beta}[Q_{\dot{\alpha}}^{A},H]\;,\;\;\;\;[Q_{\dot{\beta}}^{A},K_{\alpha\dot{\alpha}}]=\epsilon_{\dot{\beta}\dot{\alpha}}[Q_{\alpha}^{A},h]\;.
\end{equation}
The nonvanishing commutators are electric and magnetic gauge transformations. The action of supercharges on fundamental fields of the $ 5d $ $ \mathcal{N}=1 $ theory is given by \cite{5a}
\begin{eqnarray}
 \label{56}    &&[Q_{\beta}^{A},  \phi ]= -\lambda^{A}_{\beta}
\;,\;\;\;\;\;\;\;
[Q_{\dot{\beta}}^{A}, \phi ]= -\lambda_{\dot{\beta}}^{A}\;,   \\ &&\label{333}  [Q_{\dot{\beta}}^{A}, A_{\alpha\dot{\alpha}}]=2\epsilon_{\dot{\beta}\dot{\alpha}}\lambda^{A}_{\alpha}\;,\;\;\;\;\;\;\;
[Q_{\beta}^{A}, A_{\alpha\dot{\alpha}}]=2\epsilon_{\beta\alpha}\lambda_{\dot{\alpha}}^{A}\;, \\  \label{566}   && \{Q_{\beta}^{B}, \lambda_{\alpha}^{A} \}=
\epsilon^{BA} (F_{\beta\alpha}+ \epsilon_{\alpha\beta} D_{0}\phi )-\epsilon_{\alpha\beta}D^{AB}\;,\;\;\;\;\;\{Q_{\dot{\alpha}}^{B},  \lambda_{\alpha}^{A} \}=\epsilon^{AB}(D_{\alpha\dot{\alpha}}\phi +F_{\alpha\dot{\alpha}} ) \;, \\  \label{59}&& \{Q_{\dot{\beta}}^{B},  \lambda_{\dot{\alpha}}^{A} \}=\epsilon^{BA}(F_{\dot{\alpha}\dot{\beta}} +\epsilon_{\dot{\alpha}\dot{\beta}} D_{0}\phi  )+\epsilon_{\dot{\alpha}\dot{\beta}} D^{AB}\;,\;\;\;\;\;\{Q_{\alpha}^{B}, \lambda_{\dot{\alpha}}^{A} \}=\epsilon^{BA}( D_{\alpha\dot{\alpha}}\phi -F_{\alpha\dot{\alpha}})\;,
\end{eqnarray}
and 
  \begin{eqnarray}
&& [Q_{\beta}^{A},  q^{B} ]= \sqrt{2}i \epsilon^{AB}\psi_{\beta}
\;,\;\;\;\;\;\;\;
 [Q_{\dot{\beta}}^{A},  q^{B} ]= \sqrt{2} i \epsilon^{AB} \psi_{\dot{\beta}} \;, \\ \label{from} &&  
\{Q_{\beta}^{A}, \psi_{\alpha} \}=
-\sqrt{2} (i D_{0}q^{A}  +\phi q^{A}  )    \epsilon_{\alpha\beta}
\;,\;\;\;\;\;\;\;
\{Q_{\dot{\alpha}}^{A},  \psi_{\alpha} \}=\sqrt{2}i D_{\alpha\dot{\alpha}}q^{A} \;,
\\ 
 &&
\{Q_{\dot{\beta}}^{A}, \psi_{\dot{\alpha}} \}=
\sqrt{2} (-i D_{0}q^{A}  +\phi q^{A}  )    \epsilon_{\dot{\alpha}\dot{\beta}}
,\;\;\;\;\;\;\;
\{Q_{\alpha}^{A}, \psi_{\dot{\alpha}} \}=
-\sqrt{2}i D_{\alpha\dot{\alpha}}q^{A}
\end{eqnarray}
with
\begin{equation}
\sigma_{\alpha\dot{\alpha}}^{m}A_{m}=A_{\alpha\dot{\alpha}}\;,\;\;\;\frac{1}{2}\sigma^{mn}_{\alpha\beta}F_{mn}=F_{\alpha\beta}\;,\;\;\;\frac{1}{2}\bar{\sigma}^{mn}_{\dot{\alpha}\dot{\beta}}F_{mn}=F_{\dot{\alpha}\dot{\beta}}\;. 
\end{equation}
For the $ \mathcal{N}=2 $ supersymmetry, (\ref{566}) and (\ref{59}) should be modified to 
\begin{eqnarray}
\label{59a}   && \{Q_{\beta}^{B}, \lambda_{\alpha}^{A} \}=
 \Omega^{BA}F_{\beta\alpha}+(D_{0}\phi^{BA}+\frac{i}{2}[\phi_{N}^{A},\phi^{NB}])\epsilon_{\alpha\beta},\;\{Q_{\dot{\alpha}}^{B},  \lambda_{\alpha}^{A} \}=D_{\alpha\dot{\alpha}}\phi^{AB} +\Omega^{AB}F_{\alpha\dot{\alpha}} , \\ && \{Q_{\dot{\beta}}^{B}, \lambda_{\dot{\alpha}}^{A} \}=
 \Omega^{BA}F_{\dot{\beta}\dot{\alpha}}+(D_{0}\phi^{BA}-\frac{i}{2}[\phi_{N}^{A},\phi^{NB}])\epsilon_{\dot{\alpha}\dot{\beta}},\;\{Q_{\alpha}^{B}, \lambda_{\dot{\alpha}}^{A} \}= D_{\alpha\dot{\alpha}}\phi^{BA} -\Omega^{BA}F_{\alpha\dot{\alpha}}.
\end{eqnarray}

\section{Gauge transformation of the instanton operator}\label{gau1}

The instanton operator $ I(x) $ constructed in section \ref{sec} is localized at $ x $ and has $ \rho=0 $. The remaining degeneracy comes the $ SU(N) $ gauge transformation with $ 4N-5 $ parameters. The behaviour of $ I(x) $ under the gauge transformation depends on whether the $ 5d $ theory contains the Chern-Simons term or not.

Let us start with the situation when the gauge field action is of the Yang-Mills type. In this case, the Gauss constraint is 
\begin{equation}
\Lambda=D_{m}\Pi^{m}+\rho=\partial_{m}\Pi^{m}-i[A_{m},\Pi^{m}]+\rho=0\;,
\end{equation}
where $ \Pi^{m} $ is the conjugate momentum of $ A_{m} $ and $ \rho $ is the charge density of the matter fields. Local gauge transformation operator $ U(\alpha) $ with the transformation parameter $ \alpha $ is given by 
\begin{equation}
U(\alpha) =\exp \left\lbrace -i \int d^{4}y \; tr [\alpha(y) \Lambda(y)]\right\rbrace \;,
\end{equation}
where $\alpha  $ is a Lie-algebra valued function well defined everywhere. The gauge transformation matrix is $ u=e^{-i\alpha} $, 
    \begin{equation}
U^{-1} A_{m}U =uA_{m}u^{-1}+iu\partial_{m}u^{-1}\;,\;\;\;\;\;\;\;\;\;  U^{-1}\Phi U= u\Phi u^{-1}\;.
 \end{equation}
$ \mathcal{G} (G)=\{U| \; \forall \; u \in G\}$. When $ G=SU(N) $, the transformation matrix of $ I(x) $ is $ g\oplus 1_{N-2} $ with $ g $ given by (\ref{2}). Under the $ SU(N) $ transformation, $ U^{-1}I(x)U $ is another singular gauge transformation operator with the transformation matrix $  u^{-1}(g\oplus 1_{N-2} ) u$, where $ u \in SU(N) $.

When the gauge field action has an additional Chern-Simons term
\begin{equation}
S_{CS}=\frac{\kappa}{24\pi^{2}}tr \int\left( F\wedge F\wedge A+\frac{i}{2}F\wedge A\wedge A\wedge A-\frac{1}{10}A\wedge A\wedge A\wedge A\wedge A\right) 
\end{equation}
at the level $ \kappa $, the gauge transformation operator $ U(\alpha) $ becomes
\begin{equation}
U(\alpha) =\exp \left\lbrace -i \int d^{4}y \; tr [\alpha(y) \Lambda'(y)]\right\rbrace 
\end{equation}
with
 \begin{equation}
\Lambda' =- \frac{\kappa}{32\pi^{2}} \epsilon^{klmn}F_{kl}F_{mn}+\Lambda\;.
 \end{equation}
Under the action of $ I $, 
 \begin{eqnarray}\label{4.7}
\nonumber     I^{-1}[\int d^{4}y \; tr(\alpha \Lambda')] I
&=&\int d^{4}y \;tr[(g^{-1} \oplus 1_{N-2})\alpha (g\oplus 1_{N-2})\Lambda' ]-\frac{\kappa}{16\pi^{2}}\int d^{4}y \;tr(\alpha  f^{mn}f_{mn} )\\ &=& \int d^{4}y \;tr[(g ^{-1}\oplus 1_{N-2}) \alpha (g\oplus 1_{N-2})\Lambda'  ]-\frac{\kappa}{2}tr [\alpha(x)(1_{2}\oplus 0_{N-2})]\;,
\end{eqnarray} 
where the $ f^{mn}gF_{mn} g^{-1} $ term is neglected, which gives $ 0 $ when acting on the regular state. For $ \alpha $ satisfying $[ (g \oplus 1_{N-2}) ,\alpha ]=0$, $ U(\alpha) $ will have the transformation matrix
\begin{equation}\label{8}
u=e^{-i\alpha}=(e^{i\theta(N-2)},e^{i\theta(N-2)},e^{-2i\theta},\ldots,e^{-2i\theta})\times (1_{2}\oplus w_{N-2} )
\end{equation}
with $w_{N-2} \in SU(N-2)$. For such $ \alpha $, from (\ref{4.7}), we have
  \begin{equation}\label{9}
U(\alpha)I(x)U^{-1}(\alpha) =\exp \{ \frac{i\kappa}{2}tr [\alpha(x)(1_{2}\oplus 0_{N-2})]\}I(x)=e^{-i(N-2)\kappa\theta(x)}I(x)\;.
 \end{equation}
In particular, for the $ U(1) $ group with the generator 
\begin{equation}
diag(N-2,N-2,-2,\ldots,-2)\;,
\end{equation}
$ I(x) $ has the $ U(1) $ charge $(N-2)\kappa  $. This is consistent with \cite{4d}, where the instanton operator in the radial quantization formalism also carries the $ U(1) $ charge in presence of the Chern-Simons term.

 \section{SO(4) rotation of the instanton operator}\label{rot1}

 $ I(x) $ also transforms under the $ SO(4) $ rotation in $ 4d $ space. However, just as the situation for instantons, the space rotation acts as a gauge transformation without giving rise to the further degeneracy \cite{7,7as}.

$ SO(4) $ can be decomposed as $ SO(4)\cong SU(2)_{L}\times SU(2)_{R} $ with $SU(2)_{L}$ and $ SU(2)_{R}  $ acting on the $ \alpha $ and $ \dot{\alpha} $ spinor indices, respectively. When $G=SU(2)  $, from (\ref{2}), under an $ SO(4) $ rotation $ S(x) $ centered at $ x $, 
 \begin{equation}
I(x)\rightarrow I'(x)=S(x)I(x)S^{-1}(x) \;,
 \end{equation}
\begin{equation}
  g(x,y)  \rightarrow   g'(x,y)= \frac{s_{1}\sigma_{m}s^{-1}_{2}(y^{m}-x^{m})}{|y-x|} =s_{1}g(x,y)s^{-1}_{2}
 \end{equation}
with $ s_{1}  \in SU(2)_{L}$, $  s_{2} \in SU(2)_{R}$. So
 \begin{equation}\label{4}
S(x)I(x)S^{-1}(x) =V_{1}I (x)V^{-1}_{2}\;,
 \end{equation}
where $V_{1} $ and $ V_{2} $ are global $ SU(2) $  transformations with the transformation matrices $s_{1} $ and $ s_{2} $, respectively. Generically, for $G= SU(N) $, $ g \rightarrow g  \oplus 1_{N-2}$, $V_{1} $ and $ V_{2} $ are global $ SU(N) $  transformations with the transformation matrices $s_{1} \oplus 1_{N-2}$ and $ s_{2} \oplus 1_{N-2}$, respectively.

In (\ref{mn}), $ a_{n} $ and $ f_{mn} $ are only affected by $ V_{1} $. When acting on the vacuum $ \vert \Omega \rangle $, $ I(x) $ generates a chiral state $ I(x)  \vert \Omega \rangle $ also transforming under $ SU(2)_{L} $:
  \begin{equation}\label{4.9}
S(x)I(x) \vert \Omega \rangle=V_{1}I (x) \vert \Omega \rangle\;.
 \end{equation}

\section{Instanton states with the definite spin and the electric charge}\label{spec1}

With the $ SU(N) $ gauge orientation taken into account, the Hilbert space of BPS instanton states at $ x $ is spanned by $ \{UI(x)\vert \Omega \rangle|\;\forall \; U \in \mathcal{G}[SU(N)]\} $ composing a $ 4N-5 $ dimensional manifold. It is more useful to select another set of bases with the definite spin and the electric charge. In the following, we will consider three situations: the $ SU(2) $ gauge theory with no hypermultiplet, the bosonic $ SU(N) $ gauge theory for $ N\geq 3 $ and the supersymmetric $ SU(N) $ gauge theory for $ N\geq 3 $. We will study the $ SU(2) $ gauge theory with $ N_{f} $ hypermultiplets in section \ref{Nf}.

 \subsection{SU(2)}\label{SU}

When $ G=SU(2) $, $ UIU^{-1} =I$ only for $ U $ with the transformation matrix $ u=\pm 1_{2} $, so $  \{UI(x)\vert \Omega \rangle|\;\forall \; U \in \mathcal{G}[SU(2)]\}  $ compose the space $ S^{3} /\mathbb{Z}^{2}$. $ \forall \; u \in SU(2) $, the fundamental representation of $ u $ is given by $ D_{\alpha}^{a} [u]$, $ \alpha,a=1,2 $. $\{ D_{\alpha_{1}}^{a_{1}}[u]\cdots  D_{\alpha_{n}}^{a_{n}}[u] \; |\;n=0,1,\cdots\}$ is a set of complete bases on the $ SU(2) $ group manifold. $D_{\alpha}^{a} [u^{-1}]  $ is related to $D_{\alpha}^{a} [u]  $ via $ \epsilon^{pq}D_{q}^{b}[u]\epsilon_{ba}=D_{a}^{p}[u^{-1}] $ and thus is not taken into account. Consider  
   \begin{equation}\label{3}
\mathcal{I}_{\alpha_{1}\cdots \alpha_{n}}^{a_{1}\cdots a_{n}}(x)=\int DU\; D_{\alpha_{1}}^{a_{1}}[u(x)]\cdots  D_{\alpha_{n}}^{a_{n}}[u(x)]  UI(x)U^{-1}\;. 
 \end{equation}
Under the gauge transformation $ V $ with the transformation matrix $ v(y) $, 
\begin{eqnarray}
\nonumber V\mathcal{I}_{\alpha_{1}\cdots \alpha_{n}}^{a_{1}\cdots a_{n}}(x)V^{-1} &=&\int DU\; D_{\alpha_{1}}^{a_{1}}[u(x)]\cdots  D_{\alpha_{n}}^{a_{n}}[u(x)] V UI(x)U^{-1}V^{-1}
\\  &=&D_{b_{1}}^{a_{1}}[v^{-1}(x)]\cdots  D_{b_{n}}^{a_{n}}[v^{-1}(x)]
\mathcal{I}_{\alpha_{1}\cdots \alpha_{n}}^{b_{1}\cdots b_{n}}(x)\;.
\end{eqnarray}
$ \mathcal{I}_{\alpha_{1}\cdots \alpha_{n}}^{a_{1}\cdots a_{n}}(x) $ transforms as a local operator in the $ \mathbf{2}^{n}$-representation of $ SU(2) $.

Under the $ SO(4) $ rotation centered at $ x $, from (\ref{4}),
\begin{eqnarray}
\nonumber S\mathcal{I}_{\alpha_{1}\cdots \alpha_{n}}^{a_{1}\cdots a_{n}}(x)S^{-1} &=&\int DU\; D_{\alpha_{1}}^{a_{1}}[u(x)]\cdots  D_{\alpha_{n}}^{a_{n}}[u(x)]  USI(x)S^{-1}U^{-1}\\  \nonumber&=& \int DU\; D_{\alpha_{1}}^{a_{1}}[u(x)]\cdots  D_{\alpha_{n}}^{a_{n}}[u(x)]  UV_{1}I(x)V_{2}^{-1}U^{-1}
\\\nonumber  &=& D_{\alpha_{1}}^{\beta_{1}}[s_{1}^{-1}]\cdots  D_{\alpha_{n}}^{\beta_{n}}[s_{1}^{-1}]
\int DU\; D_{\beta_{1}}^{a_{1}}[u(x)]\cdots  D_{\beta_{n}}^{a_{n}}[u(x)]  UI(x)V_{2}^{-1}V_{1}U^{-1}\;.\\
\end{eqnarray}
When acting on the vacuum,
\begin{equation}
S\mathcal{I}_{\alpha_{1}\cdots \alpha_{n}}^{a_{1}\cdots a_{n}}(x)\vert \Omega \rangle =D_{\alpha_{1}}^{\beta_{1}}[s_{1}^{-1}]\cdots  D_{\alpha_{n}}^{\beta_{n}}[s_{1}^{-1}]  \mathcal{I}_{\beta_{1}\cdots \beta_{n}}^{a_{1}\cdots a_{n}}(x)\vert \Omega \rangle\;.
\end{equation}
$ \mathcal{I}_{\alpha_{1}\cdots \alpha_{n}}^{a_{1}\cdots a_{n}}(x)\vert \Omega \rangle $ is in the $ \bar{\mathbf{2}}^{n} $ representation of $ SU(2)_{L} $.

For $ W $ in the stability group with the transformation matrix $w=-1_{2} $, $WI(x)W^{-1}= I(x) $, so 
\begin{equation}\label{4.3}
 \mathcal{I}_{\alpha_{1}\cdots \alpha_{n}}^{a_{1}\cdots a_{n}}(x) 
= \int DU\; D_{\alpha_{1}}^{a_{1}}[u(x)]\cdots  D_{\alpha_{n}}^{a_{n}}[u(x)]  UWI(x)W^{-1}U^{-1}= (-1)^{n}\mathcal{I}_{\alpha_{1}\cdots \alpha_{n}}^{a_{1}\cdots a_{n}}(x)\;.
\end{equation}
$ \mathcal{I}_{\alpha_{1}\cdots \alpha_{n}}^{a_{1}\cdots a_{n}}(x)\neq 0 $ only for the even $ n $.

$ \mathcal{I}_{\alpha_{1}\cdots \alpha_{n}}^{a_{1}\cdots a_{n}}  $ can be further decomposed into $ \mathcal{I}_{\{\alpha_{1}\cdots \alpha_{k}\}}^{a_{1}\cdots a_{k}}  $ in the $ (\frac{k}{2},\frac{k}{2}) $ irreducible representation of $ SU(2)_{L} \times SU(2) $ for $ k=0,2,\cdots,n $. Therefore, the BPS spectrum of instanton states at $ x $ is 
\begin{equation}
\left\lbrace  \mathcal{I}_{\{\alpha_{1}\cdots \alpha_{2J}\}}^{a_{1}\cdots a_{2J}}(x)  \vert \Omega \rangle|J=0,1,\cdots\right\rbrace  \;,
\end{equation}
composing the complete orthogonal bases on $S^{3} /\mathbb{Z}_{2}  $.

\subsection{$SU(N)   $ without the fermionic fields}

When the gauge group is $ SU(N) $ with $ N\geq 3 $, it is possible to add a level-$ \kappa $ Chern-Simons term. We will first consider the theory with no fermionic fields. In analogy with the $ SU(2) $ case, one can construct 
\begin{equation}\label{a1}
\mathcal{I}^{b_{1}\cdots b_{m},p_{1}\cdots p_{n}}_{q_{1}\cdots q_{m},a_{1}\cdots a_{n}}(x)= \int DU\;D_{q_{1}}^{b_{1}}[u(x)]\cdots D_{q_{m}}^{b_{m}}[u(x)]D_{a_{1}}^{p_{1}}[u^{-1}(x)]\cdots D_{a_{n}}^{p_{n}}[u^{-1}(x)] UI(x)U^{-1}\;,
\end{equation}
where $ D_{q}^{b} [u]$ and $ D_{a}^{p} [u^{-1}] $ are fundamental representations of $ u $ and $ u^{-1} $, $ a,b,p,q=1,2,\cdots,N $. $  D_{a}^{p} [u^{-1}] $ is also kept although it can be written in terms of $    D_{a}^{p} [u]$.

Similar to the discussion around (\ref{4.3}), some $ \mathcal{I}^{b_{1}\cdots b_{m},p_{1}\cdots p_{n}}_{q_{1}\cdots q_{m},a_{1}\cdots a_{n}}$ in (\ref{a1}) can be $ 0 $. The transformation matrix of $ I $ is $g\oplus 1_{N-2}  $ with the stability group $ U(1) \times SU(N-2) $. The $ U(1) $ part has the generator 
\begin{equation}
diag(N-2,N-2,-2,-2,\ldots,-2)\;.
\end{equation}
The element of $   U(1) \times SU(N-2)  $ can be written as 
\begin{equation}
w=(e^{i\theta(N-2)},e^{i\theta(N-2)},e^{-2i\theta},\ldots,e^{-2i\theta})\times (1_{2}\oplus w_{N-2} )
\end{equation}
with $w_{N-2} \in SU(N-2)$. $w^{-1}(g\oplus 1_{N-2} )w=g\oplus 1_{N-2}   $. Suppose the Chern-Simons level is $ \kappa $, for $ W $ with the transformation matrix $ w $, from (\ref{8}) and (\ref{9}), 
\begin{equation}\label{iop}
WI(x)W^{-1}=e^{-i(N-2)\kappa\theta}I(x) \;,
\end{equation}
so $ \forall \; w \in  U(1) \times SU(N-2)$, the identity 
\begin{eqnarray}
\nonumber \mathcal{I}^{b_{1}\cdots b_{m},p_{1}\cdots p_{n}}_{q_{1}\cdots q_{m},a_{1}\cdots a_{n}}(x) &=& e^{i\kappa\theta(N-2)} \int DU\;D_{q_{1}}^{b_{1}}[u(x)w^{-1}]\cdots D_{q_{m}}^{b_{m}}[u(x)w^{-1}] \\  &&D_{a_{1}}^{p_{1}}[wu^{-1}(x)]\cdots D_{a_{n}}^{p_{n}}[wu^{-1}(x)] UI(x)U^{-1}
\end{eqnarray}
should hold. $  \mathcal{I}^{b_{1}\cdots b_{m},p_{1}\cdots p_{n}}_{q_{1}\cdots q_{m},a_{1}\cdots a_{n}}\neq 0 $ only when $ p_{i} ,q_{i}=1,2$ and 
\begin{equation}\label{mnk}
m-n=\kappa\;.
\end{equation}
Further properties of $  \mathcal{I}^{b_{1}\cdots b_{m},p_{1}\cdots p_{n}}_{q_{1}\cdots q_{m},a_{1}\cdots a_{n}}$ with $ p_{i} ,q_{i}=1,2 $ will be investigated in subsection \ref{Fe}.

 \subsection{$SU(N)   $ with the fermionic fields}\label{Fe}

Now consider the $ 5d $ $ SU(N) $ gauge theories with the charged fermionic fields. On the $ q $-instanton background, there will be $ n_{F}   $ normalizable fermionic zero modes. Let $ u,v =1,\cdots,N $, for the fermion $( \lambda^{\alpha} )_{v}^{u}$, $ ( \lambda_{\dot{\alpha}} )_{v}^{u} $ in the adjoint representation of $ SU(N) $, $ n_{F}=2N|q| $; for the fermion $( \psi^{\alpha} )^{u}$, $ ( \psi_{\dot{\alpha}} )^{u} $ in the fundamental representation of $ SU(N) $, $ n_{F}=|q| $. Concretely, on the $1$-instanton background and in the regular gauge, the adjoint fermionic fields $( \lambda^{\alpha} )_{v}^{u}$, $ ( \lambda_{\dot{\alpha}} )_{v}^{u} $ can be expanded as 
\begin{eqnarray}\label{zero}
\nonumber && 
(\lambda^{\alpha})^{p}_{q}(y)=-(F^{\alpha}_{\beta})^{p}_{q}(y)
\xi^{\beta}+\sigma^{\beta\dot{\beta}}_{\nu}y^{\nu}
(F^{\alpha}_{\beta})^{p}_{q}(y)\bar{\eta}_{\dot{\beta}}+\cdots\;,\;\;\;\;\;\;\; (\lambda^{\alpha})^{a}_{b}(y)=\cdots\;,\;\;\;\;\;\;\;   ( \lambda_{\dot{\alpha}} )_{v}^{u}(y) =\cdots\;,\\     \nonumber&& (\lambda^{\alpha})_{q}^{ a}(y)=\frac{\rho}{(|x-y|^{2}+\rho^{2})^{3/2}}\delta^{\alpha}_{q}\mu^{a}+\cdots\;,\;\;\;\;\;\;\;(\lambda^{\alpha})_{b}^{p}(y)=\frac{\rho}{(|x-y|^{2}+\rho^{2})^{3/2}}\epsilon^{\alpha p}\bar{\mu}_{b}+\cdots\;,\\
\end{eqnarray}
where $ p,q=1,2 $, $ a,b=3,\cdots,N $, $ F^{\alpha}_{\beta} $ is the field strength of the instanton, $ x $ and $ \rho $ are the position and the scale size of the instanton, $( \xi^{\beta} ,\bar{\eta}_{\dot{\beta}}  ,\mu^{a}  ,\bar{\mu}_{a} ) $ are $ 2N $ zero modes, and ``$ \cdots $" represents the nonzero modes. $ \xi^{\beta} $, $ \bar{\eta}_{\dot{\beta}} $, $ \mu^{a}  $ and $ \bar{\mu}_{a} $ have the conformal dimensions $ -1/2 $, $1/2  $, $-1/2  $ and $  -1/2$, respectively. The fermionic fields $( \psi^{\alpha} )^{u}$, $ ( \psi_{\dot{\alpha}} )^{u} $ in the fundamental representation can be expanded as 
\begin{equation}\label{zeroo}
(\psi^{\alpha})^{p}(y)=\frac{\rho}{(|y-x|^{2}+\rho^{2})^{3/2}}\epsilon^{\alpha p}K+\cdots\;,\;\;\;\;\;\;\; (\psi^{\alpha})^{a}(y)=\cdots\;,\;\;\;\;\;\;\;  ( \psi_{\dot{\alpha}} )^{u} (y)=\cdots\;.
\end{equation}
$ K $ has the conformal dimension $ 0 $.

When $ \rho\rightarrow 0 $, zero modes should still persist but become local. From (\ref{zero}) and (\ref{zeroo}), the zero mode operators on the small instanton background can be solved as 
\begin{eqnarray}
 && 
\xi_{\alpha}=-\frac{1}{2\pi^{2}}\lim_{\rho\rightarrow 0} \int d^{4}y \; \frac{\rho^{2}}{(|y-x|^{2}+\rho^{2})^{2}}(\lambda_{\beta})_{\alpha}^{\beta}(y) \propto  (\lambda_{\beta})_{\alpha}^{\beta}(x)  \;,\label{909}\\    && \mu^{a}=\frac{1}{\pi^{2}}\lim_{\rho\rightarrow 0}\int d^{4}y \; \frac{\rho}{(|y-x|^{2}+\rho^{2})^{3/2}}(\lambda^{\beta})^{a}_{\beta }(y) \propto   (\lambda^{\beta})^{a}_{\beta }(x) \;,\\     &&\label{909k}\bar{\mu}_{a}=\frac{1}{\pi^{2}}\lim_{\rho\rightarrow 0}\int d^{4}y \; \frac{\rho}{(|y-x|^{2}+\rho^{2})^{3/2}}(\lambda_{\beta})^{\beta}_{ a}(y)\propto (\lambda_{\beta})^{\beta}_{ a}(x) \;,\\     &&K=\frac{1}{\pi^{2}}\lim_{\rho\rightarrow 0}\int d^{4}y \; \frac{\rho}{(|y-x|^{2}+\rho^{2})^{3/2}}(\psi_{\beta})^{\beta}(y)\propto (\psi_{\beta})^{\beta}(x)
\;.
\end{eqnarray}
(\ref{909}) is just the zero mode in (\ref{90909}) generated by the supercharge.

For a $ U(1) $ charge operator $ Q $ with $ Q \vert \Omega \rangle=0  $, on an instanton background, there will be quantum corrections $ \Delta  Q$ to $Q  $ encoded in the normal ordering constant of the fermionic zero modes \cite{8as}. The action of $ I(x) $ creates the small instanton background with $ q=1 $, so we will have $I^{-1}(x) QI(x) \vert \Omega \rangle= \Delta Q \vert \Omega \rangle $, or equivalently, $ QI(x) \vert \Omega \rangle= \Delta QI(x) \vert \Omega \rangle $. The instanton operator $ I(x) $ with the transformation matrix $g\oplus 1_{N-2}  $ breaks the $ SU(N) $ group to $ U(1) \times SU(N-2) $. The generator of the $ U(1) $ group is
\begin{equation}
diag(N-2,N-2,-2,\ldots,-2)\;.
\end{equation}
$ I(x) $ is invariant under $ SU(N-2) $ but may carry the $ U(1) $ charge in presence of the fermionic fields or the Chern-Simons term.

Consider a $ 5d $ $ \mathcal{N}=1 $ gauge theory with the Chern-Simons level $ \kappa $. The theory contains a symplectic-Majorana fermion $ \lambda^{A} $ in the adjoint representation of $ SU(N) $ and $ N_{f} $ complex fermions $ \psi_{s} $ in the fundamental representation of $ SU(N) $. $\bar{\lambda}_{A}=(\lambda^{T})^{B}\epsilon_{BA}C  $, $ A,B=1,2 $, $ s=1,\cdots,N_{f} $. $ U(1) $ charges of the fermionic zero modes are given by 
\begin{equation}
\begin{tabular}{|l|c|c|c|c|c|}
     \hline
       $\text{Zero mode} $  & $\xi^{A}_{\alpha} $   &$ \mu^{Aa}$ & $\bar{\mu}^{A}_{a} $& $K_{ s} $               &  $\bar{K}_{s} $         \\\hline
            $Q_{U(1)}$   &$\;\;\;\;0\;\;\;\;$ & $\;\;-N\;\;$& $\;\;\;N\;\;\;$ & $N-2 $ &  $2-N$  \\
            \hline
   \end{tabular}
\end{equation}
\noindent  The vacuum annihilated by the different combinations of the complete anti-commutative fermionic zero modes carries the different $ U(1) $ charges. With (\ref{iop}) taken into account, $\Delta Q_{U(1)}$ carried by the different $  I(x) \vert \Omega \rangle $ are given by 
\begin{equation}\label{tab}
\begin{tabular}{|l|c|}
     \hline
       $A=1,2\; ;\;  a=3,\cdots,N \;;\; s=1,\cdots,N_{f} $         &  $\Delta Q_{U(1)}$         \\\hline
            $\bar{\mu}^{A}_{a}   I(x) \vert \Omega \rangle =0\;,\;\; \bar{K}_{s}  I(x) \vert \Omega \rangle =0$               & $ (N-2)(\kappa+ N-\frac{1}{2}N_{f})  $   \\
       $ \mu^{Aa} I(x) \vert \Omega \rangle =0\;,\;\; \bar{K}_{s}  I(x) \vert \Omega \rangle =0$               & $(N-2)(\kappa-N-\frac{1}{2}N_{f})   $      \\
          $\bar{\mu}^{A}_{a}   I(x) \vert \Omega \rangle =0\;,\;\;K_{ s} I(x) \vert \Omega \rangle =0$               & $ (N-2)(\kappa+N+\frac{1}{2}N_{f}) $          \\
         $ \mu^{Aa}  I(x) \vert \Omega \rangle =0\;,\;\;K_{ s}     I(x) \vert \Omega \rangle =0$               & $  (N-2)(\kappa-N+\frac{1}{2}N_{f})   $        \\
            \hline
   \end{tabular}
\end{equation}
All these states are $ Sp(1)_{R} $ singlets. On the other hand, $ I(x) \vert \Omega \rangle $ satisfying 
\begin{equation}
\mu^{1a}  I(x) \vert \Omega \rangle =\bar{\mu}_{a}^{1}  I(x) \vert \Omega \rangle =\bar{K}_{s}  I(x) \vert \Omega \rangle =0
\end{equation}
carries the $ U(1) $ charge $  (N-2)(\kappa-\frac{1}{2}N_{f})    $ as well as the $ Sp(1)_{R} $ charge $ (N-2) $.

For a $ 5d $ $ \mathcal{N}=2 $ gauge theory containing the symplectic-Majorana fermion $ \lambda^{A} $ in the adjoint representation of $ SU(N) $ with $\bar{\lambda}_{A}=(\lambda^{T})^{B}\Omega_{BA}C  $, $A,B=1,2,3,4  $, $ I(x) \vert \Omega \rangle $ satisfying $\bar{\mu}^{A}_{ a}  I(x) \vert \Omega \rangle =0 $ carries the $ U(1) $ charge $ 2N(N-2)  $; $ I(x) \vert \Omega \rangle $ satisfying $ \mu^{Aa}  I(x) \vert \Omega \rangle =0 $ carries the $ U(1) $ charge $-2N (N-2)  $; $ I(x) \vert \Omega \rangle $ satisfying 
\begin{equation}
\mu^{1a}   I(x) \vert \Omega \rangle =\bar{\mu}^{1}_{ a}  I(x) \vert \Omega \rangle=\mu^{3a}   I(x) \vert \Omega \rangle =\bar{\mu}^{3}_{ a} I(x) \vert \Omega \rangle=0
\end{equation}
carries the $ U(1) $ charge $0 $ but has the $ Sp(2)_{R} $ charge $ (N-2,N-2) $.

In presence of the fermionic fields, instead of (\ref{mnk}), the instanton operator in the $ \mathbf{N}^{m} \times \bar{ \mathbf{N}}^{n} $ representation of $ SU(N) $ exists only when 
\begin{equation}\label{mnn}
m-n=\frac{\Delta Q_{U(1)}}{N-2}\;,
\end{equation}
where $ \Delta Q_{U(1)} $ is the $ U(1) $ charge of $ I(x)\vert \Omega\rangle $. When (\ref{mnn}) is satisfied, such operator can be constructed as 
\begin{equation}\label{6.25}
\mathcal{I}^{b_{1}\cdots b_{m},p_{1}\cdots p_{n}}_{q_{1}\cdots q_{m},a_{1}\cdots a_{n}}(x)= \int DU\;D_{q_{1}}^{b_{1}}[u(x)]\cdots D_{q_{m}}^{b_{m}}[u(x)]D_{a_{1}}^{p_{1}}[u^{-1}(x)]\cdots D_{a_{n}}^{p_{n}}[u^{-1}(x)] UI(x)U^{-1}\;,
\end{equation}
where $ a_{i} ,b_{i}=1,2,\ldots,N$, $ p_{i},q_{i} =1,2$. From (\ref{6.25}), 
\begin{equation}
\mathcal{I}^{b_{1}\cdots b_{i}\cdots b_{j}\cdots b_{m},p_{1}\cdots p_{n}}_{q_{1} \cdots q_{i}\cdots q_{j} \cdots q_{m},a_{1}\cdots a_{n}}=\mathcal{I}^{b_{1}\cdots b_{j}\cdots b_{i}\cdots b_{m},p_{1}\cdots p_{n}}_{q_{1} \cdots q_{j}\cdots q_{i} \cdots q_{m},a_{1}\cdots a_{n}}\;,\;\;\;\;\;\;\;\mathcal{I}^{b_{1} \cdots b_{m},p_{1}\cdots p_{i}\cdots p_{j}\cdots p_{n}}_{q_{1}  \cdots q_{m},a_{1}\cdots a_{i}\cdots a_{j}\cdots a_{n}}=\mathcal{I}^{b_{1} \cdots b_{m},p_{1}\cdots p_{j}\cdots p_{i}\cdots p_{n}}_{q_{1}  \cdots q_{m},a_{1}\cdots a_{j}\cdots a_{i}\cdots a_{n}}\;.
\end{equation}
Under the gauge transformation $ V $,
\begin{equation}
V \mathcal{I}^{b_{1}\cdots b_{m},p_{1}\cdots p_{n}}_{q_{1}\cdots q_{m},a_{1}\cdots a_{n}}(x)V^{-1}=D_{c_{1}}^{b_{1}}[v^{-1}(x)]\cdots D_{c_{m}}^{b_{m}}[v^{-1}(x)]D_{a_{1}}^{d_{1}}[v(x)]\cdots D_{a_{n}}^{d_{n}}[v(x)]\mathcal{I}^{c_{1}\cdots c_{m},p_{1}\cdots p_{n}}_{q_{1}\cdots q_{m},d_{1}\cdots d_{n}}(x)\;.
\end{equation}
Under the action of the $ SO(4)\cong SU(2)_{L}\times SU(2)_{R} $ rotation centered at $ x $, 
\begin{equation}
S\mathcal{I}^{b_{1}\cdots b_{m},p_{1}\cdots p_{n}}_{q_{1}\cdots q_{m},a_{1}\cdots a_{n}}(x)\vert \Omega \rangle=D_{q_{1}}^{t_{1}}[s^{-1}_{1}]\cdots D_{q_{m}}^{t_{m}}[s^{-1}_{1}]
D_{r_{1}}^{p_{1}}[s_{1}]\cdots D_{r_{n}}^{p_{n}}[s_{1}]\mathcal{I}^{b_{1}\cdots b_{m},r_{1}\cdots r_{n}}_{t_{1}\cdots t_{m},a_{1}\cdots a_{n}}(x)\vert \Omega \rangle\;.
\end{equation}
$ \mathcal{I}^{b_{1}\cdots b_{m},p_{1}\cdots p_{n}}_{q_{1}\cdots q_{m},a_{1}\cdots a_{n}}(x)\vert \Omega \rangle $ is a chiral state in the $ \mathbf{2}^{n} \times \bar{ \mathbf{2}}^{m} $ representation of $ SU(2)_{L} $ and the $ \mathbf{N}^{m} \times \bar{ \mathbf{N}}^{n} $ representation of $ SU(N) $. $ \mathcal{I}^{b_{1}\cdots b_{m},p_{1}\cdots p_{n}}_{q_{1}\cdots q_{m},a_{1}\cdots a_{n}}(x)\vert \Omega \rangle $ can be further decomposed into the irreducible representations of $SU(2)_{L}   $ with the spin taking values in $ (\frac{1}{2}, \frac{3}{2},\cdots,\frac{m+n}{2})$ for the odd $ m+n $, and $ (0, 1,\cdots,\frac{m+n}{2})$ for the even $ m+n $.

Now consider the $ 5d $ $ \mathcal{N} =1$ $ SU(N) $ gauge theory with $ N_{f} =\kappa=0 $. From $ I(x) \vert \Omega \rangle $ satisfying
\begin{equation}
\mu^{Aa}  I(x) \vert \Omega \rangle =0\;,\;\;\;\;A=1,2\; ;\;  a=3,\cdots,N \;,
\end{equation}
we may construct $ \mathcal{I}^{p_{1}\cdots p_{N}}_{a_{1}\cdots a_{N}}(x) $ in the $\bar{ \mathbf{N}}^{N}$ representation of $ SU(N) $. The minimal gauge invariant operator built from $  \mathcal{I}^{p_{1}\cdots p_{N}}_{a_{1}\cdots a_{N}}(x) $ is a scalar
\begin{equation}
\tilde{\Phi}^{A_{3}\cdots A_{N}}=\epsilon^{a_{1} a_{2}a_{3}\cdots a_{N}}(\lambda^{A_{3}}_{p_{3}})^{b_{3}}_{a_{3}} \cdots  (\lambda_{p_{N}}^{A_{N}})^{b_{N}}_{a_{N}}\mathcal{I}^{12p_{3}\cdots p_{N}}_{a_{1}a_{2}b_{3}\cdots b_{N}} \;.
\end{equation}
In fact, since $   I(x) \vert \Omega \rangle$ carries the $ U(1) $ charge $ -N(N-2)$, the $ U(1) \times SU(N-2) $ singlet must be obtained via the action of $ N-2 $ zero modes $\bar{\mu}^{A}_{ a}(x)  $, i.e.
\begin{equation}
\epsilon^{12a_{3}\cdots a_{N}}
\bar{\mu}^{A_{3}}_{a_{3}}\cdots \bar{\mu}^{A_{N}}_{a_{N}}   I(x)\vert \Omega\rangle\;.
\end{equation}
The integration over the $SU(N)  $ transformation gives the gauge singlet
\begin{eqnarray}\label{sun}
\nonumber && 
\Phi^{A_{3}\cdots A_{N}} (x)\\     \nonumber&= &  \int DU\; \epsilon^{12a_{3}\cdots a_{N}}
  U   \bar{\mu}^{A_{3}}_{a_{3}}\cdots \bar{\mu}^{A_{N}}_{a_{N}}           I(x)U^{-1}\\     \nonumber&\propto &    (\lambda_{p_{3}}^{A_{3}})^{c_{3}}_{b_{3}}(x) \cdots  (\lambda_{p_{N}}^{A_{N}})^{c_{N}}_{b_{N}} (x)\\     \nonumber&&\int DU\; \epsilon^{12 a_{3}\cdots a_{N}} D^{p_{3}}_{c_{3}}[u^{-1}(x)] \cdots     D_{c_{N}}^{p_{N}}[u^{-1}(x)]
D^{b_{3}}_{a_{3}}[u(x)]  \cdots D^{b_{N}}_{a_{N}}[u(x)]   UI(x)U^{-1} \\     \nonumber&=& \epsilon^{b_{1} b_{2}b_{3}\cdots b_{N}}(\lambda_{p_{3}}^{A_{3}})^{c_{3}}_{b_{3}}(x) \cdots  (\lambda_{p_{N}}^{A_{N}})^{c_{N}}_{b_{N}} (x)\\     \nonumber&&\int DU\;
D^{1}_{b_{1}}[u^{-1}(x)]   D^{2}_{b_{2}}[u^{-1}(x)]    D^{p_{3}}_{c_{3}}[u^{-1}(x)] \cdots     D_{c_{N}}^{p_{N}}[u^{-1}(x)]UI(x)U^{-1}\\     &=&\tilde{\Phi}^{A_{3}\cdots A_{N}} (x)\;,
\end{eqnarray}
where we have used (\ref{909k}).

$ \Phi^{A_{3}\cdots A_{N}}=\Phi^{\{A_{3}\cdots A_{N}\}} $ is an $ (N-1) $-dimensional totally symmetric representation of $ Sp(1)_{R} $. The highest weight state can be selected as $ \Phi^{1\cdots 1} $. The action of $ Q_{1}^{1}Q_{2}^{1} $ on $  \Phi^{1\cdots 1}  $ gives
\begin{eqnarray}
\nonumber \rho^{111\cdots 1}&=& 
Q_{1}^{1}Q_{2}^{1}\Phi^{1\cdots 1}(x)\\   \nonumber&=&\frac{16}{R^{2}_{5}N(N-1)} \int DU\; \epsilon^{a_{1}\cdots a_{N}}
U \bar{\mu}^{1}_{a_{1}}\cdots \bar{\mu}^{1}_{a_{N}} I(x)U^{-1}\\     \nonumber&\propto&\epsilon^{a_{1} \cdots a_{N}}(\lambda_{p_{1}}^{1})_{a_{1}}^{b_{1}}(x) \cdots  (\lambda_{p_{N}}^{1})_{a_{N}}^{b_{N}}(x)\mathcal{I}^{p_{1}\cdots p_{N}}_{b_{1}\cdots b_{N}} (x)\;.
\end{eqnarray}
From (\ref{k1}), (\ref{qwabb}), (\ref{qwab}) and (\ref{566}), we have
\begin{equation}
Q_{\alpha}^{1}\rho^{111\cdots 1}=Q_{\dot{\alpha}}^{1}\rho^{111\cdots 1}=0\;.
\end{equation}
$ \rho^{111\cdots 1} $ is a $ 1/2 $ BPS scalar with the $ Sp(1)_{R} $ charge $ \frac{N}{2} $. Successive actions of $Q_{\alpha}^{2}  $ and $ Q_{\dot{\alpha}}^{2} $ on $ \rho^{111\cdots 1}  $ generate a supermultiplet with $ 16 $ states. In the radial quantization formalism, when the symmetry is not enhanced, the existence of such short supermultiplet is also addressed in \cite{Hoo3}.

When $ N=2 $, the symmetry enhancement is possible. From 
\begin{equation}\label{6.40}
\mathcal{I}(x)=  \int DU\;  U      I(x)U^{-1}\;,
\end{equation}
we get 
\begin{equation}
\rho^{11}(x)=
Q_{1}^{1}Q_{2}^{1}\mathcal{I}(x)\;,
\end{equation}
which is the highest weight state of an $ Sp(1)_{R} $ triplet scalar $  \rho^{\{AB\}}$, and is also the primary state of the broken current supermultiplet introduced in \cite{Hoo3}, whose existence indicates the symmetry enhancement in UV. $ Q_{\alpha}^{1}\rho^{11}=Q_{\dot{\alpha}}^{1}\rho^{11}=0 $. The supermultiplet contains $ 16 $ states. Similarly, from the anti-instanton, we can also get $ \rho^{\dagger 11} $ carrying the instanton number $ -1 $. In $ 5d $ $ \mathcal{N} =1$ gauge theories, the $ U(1) $ instanton number current belongs to the $ U(1) $ current multiplet whose primary state is a scalar \cite{tw13}
\begin{equation}
L^{11}=\frac{1}{4\pi^{2}}tr \left(\epsilon^{\alpha\beta}\lambda_{\alpha}^{1}\lambda^{1}_{\beta} +\epsilon_{\dot{\alpha}\dot{\beta}}\lambda^{1\dot{\alpha}}   \lambda^{1\dot{\beta}}+D^{11}\phi\right) +\cdots
\end{equation}
$ \rho^{11} $, $  \rho^{\dagger 11}  $ and $ L^{11} $ furnish the adjoint representation of $ SU(2) $. Away from the UV fixed point, only the $ U(1) $ current is conserved.

When $ N=2 $ and $ N_{f}\neq 0 $, $ \mathcal{I}(x) $ is extended to 
\begin{equation}
\mathcal{I}_{s_{1}\cdots s_{k}}(x)=  \int DU\;  U  K_{s_{1}}\cdots K_{s_{k}}    I(x)U^{-1}
\end{equation}
with $ s_{i}=1,2,\cdots,N_{f} $ and $ k $ even, since as will be shown in (\ref{9.2}), $ \mathcal{I}_{s_{1}\cdots s_{k}}(x)=0 $ for the odd $ k $. Let 
\begin{equation}
\rho_{s_{1}\cdots s_{k}}^{11}(x)=
Q_{1}^{1}Q_{2}^{1}\mathcal{I}_{s_{1}\cdots s_{k}}(x)\;.
\end{equation}
From (\ref{k1}), (\ref{qwabb}), (\ref{qwab}) and (\ref{9.10a}), 
\begin{equation}
 Q_{\alpha}^{1}\rho_{s_{1}\cdots s_{k}}^{11}=Q_{\dot{\alpha}}^{1}\rho_{s_{1}\cdots s_{k}}^{11}=0 \;.
\end{equation}
The related $ Sp(1)_{R} $ triplet scalar $\rho_{s_{1}\cdots s_{k}}^{\{AB\}}  $ is in the spinor representation of $ SO(2N_{f}) $ with the positive chirality, while the generated broken current supermultiplet could make the $ U(1) \times SO(2N_{f})$ symmetry enhance to $ E_{N_{f}+1}  $ \cite{Hoo3}.

When $ N\geq 3 $, from (\ref{mnn}), the gauge invariant operator $ \mathcal{I}(x) $ with $ m=n=0 $ exists only when $\Delta Q_{U(1)} =0  $. According to (\ref{tab}), $\Delta Q_{U(1)} =0  $ yields $ \pm \kappa=N-\frac{1}{2}N_{f}  $ or $ \pm \kappa=N+\frac{1}{2}N_{f}  $, while the latter is excluded since $|\kappa|\leq N-\frac{1}{2}N_{f}  $ is required to have an UV fixed point \cite{tw10ac}. So the symmetry enhancement occurs when $  \pm \kappa=N-\frac{1}{2}N_{f}     $ and the enhancement pattern can only be $U(1) \times SO(2N_{f})\rightarrow  SU(2) \times SO(2N_{f}) $ since the broken current supermultiplet is a flavor singlet. The conclusion matches with \cite{Hoo3}.

Finally, for the $ 5d $ $ \mathcal{N} =2$ $ SU(N) $ gauge theory, from $ I(x) \vert \Omega \rangle $ satisfying
\begin{equation}
\mu^{Aa}  I(x) \vert \Omega \rangle =0\;,\;\;\;\;A=1,2,3,4\; ;\;  a=3,\cdots,N \;,
\end{equation}
we may construct $ \mathcal{I}^{p_{1}\cdots p_{2N}}_{a_{1}\cdots a_{2N}}(x) $ in the $\bar{ \mathbf{N}}^{2N}$ representation of $ SU(N) $. Instead of (\ref{sun}), the minimal gauge invariant operator built from $  \mathcal{I}^{p_{1}\cdots p_{2N}}_{a_{1}\cdots a_{2N}}(x) $ is
\begin{eqnarray}
\nonumber && 
\Phi^{A_{3}\cdots A_{N},B_{3}\cdots B_{N}} (x)\\     \nonumber&= &  \int DU\; \epsilon^{12a_{3}\cdots a_{N}}\epsilon^{12b_{3}\cdots b_{N}}
  U   \bar{\mu}^{A_{3}}_{a_{3}}\cdots \bar{\mu}^{A_{N}}_{a_{N}}   \bar{\mu}^{B_{3}}_{b_{3}}\cdots \bar{\mu}^{B_{N}}_{b_{N}}                I(x)U^{-1}\\     \nonumber&\propto &  \epsilon^{a_{1} a_{2}a_{3}\cdots a_{N}}\epsilon^{b_{1} b_{2}b_{3}\cdots b_{N}}(\lambda_{p_{3}}^{A_{3}})^{c_{3}}_{a_{3}}(x) \cdots  (\lambda_{p_{N}}^{A_{N}})^{c_{N}}_{a_{N}} (x)(\lambda_{q_{3}}^{B_{3}})^{d_{3}}_{b_{3}}(x) \cdots  (\lambda_{q_{N}}^{B_{N}})^{d_{N}}_{b_{N}} (x)\mathcal{I}^{12p_{3}\cdots p_{N}12q_{3}\cdots q_{N}}_{a_{1}a_{2}c_{3}\cdots c_{N}b_{1}b_{2}d_{3}\cdots d_{N}} (x)\;\\
\end{eqnarray}
with
\begin{equation}\label{6.46}
 \Phi^{A_{3}\cdots A_{N},B_{3}\cdots B_{N}}= \Phi^{\{A_{3}\cdots A_{N}\},B_{3}\cdots B_{N}}= \Phi^{A_{3}\cdots A_{N},\{B_{3}\cdots B_{N}\}}=- \Phi^{B_{3}A_{4}\cdots A_{N},A_{3}B_{4}\cdots B_{N}}\;.
\end{equation}
As is shown in \cite{Hoo3}, $  \Phi^{A_{3}\cdots A_{N},B_{3}\cdots B_{N}} $ is in the $\mathbf{6}^{N-2} $ symmetric representation of $SO(6)$ and could be decomposed into $ V_{0} \oplus V_{1} \cdots \oplus V_{N-2}$ with $ V_{k} $ in the $k_{\text{th}}$ symmetric traceless representation of $SO(5)$.

Now consider $ \Phi^{1\cdots 1,3 \cdots 3} $ which is the highest weight state in $V_{N-2} $. The action of $ Q $ gives an $N_{\text{th}}$ symmetric traceless scalar 
\begin{equation}
\rho^{111\cdots 1,333\cdots  3}=\epsilon^{\alpha\rho}\epsilon^{\beta\sigma}Q_{\{\alpha}^{1}Q_{\beta\}}^{3}Q_{\{\rho}^{1}Q_{\sigma\}}^{3}\Phi^{1\cdots 1,3\cdots  3}
\end{equation}
with
\begin{equation}
Q_{\alpha}^{1}\rho^{111\cdots 1,333\cdots  3}=Q_{\alpha}^{3}\rho^{111\cdots 1,333\cdots  3}=Q_{\dot{\alpha}}^{1}\rho^{111\cdots 1,333\cdots  3}=Q_{\dot{\alpha}}^{3}\rho^{111\cdots 1,333\cdots  3}=0\;,
\end{equation}
where we have used (\ref{kk4k}), (\ref{kk4}), (\ref{k1}) and (\ref{9.8i}), ignoring the central charges in (\ref{kk4k}) and (\ref{kk4}) which vanish when acting on gauge invariant operators. Starting from $ \rho^{111\cdots 1,333\cdots  3} $, successive actions of $Q_{\alpha}^{2}  $, $ Q_{\alpha}^{4} $, $ Q_{\dot{\alpha}}^{2} $ and $ Q_{\dot{\alpha}}^{4} $ give a supermultiplet with $ 256 $ states, which is the KK mode of the rank-$ N $ short multiplet in the $ 6d $ $ (2,0) $ theory \cite{Hoo3}. The generic $  \rho^{A_{1}\cdots A_{N},B_{1}\cdots B_{N}} $ in $V_{N}  $ can be ontained through the action of the $  Sp(2)_{R} $ charges on $\rho^{111\cdots 1,333\cdots  3}  $.

However, it seems that we cannot get KK modes of the rank-$ k $ short multiplets when $ k<N $. Operators in $ V_{0} \oplus V_{1} \cdots \oplus V_{N-2} $ all have the same classical conformal dimension, so those in $ V_{k} $ with $ k <N-2 $ are konishi-like operators \cite{1b}. For example, when $ N=3 $, $ V_{1} =\{ \Phi^{A,B} +\frac{1}{4}\Omega^{AB}\Omega_{CD}\Phi^{C,D}|A,B=1,2,3,4\} $, $  V_{0} =\{ \Omega_{CD}\Phi^{C,D}\}  $. Supermultiplets built on $ V_{k} $ for $ k <N-2 $ may be the KK modes of the konishi multiplets in the $ 6d $ $ (2,0) $ theory.

\section{Instanton states and $ (p,q) $ string webs inside the 5-brane webs}\label{web1}

$ \mathcal{N}=1 $ $ 5d $ gauge theories can also be realized on the $ (p,q) $ 5-brane webs in Type IIB string theory \cite{Hoo, Hoo2}. In the planar web diagram, 5-branes are oriented preserving $1/4$ of the background supersymmetries with charges summing to zero at each vertex. The relative positions of the external 5-branes correspond to parameters of the field theory, like the coupling constant. Deformations of the web with the planar positions of the external 5-branes fixed correspond to the moduli of the theory, such as the Coulomb branch moduli. The $5d$ SCFT is related to the singular web with all external 5-brane lines meeting at one point. BPS states are encoded as the $ (p,q) $ string webs inside the 5-brane webs \cite{tw15}. Quantization of a string web with $ n_{X} $ external legs gives a supermultiplet with the highest spin $ J=\frac{n_{X}}{2} $.

In this section, we will consider the brane webs for the $ 5d $ $ \mathcal{N}=1 $ $ SU(N) $ gauge theory with $ N_{f} =0$ and the Chern-Simons level $ \kappa $. We will determine the spin and the electric charge for all of the string webs with the instanton number $ 1 $, and show that for each string web, one can always find an instanton state carrying the same spin and the same electric charge.

For the $ 5d $ $ \mathcal{N} =1$ $ SU(N)$ gauge theory with $ N_{f} =0$, in the Coulomb branch, the related $ N $ $ D5 $ branes are separated at the positions $ (\phi_{1} ,\cdots,\phi_{N})$ with $ \sum^{N}_{i=1} \phi_{i}=0$, $ \phi_{1} \geq \cdots \geq\phi_{N}$. The distance between the $ k_{\text{th}} $ and the $ (k+1)_{\text{th}} $ $ D5 $ branes is $ \phi_{k,k+1} =\phi_{k}-\phi_{k+1}$. A BPS state characterized by the electric charge $Q_{E}= (Q_{e1} ,\cdots,Q_{eN})$ and the instanton number $ q $ has the mass \cite{L1} 
\begin{equation}\label{mass}
M=|qm_{0}- \sum^{N}_{i=1} Q_{ei}\phi_{i}|\;,
\end{equation}
where $ m_{0}=1/R_{5} $ is the instanton mass and $ \kappa $ is absorbed in $ Q_{ei} $. The BPS bound follows from (\ref{qwa}) and (\ref{zab}).

On the other hand, for the vacuum satisfying $ \mu^{Aa}  I(x) \vert \Omega \rangle =0  $, from (\ref{tab}) and (\ref{mnn}), the instanton states $ \mathcal{I}^{b_{1}\cdots b_{m},p_{1}\cdots p_{n}}_{q_{1}\cdots q_{m},a_{1}\cdots a_{n}}(x) \vert \Omega \rangle $ in the $ \mathbf{N}^{m} \times \bar{ \mathbf{N}}^{n} $ representation of $ SU(N) $ and the $ \mathbf{2}^{n} \times \bar{ \mathbf{2}}^{m} $ representation of $ SU(2)_{L} $ exist when $ m-n=\kappa -N$. $ \mathcal{I}^{b_{1}\cdots b_{m},p_{1}\cdots p_{n}}_{q_{1}\cdots q_{m},a_{1}\cdots a_{n}}(x) $ could be decomposed into the irreducible representations of $ SU(2)_{L} $ with
\begin{numcases}{J =} 
0,1,\cdots,\frac{m+n}{2}\;\;\;\;\;\;\text{when $ m+n $ is even}\notag \\
\frac{1}{2},\frac{3}{2},\cdots,\frac{m+n}{2} \;\;\;\;\;\;\text{when $ m+n $ is odd}\;.\notag
\end{numcases}
$\mathcal{I}^{b_{1}\cdots b_{m},p_{1}\cdots p_{n}}_{q_{1}\cdots q_{m},a_{1}\cdots a_{n}}(x) $ is a $ 1/2 $ BPS operator with 
\begin{equation}\label{Qw1}
Q_{\dot{\alpha}}^{A}\mathcal{I}^{b_{1}\cdots b_{m},p_{1}\cdots p_{n}}_{q_{1}\cdots q_{m},a_{1}\cdots a_{n}}(x) =0\;.
\end{equation}
The electric charge of $  \mathcal{I}^{b_{1}\cdots b_{m},p_{1}\cdots p_{n}}_{q_{1}\cdots q_{m},a_{1}\cdots a_{n}} $ is given by
\begin{equation}
 (Q_{e1} ,\cdots,Q_{eN})=\sum^{n}_{i=1}(\underbrace{0,\cdots,0}_{a_{i}-1},-1,0,\cdots,0)+\sum^{m}_{i=1}(\underbrace{0,\cdots,0}_{b_{i}-1},1,0,\cdots,0)\;.
\end{equation}
$ \sum^{N}_{i=1} Q_{ei}=m-n$. From (\ref{Qw1}), (\ref{qwa}) and (\ref{511}), we have 
\begin{equation}
P_{0}\mathcal{I}^{b_{1}\cdots b_{m},p_{1}\cdots p_{n}}_{q_{1}\cdots q_{m},a_{1}\cdots a_{n}}(x)  \vert \Omega \rangle=(Q_{I}m_{0}+Z)\mathcal{I}^{b_{1}\cdots b_{m},p_{1}\cdots p_{n}}_{q_{1}\cdots q_{m},a_{1}\cdots a_{n}}(x) \vert \Omega \rangle\;.
\end{equation}
In the Coulomb branch, $ \mathcal{I}^{b_{1}\cdots b_{m},p_{1}\cdots p_{n}}_{q_{1}\cdots q_{m},a_{1}\cdots a_{n}} $ does not commute with $ Z $ and thus acquires the mass
\begin{equation}
M=| m_{0}-\sum^{N}_{i=1} Q_{ei}\phi_{i}|\;.
\end{equation}

$\;$

  \begin{figure}[H]
\centering
\subfloat[$ \kappa=0 $]{\includegraphics[scale=0.3]{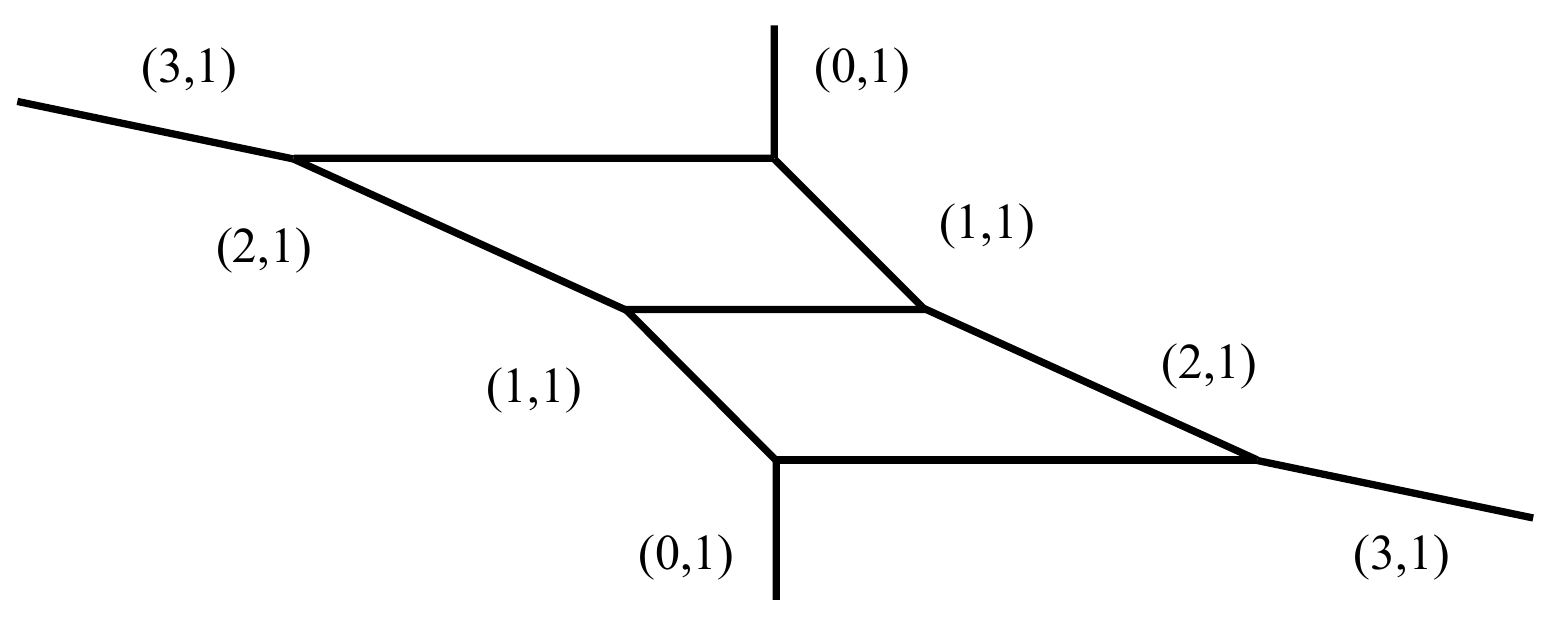}}
\subfloat[$ \kappa=1 $]{\includegraphics[scale=0.3]{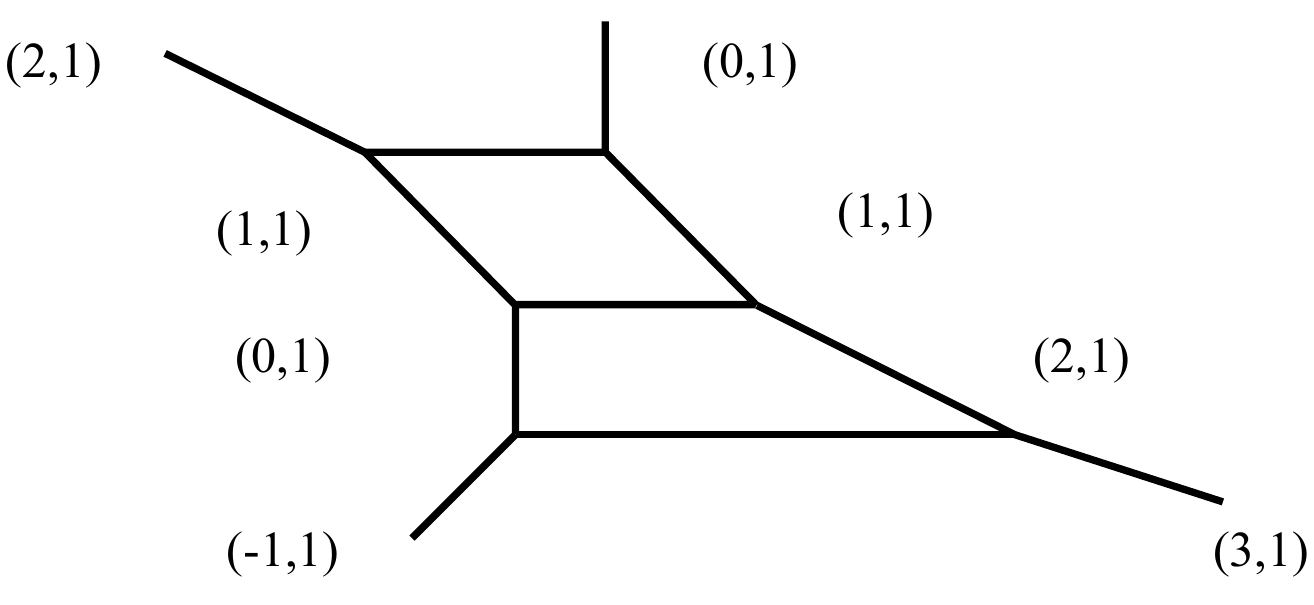}}\\
\subfloat[$ \kappa=2 $]{\includegraphics[scale=0.3]{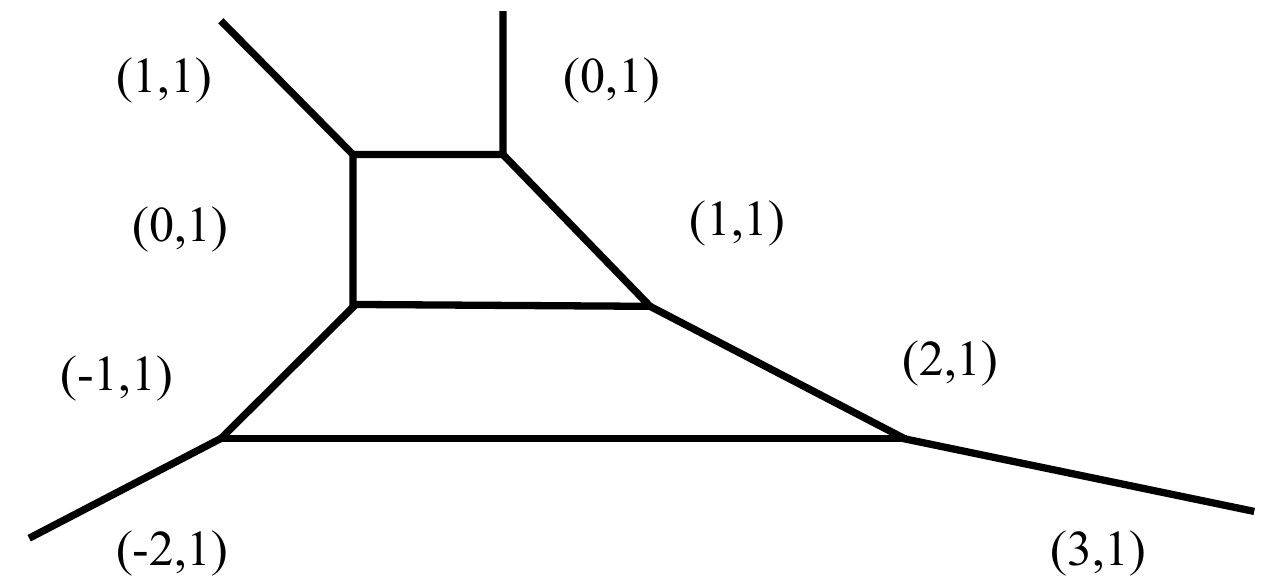}}\;\;\;
\subfloat[$ \kappa=3 $]{\includegraphics[scale=0.3]{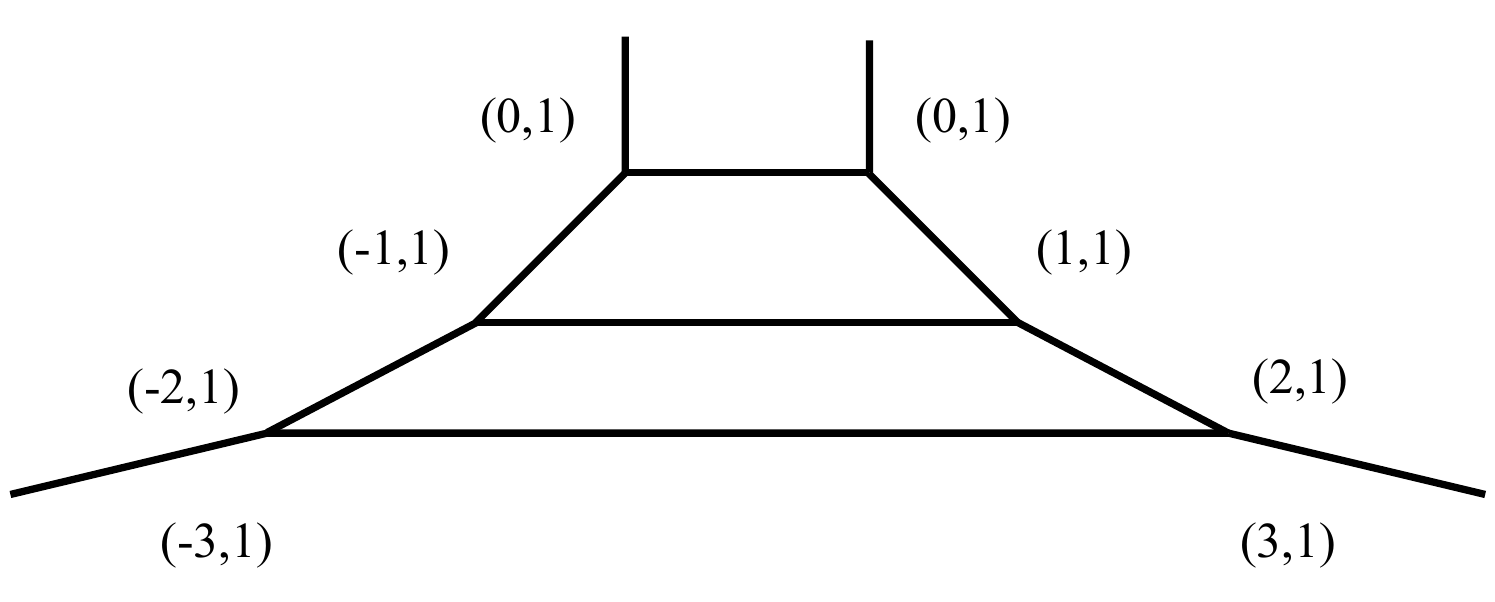}}
\captionsetup{font={footnotesize,stretch=1.25},justification=raggedright}
\caption{5-brane webs with $ N=3 $, $ t=0 $, $ \kappa=0,1,2,3 $}
\label{F6}
\end{figure}

The 5-brane web configuration of this theory is composed by four external legs with charges $ (N-\kappa+t,1) $, $ (t,1) $, $ (t-\kappa,1) $ and $ (t+N,1) $, $ N $ $ D5 $ brane segments and $ 2(N-1) $ $ (p,q) $ brane segments. $ t \in \mathbb{Z} $, and the webs related with the different values of $ t $ are equivalent. The charge conservation condition is imposed at each vertex. The brane web configuration is characterized by the parameter $ m_{0} $ and the Coulomb branch moduli $\phi_{k,k+1}$ with $ k=1,\cdots,N-1$. The brane web contains $ N-1 $ faces with each face surrounded by four 5-brane segments. Charges of the left and the right 5-brane segments on each face are given by

$ \; $

\begin{equation}
\begin{tabular}{|l|c|c|c|}
     \hline
           Face     &  Left    & Right     \\\hline
            1               & $(N-\kappa-1+t,1)$ & $(1+t,1) $  \\
             $\cdots$            & $\cdots$     & $\cdots$    \\
           $k$               & $ (N-\kappa-k+t,1) $     & $  (k+t,1)$   \\
          $\cdots$               & $\cdots$       & $\cdots$       \\
         $N-1$               & $(1-\kappa+t,1)  $        & $ (N-1+t,1)$   \\
            \hline
   \end{tabular}
\end{equation}

 $ \; $

$\;$

\noindent See Figure. \ref{F6} for the example of 5-brane webs with $ N=3 $, $ t=0 $, $ \kappa=0,1,2,3 $. 

BPS states are realized as the string webs ending on the 5-branes. The simplest BPS states are $ N-1 $ spin $ 1 $ W bosons corresponding to the F-strings connecting the adjacent $D5$ branes. The mass $ M $ and the electric charge $Q_{E}  $ are given by (the tensions are in string unit, and the Type IIB string coupling is taken to be $ \tau=i $)

$\;$

\begin{equation}\label{Wb}
\begin{tabular}{|l|c|c|c|}
     \hline
                &  $M$    & $Q_{E}$      \\\hline
            1               & $ \phi_{1,2}$ & $(1,-1,0,\cdots,0) $  \\
             $\cdots$            & $\cdots$     & $\cdots$    \\
           $k$               & $ \phi_{k,k+1} $     & $  (\underbrace{0,\cdots,0}_{k-1}, 1,-1  ,0,\cdots,0)$   \\
          $\cdots$               & $\cdots$       & $\cdots$       \\
         $N-1$               & $ \phi_{N-1,N}  $        & $ (0,\cdots,0,1,-1)$   \\
            \hline
   \end{tabular}
\end{equation}

$\;$

$\;$

\noindent On each face, an infinite number of string webs carrying the instanton number $ q $ can also be constructed. The brane web configurations with the different values of $ t $ are equivalent, while the spectra of string webs are also $ t $-independent. For a string web with the mass $ M $, its instanton correspondence must have $ q $ and the electric charge $ Q_{E} $ satisfying
  \begin{equation}\label{106}
   M=|qm_{0}-\sum^{N}_{i=1} Q_{ei}\phi_{i}|\;,\;\;\;\;\;\;\;\;
  \sum^{N}_{i=1} Q_{ei}=\kappa-N\;.
   \end{equation}
For the given $ M $, the solution to (\ref{106}) is unique and may not be integers, but as we will see later, for all of the string webs, the obtained $  Q_{ei} $ are integers.

In the following, for the $ q=1 $ class, we will list the spin and the electric charge of the minimum string web on the $ k_{\text{th}} $ face with $ k=1,2,\cdots,N-1 $. Generic string webs are bound states of the minimum web and the F-strings. We will show that for each string web, one can always find an instanton state carrying the same spin and the same charge.

\begin{enumerate}[(1)]

\item When $ k\leq \frac{N-\kappa}{2} $, the minimum string web contains $ N-\kappa-2k +2$ external legs, and thus has the highest spin $J=1-k+ \frac{1}{2}( N-\kappa ) $. The electric charge obtained from $ M$ and (\ref{106}) is 
\begin{equation}
Q_{E}=(\underbrace{-2,\cdots,-2}_{k-1}, \kappa-N+2k-2  ,0,\cdots,0)\;.
\end{equation}
From $ \mathcal{I}^{b_{1}\cdots b_{m},p_{1}\cdots p_{n}}_{q_{1}\cdots q_{m},a_{1}\cdots a_{n}} $, instanton states with the desired $ Q_{E} $ and $ J $ can be constructed. For example, 
\begin{itemize}
\item [(a)] when $ k=1 $, $ \mathcal{I}^{p_{1}\cdots p_{N-\kappa}}_{1\cdots 1} $ has $Q_{E} =(\kappa-N  ,0,\cdots,0) $ and $J=\frac{1}{2}( N-\kappa ) $;

\item [(b)] when $ k=2 $, $ \mathcal{I}^{[p_{1}p_{2}]\{p_{3}\cdots p_{N-\kappa}\}}_{[12]\{12\cdots 2\}} $ has $Q_{E} = (-2, \kappa-N+2  ,0,\cdots,0)  $ and $J=\frac{1}{2}( N-\kappa ) -1 $;

\item [(c)] when $ k=3 $, $ \mathcal{I}^{[p_{1}p_{2}][p_{3}p_{4}]p_{5}\cdots p_{N-\kappa}}_{[12][12]3\cdots 3} $ has $Q_{E} =  (-2, -2,\kappa-N+4  ,0,\cdots,0)  $ and $ J=\frac{1}{2}( N-\kappa ) -2 $;

\item [] $\cdots$
\end{itemize}
\noindent The mass of the minimum string web equals the length of the $ k_{\text{th}} $ $ D5 $ brane segment. The minimum web can also be continuously deformed into the D-string in the $ k_{\text{th}} $ $ D5 $ brane.

\item When $ k> \frac{N-\kappa}{2} $, the minimum string web contains $ \kappa+2k-N +2$ external legs, and thus has the highest spin $J=1+k-\frac{1}{2}( N-\kappa ) $. The electric charge is  
\begin{equation}
 Q_{E}=(\underbrace{-2,\cdots,-2}_{k},\kappa- N+2k  ,0,\cdots,0)\;.
\end{equation}
\begin{itemize}
\item [(a)]  When $ k=1 $ and $ N-\kappa =0$, $ \mathcal{I}^{22,\{p_{1}p_{2}}_{q_{1}q_{2},11}\epsilon^{r_{1}\underline{q_{1}}} \epsilon^{r_{2}\}q_{2}} $ has $Q_{E} =(-2,2  ,0,\cdots,0) $ and $ J=2 $, where $ \underline{q_{1}} $ means $ q_{1} $ is excluded in the totally symmetric permutation.

\item [(b)] When $ k=1 $ and $ N-\kappa =1$, $ \mathcal{I}^{2,\{p_{1}p_{2}}_{q_{1},11}\epsilon^{r_{1}\}q_{1}}  $ has $Q_{E} = (-2, 1  ,0,\cdots,0) $ and $ J=\frac{3}{2}$.

\item [(c)] When $ k=2$ and $ N-\kappa <4$, $ \mathcal{I}^{3\cdots 3,[p_{1}p_{2}]\{p_{3}p_{4}}_{q_{1}\cdots q_{\kappa-N+4},[12]21}\epsilon^{r_{1}\underline{q_{1}}}\cdots \epsilon^{r_{\kappa-N+4}\}q_{\kappa-N+4}} $ has $Q_{E}=(-2,-2, \kappa-N+4  ,0,\cdots,0)$ and $ J=3-\frac{1}{2}( N-\kappa )$.

\item [] $\cdots$
\end{itemize}
\noindent The mass of the minimum string web equals the length of the $  (k+1)_{\text{th}} $ $ D5 $ brane segment. The minimum web can also be continuously deformed into the D-string in the $  (k+1)_{\text{th}} $ $ D5 $ brane.

\end{enumerate}

$\;$

\begin{figure}[H]
\centering
\includegraphics[scale=0.5]{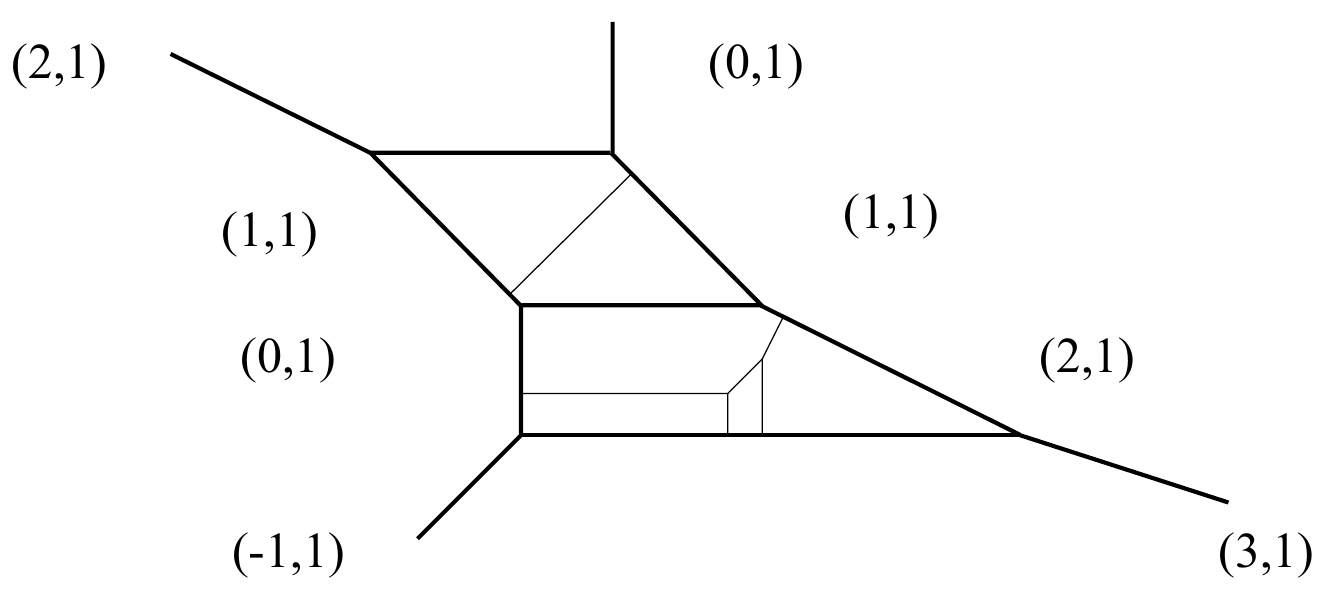}
\captionsetup{font={footnotesize,stretch=1.25},justification=raggedright}
\caption{Minimum string webs in a 5-brane web with $ N=3 $, $ \kappa=1$. The first one could be continuously deformed into the D-string embedded in the first and the second $ D5 $ branes, and has $ q=1 $, $ J=1$, $ Q_{E}=(-2,0,0) $ and $ M=m_{0} +\frac{4}{3}\phi_{1,2}+\frac{2}{3}\phi_{2,3}$. $ \mathcal{I}^{p_{1} p_{2}}_{1 1}   $ carries the same spin and charge. The second one could be continuously deformed into the D-string embedded in the third $ D5 $ brane, and has $ q=1 $, $ J=2 $, $Q_{E} =(-2,-2,2)$ and $ M=m_{0} +\frac{4}{3}\phi_{1,2}+\frac{8}{3}\phi_{2,3}$. $  \mathcal{I}^{33,[p_{1}p_{2}]\{p_{3}p_{4}}_{q_{1} q_{2},[12]21}\epsilon^{r_{1}\underline{q_{1}}} \epsilon^{r_{2}\}q_{2}}  $ carries the same spin and charge. }
\label{FF7}
\end{figure}

\begin{figure}[H]
\centering
\includegraphics[scale=0.5]{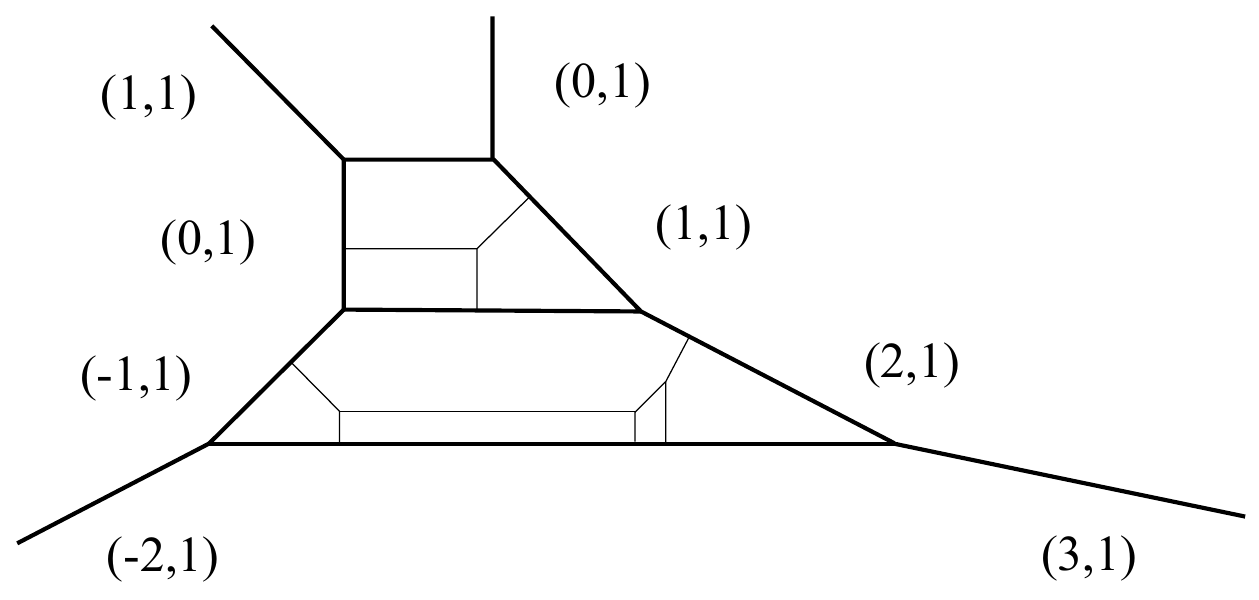}
\captionsetup{font={footnotesize,stretch=1.25},justification=raggedright}
\caption{Minimum string webs in a 5-brane web with $ N=3 $, $ \kappa=2$. The first one could be continuously deformed into the D-string embedded in the second $ D5 $ brane, and has $ q=1 $, $ J=\frac{3}{2} $, $ Q_{E}=(-2,1,0) $ and $ M=m_{0} +\frac{5}{3}\phi_{1,2}+\frac{1}{3}\phi_{2,3}$. $ \mathcal{I}^{2,\{p_{1}p_{2}}_{q_{1},11}\epsilon^{r_{1}\}q_{1}}    $ carries the same spin and charge. The second one could be continuously deformed into the D-string embedded in the third $ D5 $ brane, and has $ q=1 $, $ J=\frac{5}{2} $, $Q_{E} =(-2,-2,3)$ and $ M=m_{0} +\frac{5}{3}\phi_{1,2}+\frac{10}{3}\phi_{2,3}$. $ \mathcal{I}^{33 3,[p_{1}p_{2}]\{p_{3}p_{4}}_{q_{1}q_{2} q_{3},[12]21}\epsilon^{r_{1}\underline{q_{1}}}\epsilon^{r_{2}\underline{q_{2}}} \epsilon^{r_{3}\}q_{3}}   $ carries the same spin and charge. }
\label{FF8}
\end{figure}

Figure. \ref{FF7} and \ref{FF8} are examples of the minimum string webs inside the 5-brane webs with $ N=3 $ and $ \kappa=1,2$.

Since the minimum string webs can be deformed into the D-strings. D-strings in $ D5 $ branes can be taken as the basic constitutions of the instanton string webs. The mass and the electric charge of the D-strings in $N$ $D5$ branes are given by 
$ \; $

\begin{equation}
\begin{tabular}{|l|c|c|c|}
     \hline
                &  $M$    & $Q_{e}$      \\\hline
            1               & $m_{0}+ \sum^{N-1}_{i=1} \frac{(N-i)(N-\kappa)}{N}  \phi_{i,i+1}$ & $(\kappa- N,0,\cdots,0) $  \\
             $\cdots$            & $\cdots$     & $\cdots$    \\
           $k$               & $ m_{0}+ \sum^{k-1}_{i=1} \frac{i(N+\kappa)}{N}\phi_{i,i+1}+\sum^{N-1}_{i=k} \frac{(N-i)(N-\kappa)}{N}   \phi_{i,i+1}  $     & $  (\underbrace{-2,\cdots,-2}_{k-1}, \kappa- N+2k-2  ,0,\cdots,0)$   \\
          $\cdots$               & $\cdots$       & $\cdots$       \\
         $N$               & $ m_{0}+ \sum^{N-1}_{i=1} \frac{i(N+\kappa)}{N}\phi_{i,i+1}  $        & $ (-2,\cdots,-2,\kappa+N-2)$   \\
            \hline
   \end{tabular}\nonumber
\end{equation}

$ \; $

$ \; $

\noindent From the equivalent string web, the spin of the D-string state in the $k_{\text{th}} $ $ D5 $ brane is determined as
\begin{numcases}{J =} 
1-k+\frac{N-\kappa}{2}\;,\;\;\;\;\;\;\text{when $ k \leq \frac{N-\kappa}{2} $}\notag \\
k-\frac{N-\kappa}{2}\;,\;\;\;\;\;\;\text{when $  k > \frac{N-\kappa}{2}  $}\notag
\end{numcases}
\begin{itemize}
\item[(1)] When $N-\kappa  $ is positive and odd, the mass and the spin of the D-string state both take the minimum value for $k=\frac{N-\kappa+1}{2}   $. The corresponding electric charge and the spin are given by 
\begin{equation}\label{1q}
 Q_{E}  = (\underbrace{-2,\cdots,-2}_{\frac{1}{2} (N-\kappa-1)}, -1  ,0,\cdots,0) \;,\;\;\;\;\;\;\; J=\frac{1}{2} \;,
\end{equation}
which are the same as the charge and the spin of  $ \mathcal{I}^{[p_{1}p_{2}][p_{3}p_{4}]\cdots [p_{N-\kappa-2}p_{N-\kappa-1}]p_{N-\kappa}}_{[12][23]\cdots [\frac{1}{2} (N-\kappa-1)\frac{1}{2} (N-\kappa+1)]1} $.

\item[(2)] When $N-\kappa  $ is positive and even, the mass and the spin of the D-string state both take the minimum value for $ k=\frac{1}{2} (N-\kappa)  $ and $ k=\frac{1}{2} (N-\kappa)  +1$. The corresponding electric charge and the spin are given by
\begin{equation}\label{2q}
 Q_{E}  =  (\underbrace{-2,\cdots,-2}_{\frac{1}{2} (N-\kappa) } ,0,\cdots,0)   \;,\;\;\;\;\;\;\; J=1\;,
\end{equation}
which are the same as the charge and the spin of $ \mathcal{I}^{[p_{1}p_{2}][p_{3}p_{4}]\cdots [p_{N-\kappa-3}p_{N-\kappa-2}]\{p_{N-\kappa-1}p_{N-\kappa}\}}_{[12][23]\cdots [\frac{1}{2} (N-\kappa-2) \frac{1}{2} (N-\kappa) ]\{\frac{1}{2} (N-\kappa) 1\}} $.

\item[(3)] When $N-\kappa \leq 0 $, the mass and the spin of the D-string state both take the minimum value for $k=1$. The corresponding electric charge and the spin are given by 
\begin{equation}\label{3q}
 Q_{E}  = (\kappa-N ,0,\cdots,0) \;,\;\;\;\;\;\;\; J=1-\frac{N-\kappa}{2} \;,
\end{equation}
which are the same as the charge and the spin of  $ \epsilon_{p_{1}\{r_{1}} \mathcal{I}^{1\cdots 1,p_{1}}_{q_{1}\cdots q_{1+\kappa-N}\},1} $.

\end{itemize}

 (\ref{Wb}), (\ref{1q}), (\ref{2q}) and (\ref{3q}) are basic components, from which, all of the string webs with $ q=1 $ can be constructed as the bound states. For the arbitrary string web, one can always find the suitable $ \mathcal{I}^{b_{1}\cdots b_{m},p_{1}\cdots p_{n}}_{q_{1}\cdots q_{m},a_{1}\cdots a_{n}} $ carrying the same spin and the electric charge.

\begin{figure}[h]
\centering
\subfloat[F-string]{\includegraphics[scale=0.5]{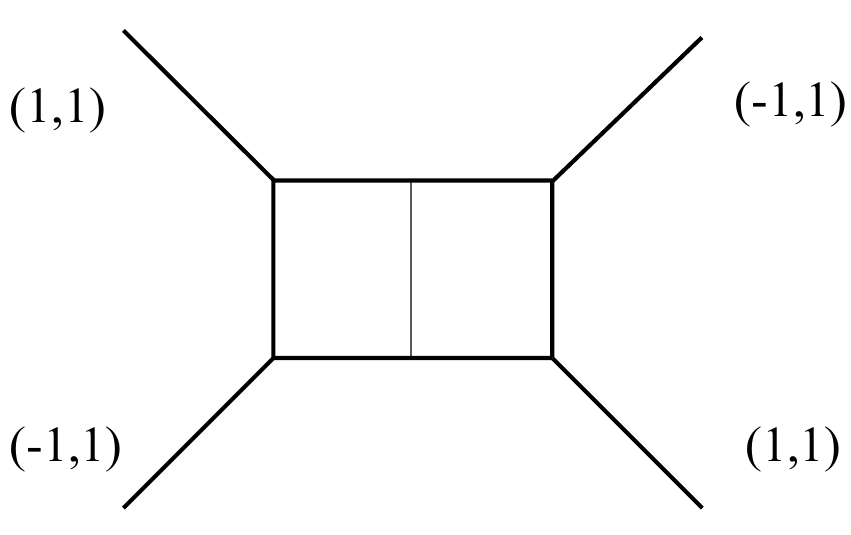}}\;\;\;\;\;\;\;
\subfloat[D-string]{\includegraphics[scale=0.5]{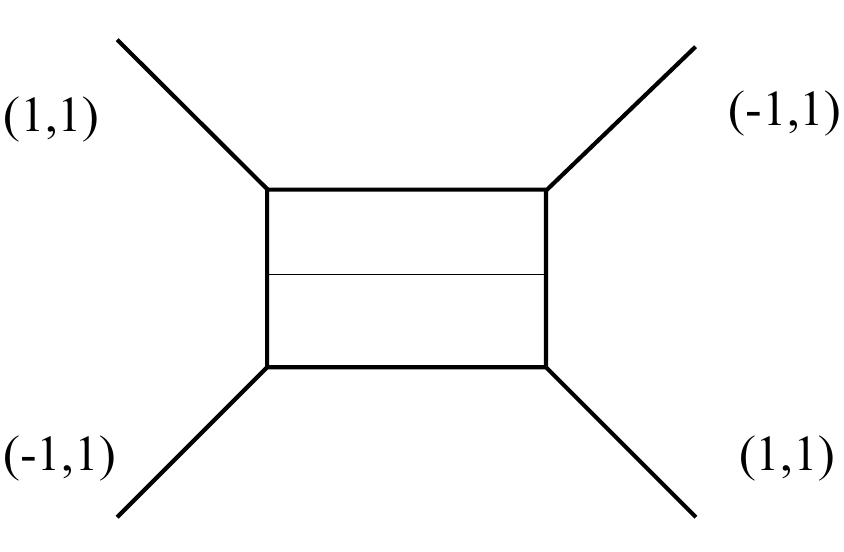}}
\captionsetup{font={footnotesize,stretch=1.25},justification=raggedright}
\caption{5-brane web for $ N=2 $}
\label{g6}
\end{figure}

\begin{figure}[h]
\centering
\subfloat[]{\includegraphics[scale=0.5]{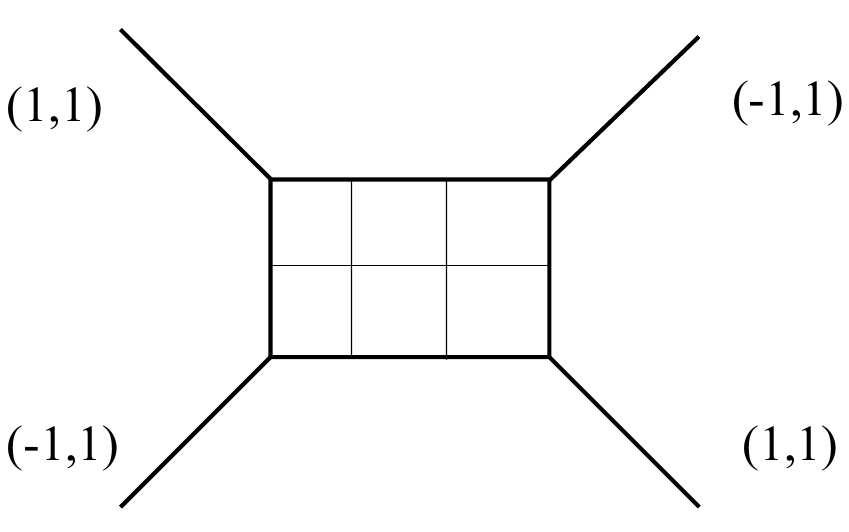}}\;\;\;\;\;\;\;
\subfloat[]{\includegraphics[scale=0.5]{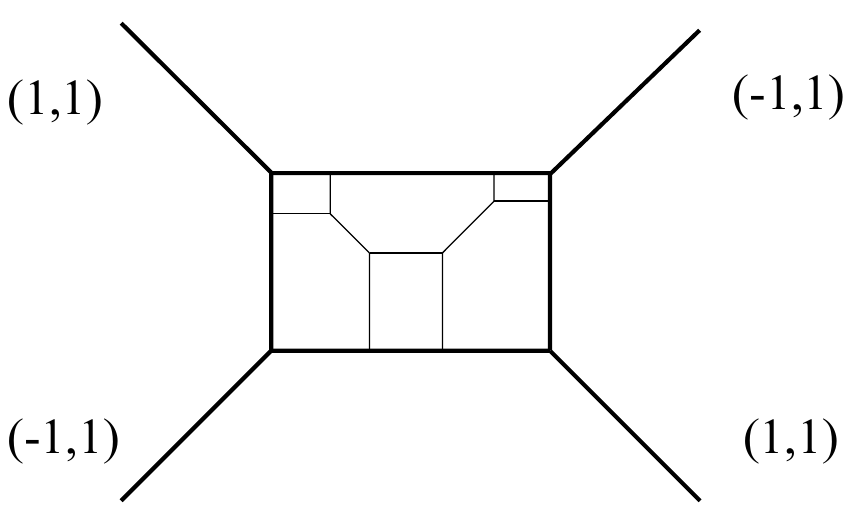}}
\captionsetup{font={footnotesize,stretch=1.25},justification=raggedright}
\caption{String web as the bound state of a D-string and two F-strings}
\label{g66}
\end{figure}

The situation for $ N=2 $ is especially simple, where the basic elements are F-string and D-string in Figure. \ref{g6}. In the $ q=1 $ class, the generic string webs are bound states of a D-string and $ l $ F-strings (see Figure. \ref{g66} for an example of $ l=2 $). The corresponding spin and the electric charge are $ J=l+1 $, $ Q_{E} =(-l-1,l+1)$, which are the same as those of $ \mathcal{I}_{q_{1}\cdots q_{l+1},1\cdots 1}^{2\cdots 2,\{p_{1}\cdots p_{l+1}}\epsilon^{r_{1}\underline{q_{1}}} \cdots \epsilon^{r_{l+1}\}q_{l+1}} \sim \mathcal{I}^{2\cdots 2}_{q_{1}\cdots q_{2l+2}}$. BPS string webs are in one-to-one correspondence with the BPS spectrum of the instanton states analysed in subsection \ref{SU}.

\section{The instanton supermultiplet}\label{mul1}

In the brane web construction of $ 5d $ $ \mathcal{N}=1 $ theories, the quantization of a string web containing $ 2J $ external legs and preserving $ Q_{\alpha}^{A} $ supercharges gives a $ 1/2 $ BPS supermultiplet in the 
\begin{equation}\label{8.121}
 (J-\frac{1}{2},0)\otimes \left[  (0,\frac{1}{2})\oplus 2(0,0)\right]  =(J-\frac{1}{2},\frac{1}{2})\oplus 2(J-\frac{1}{2},0)
\end{equation}
representation of the $ SU(2)_{L} \otimes SU(2)_{R} $ group \cite{tw10acd,tw15}. The highest spin component is in the $ (J-\frac{1}{2},\frac{1}{2}) $ representation. We have seen that the irreducible decomposition of $\mathcal{I}^{b_{1}\cdots b_{m},p_{1}\cdots p_{n}}_{q_{1}\cdots q_{m},a_{1}\cdots a_{n}} $ gives a series of operators carrying the same spin and the same electric charge as the highest spin field of the string web supermultiplet. However, since the instanton states are chiral, the resembled highest spin field is of the $(J,0) $ type. In this section, we will study the $ 1/2 $ BPS chiral supermultiplet in the   
\begin{equation}\label{8.122}
  (J,0)\oplus 2(J-\frac{1}{2},0)\oplus (J-1,0)
\end{equation}
representation of $ SU(2)_{L} \otimes SU(2)_{R} $, and show that (\ref{8.121}) and (\ref{8.122}) are equivalent, sharing the same gauge invariant components. We will also construct the chiral supermultiplets from $\mathcal{I}^{b_{1}\cdots b_{m},p_{1}\cdots p_{n}}_{q_{1}\cdots q_{m},a_{1}\cdots a_{n}} $, which could act as the operator realization of (\ref{8.122}).

\subsection{$ 1/2 $ BPS chiral multiplet in $ 5d $ $ \mathcal{N}=1 $ gauge theory}

In the $ 5d $ $ \mathcal{N}=1 $ gauge theory, for a $ 1/2 $ BPS chiral multiplet, the highest weight state of $ SU(2)_{L} $ is $ B_{1\cdots 1} $ satisfying 
\begin{equation}\label{111}
Q^{A}_{\dot{\beta}} B _{1\cdots 1}   =0\;,\;\;\;\;\;\;\;\;Q^{A}_{1} B _{1\cdots 1} =0\;,\;\;\;\;\;\;\;\;A=1,2\;.
\end{equation}
Descendants generated by supercharges are
\begin{equation}\label{222}
 B _{1\cdots 1} \rightarrow  Q_{2}^{A} B _{1\cdots 1}  \rightarrow Q_{2}^{1}Q_{2}^{2} B _{1\cdots 1}\;.
\end{equation}
The $ SU(2)_{L} $ transformation of (\ref{111}) gives 
\begin{eqnarray}
&&\label{111a} Q^{A}_{\dot{\beta}} B _{ \alpha_{1}\alpha_{2} \cdots\alpha_{2J}       }   =0\;,\\   &&\label{111b}Q^{A}_{\{ \beta} B _{\alpha_{1}\alpha_{2} \cdots\alpha_{2J}       \}} =0 \;,
\end{eqnarray}
where $B _{ \alpha_{1}\alpha_{2} \cdots\alpha_{2J}       }   =B _{ \{\alpha_{1}\alpha_{2} \cdots\alpha_{2J}     \}  }     $. From (\ref{111a}), (\ref{111b}), (\ref{qwabb}) and (\ref{qwab}), the supersymmetry transformation rule is obtained as 
\begin{eqnarray}
 && \label{rule1}Q^{A}_{ \beta} B _{\alpha_{1}\alpha_{2} \cdots\alpha_{2J}       } =2J\sum^{2J}_{i=1}\epsilon_{\alpha_{i}\beta }  \Psi^{A}_{\alpha_{1}\cdots \alpha_{i-1} \alpha_{i+1 }\cdots\alpha_{2J}   }\;,\\  &&Q^{B}_{ \beta} \Psi ^{A}_{\alpha_{1}\alpha_{2} \cdots\alpha_{2J-1}       } =\frac{i}{2J^{2}}\epsilon^{BA}H B _{\beta \alpha_{1}\alpha_{2} \cdots\alpha_{2J-1}      } + \frac{i}{J^{2}}\epsilon^{BA}\sum^{2J-1}_{i=1}\epsilon_{\alpha_{i}\beta  }H\Phi_{\alpha_{1}\cdots \alpha_{i-1} \alpha_{i+1 }\cdots\alpha_{2J-1}  }\;,\\  &&\label{8.7}Q^{A}_{ \beta} H\Phi_{\alpha_{1}\alpha_{2} \cdots\alpha_{2J-2}       } =-JH\Psi^{A}_{\beta \alpha_{1}\alpha_{2} \cdots\alpha_{2J-2}      }
\end{eqnarray}
and
\begin{eqnarray}
&&\label{88} Q^{A}_{ \dot{\beta}} B _{\alpha_{1}\alpha_{2} \cdots\alpha_{2J}       } =0\;,\\  &&Q_{\dot{\beta}}^{B}\Psi ^{A}_{\alpha_{1}\alpha_{2} \cdots\alpha_{2J-1}       } =\frac{i}{J(2J+1)}\epsilon^{BA}\epsilon^{\beta\alpha}K_{\beta\dot{\beta}}B_{\alpha\alpha_{1}\alpha_{2} \cdots\alpha_{2J-1} }\;,\\  &&\label{8.10}Q_{\dot{\beta}}^{A} H\Phi_{\alpha_{1}\cdots \alpha_{2J-2}  }=-J\epsilon^{\beta\alpha}K_{\beta\dot{\beta}} \Psi ^{A}_{\alpha\alpha_{1}\alpha_{2} \cdots\alpha_{2J-2}       } \;,
\end{eqnarray}
where 
\begin{equation}
H=P_{0}-P_{5}+Z\;,\;\;\;\;\;\;K_{\alpha\dot{\alpha}}=P_{\alpha\dot{\alpha}}+Z_{\alpha\dot{\alpha}}\;.
\end{equation}
Since $ [Q^{A}_{ \beta}, H]=0 $, (\ref{8.7}) also indicates
\begin{equation}
Q^{A}_{ \beta} \Phi_{\alpha_{1}\alpha_{2} \cdots\alpha_{2J-2}       } =-J\Psi^{A}_{\beta \alpha_{1}\alpha_{2} \cdots\alpha_{2J-2}      }\;.
\end{equation}
With $Q_{\dot{\beta}}^{A} \Phi_{\alpha_{1}\cdots \alpha_{2J-2}  } \equiv -J\Psi^{A}_{\dot{\beta}\alpha_{1}\cdots \alpha_{2J-2}  }  $, (\ref{8.10}) can be regarded as the equation of motion
\begin{equation}
H\Psi^{A}_{\dot{\beta}\alpha_{1}\cdots \alpha_{2J-2}  }-\epsilon^{\beta\alpha}K_{\beta\dot{\beta}} \Psi ^{A}_{\alpha\alpha_{1}\alpha_{2} \cdots\alpha_{2J-2}       }-\frac{i}{J}[Q_{\dot{\beta}}^{A} ,H]\Phi_{\alpha_{1}\cdots \alpha_{2J-2}  }=0\;.
\end{equation}

From (\ref{88}) and (\ref{qwa}), we also have 
\begin{equation}
hB _{\alpha_{1}\alpha_{2} \cdots\alpha_{2J}       } =(P_{0}+P_{5}-Z)B _{\alpha_{1}\alpha_{2} \cdots\alpha_{2J}       } =0\;.
\end{equation}
Since $ Z B _{\alpha_{1}\alpha_{2} \cdots\alpha_{2J}       } =0$, the chirality condition (\ref{88}) leads to the self-duality condition 
\begin{equation}
(P_{0}+P_{5})B _{\alpha_{1}\alpha_{2} \cdots\alpha_{2J}       } =0\;,
\end{equation}
which is also the BPS bound saturated by instantons.

The chiral supermultiplet consists of fields $ B $, $ \Psi $ and $ \Phi $ with $8J$ components. Representations of the supermultiplet under the $Sp(1)_{R}  $ R-symmetry group and the $  SU(2)_{L} \times SU(2)_{R} $ rotation group are given by 

$ \; $
\begin{equation}
\begin{tabular}{|l|c|c|c|}
     \hline
           Field     &  $Sp(1)_{R}$    & $ SU(2)_{L} \times SU(2)_{R} $     \\\hline
            $B_{\alpha_{1}\alpha_{2} \cdots\alpha_{2J}      }$              & $0$ & $(J,0) $  \\\hline
             $\Psi^{A}_{ \alpha_{1}\alpha_{2} \cdots\alpha_{2J-1}       }$            & $\frac{1}{2}$     & $(J-\frac{1}{2},0)$    \\\hline
           $\Phi_{\alpha_{1}\alpha_{2} \cdots\alpha_{2J-2}    }$               & $0$     & $(J-1,0)$   \\
            \hline
   \end{tabular}
\end{equation}

$ \; $

\noindent When $ J=\frac{1}{2} $, we have the hypermultiplet with the component fields $( B _{\alpha } ,\Psi^{A})  $.
\begin{align}
 &\label{ali22}  Q^{A}_{ \beta} B _{\alpha } =\epsilon_{\alpha\beta }  \Psi^{A} \;,& Q^{A}_{\dot{\beta}} B _{\alpha }&=0 \;,\\ &\label{ali33} Q^{B}_{ \beta} \Psi ^{A}=2i\epsilon^{BA}H B _{\beta      }\;, &Q_{\dot{\beta}}^{B}\Psi ^{A} &=i\epsilon^{BA}\epsilon^{\beta\alpha}K_{\beta\dot{\beta}}B_{\alpha }\;.
\end{align}
When $ J=1 $, we have the tensor multiplet with the component fields $( B _{\alpha \beta} ,\Psi_{\alpha}^{A},\Phi)  $.
\begin{align}
 &\label{ali}Q^{A}_{\gamma} B _{\alpha\beta    } =2(\epsilon_{\alpha\gamma }  \Psi^{A}_{\beta }+\epsilon_{\beta\gamma }  \Psi^{A}_{\alpha })\;, &Q^{A}_{\dot{\gamma}} B _{\alpha \beta}&=0 \;,\\ \label{alij} &Q^{B}_{ \beta} \Psi ^{A}_{\alpha    } =\frac{i}{2}\epsilon^{BA}H B _{\alpha\beta   } + i\epsilon^{BA}\epsilon_{\alpha\beta  }H\Phi\;,&Q_{\dot{\beta}}^{B}\Psi ^{A}_{\alpha       }& =\frac{i}{3}\epsilon^{BA}\epsilon^{\beta\gamma}K_{\beta\dot{\beta}}B_{\alpha\gamma }\;,\\ &\label{alii} Q^{A}_{ \beta}H \Phi =-H\Psi_{\beta}^{A}\;,&Q_{\dot{\beta}}^{A} H\Phi&=-\epsilon^{\beta\alpha}K_{\beta\dot{\beta}} \Psi ^{A}_{\alpha       }\;.
\end{align}
The situation for $ \mathcal{N}=2 $ is in Appendix \ref{B}. Starting from $B _{ \alpha_{1}\alpha_{2} \cdots\alpha_{2J}       }    $ satisfying (\ref{111a}) and (\ref{111b}), we do not get the multiplet $ (B,\Psi,\Phi) $ unless $ J=1 $, for which, the multiplet is $ (B_{\alpha\beta},\Psi_{\alpha}^{A} ,\Phi^{AB})$.

Now let us return to the non-chiral supermultiplet (\ref{8.121}). The highest spin field is $ A_{\alpha_{1}\alpha_{2} \cdots\alpha_{2J-1}  \dot{\alpha}    } =A_{\{\alpha_{1}\alpha_{2} \cdots\alpha_{2J-1}\}  \dot{\alpha}    }$ satisfying 
\begin{equation}
Q_{\beta}^{A}A_{\alpha_{1}\alpha_{2} \cdots\alpha_{2J-1}  \dot{\alpha}    }=0\;,\;\;\;\;\;\;\;Q_{\{\dot{\beta}}^{A}A_{\alpha_{1} \alpha_{2} \cdots\alpha_{2J-1} \dot{\alpha}  \}  }=0\;,
\end{equation}
from which, (\ref{qwabb}) and (\ref{qwa}) give the supersymmetry transformation rule  
\begin{eqnarray}
 && Q_{\dot{\beta}}^{A}A_{\alpha_{1} \alpha_{2} \cdots\alpha_{2J-1} \dot{\alpha}    }=2\epsilon_{\dot{\beta}\dot{\alpha}} \Psi^{A}_{\alpha_{1} \cdots\alpha_{2J-1}   }\;,\\  \label{826t}&&Q^{B}_{\dot{\beta}} \Psi ^{A}_{\alpha_{1}\alpha_{2} \cdots\alpha_{2J-1}       } =
 i\epsilon^{AB}hA_{\alpha_{1} \alpha_{2} \cdots\alpha_{2J-1} \dot{\beta}    }=\frac{i}{J(2J+1)}\epsilon^{BA}\epsilon^{\beta\alpha}K_{\beta\dot{\beta}}B_{\alpha\alpha_{1}\alpha_{2} \cdots\alpha_{2J-1} }\;,
\end{eqnarray}
and 
\begin{eqnarray}
 Q^{A}_{\beta} A_{\alpha_{1}\alpha_{2} \cdots\alpha_{2J-1}  \dot{\alpha}    } &=& 0\;,\\ \label{826tt} Q_{\beta}^{B}\Psi ^{A}_{\alpha_{1}\alpha_{2} \cdots\alpha_{2J-1}       } &=&\frac{i}{2}\epsilon^{AB}\epsilon^{\dot{\alpha}\dot{\beta}}K_{\beta\dot{\beta}}A_{\alpha_{1}\alpha_{2} \cdots\alpha_{2J-1}  \dot{\alpha}  }\\ \label{826ttt} &=&\frac{i}{2J^{2}}\epsilon^{BA}H B _{\beta \alpha_{1}\alpha_{2} \cdots\alpha_{2J-1}      } + \frac{i}{J^{2}}\epsilon^{BA}\sum^{2J-1}_{i=1}\epsilon_{\alpha_{i}\beta  }H\Phi_{\alpha_{1}\cdots \alpha_{i-1} \alpha_{i+1 }\cdots\alpha_{2J-1}  }
 \;. 
\end{eqnarray}
So the gauge invariant components of $ (B,\Psi,\Phi) $ and $ (A,\Psi) $ can be identified with each other. (\ref{8.121}) and (\ref{8.122}) are equivalent. The non-chiral supermultiplet contains $ 8J $ components with representations under the $Sp(1)_{R}  $ group and the $  SU(2)_{L} \times SU(2)_{R} $ group given by 

$ \; $
\begin{equation}
\begin{tabular}{|l|c|c|c|}
     \hline
           Field     &  $Sp(1)_{R}$    & $ SU(2)_{L} \times SU(2)_{R} $     \\\hline
            $A_{\alpha_{1}\alpha_{2} \cdots\alpha_{2J-1}  \dot{\alpha}    }$              & $0$ & $(J-\frac{1}{2},\frac{1}{2}) $  \\\hline
             $\Psi^{A}_{ \alpha_{1}\alpha_{2} \cdots\alpha_{2J-1}       }$            & $\frac{1}{2}$     & $(J-\frac{1}{2},0)$    \\\hline
   \end{tabular}
\end{equation}

$ \; $

The $ 5d $ tensor multiplet can also be viewed as the KK modes of the $ 6d $ selfdual tensor multiplet \cite{6}. Here we give a summary of the results, demonstrating the similarity between $ I(x) $ and $ B_{\alpha\beta} $.

Consider a $ 6d $ selfdual 2-form $ B_{MN} $ with the 3-form field strength $ H_{LMN} $ obtained via the action of the covariant derivative $ D_{L} $, i.e. $ D_{L} B_{MN}=H_{LMN}$, where $L, M,N=0,1,\cdots,5 $. The self-duality condition is taken to be
\begin{equation}\label{self}
H^{IJK}=\frac{1}{6}\epsilon^{IJKLMN}H_{LMN}
\end{equation}
with $ \epsilon^{012345}= 1$, which also gives the equation of motion of the theory. The $ x^{5} $ direction is compact with the radius $ R_{5} $. Let $ m,n,l,k=1,\cdots,4 $, (\ref{self}) can be decomposed into
\begin{equation}\label{sd1}
H_{0mn}=-\frac{1}{2}\epsilon_{mnlk}H_{5lk}\;,\;\;\;\;\;\;H_{lmn}=\epsilon_{lmnk}H_{05k}\;,
\end{equation}
where $ \epsilon_{1234} =1$. In temporal gauge, $ B_{0m}=0 $, $ B_{mn} $ and $ B_{5m} $ can be expanded into the KK modes. If $ B^{(q)}_{mn} $ and $ B^{(q)}_{5m} $ are KK modes with the $ x^{5} $ momentum $ \frac{q}{R_{5}} $, then $ B^{(q)}_{mn}=B^{*(-q)}_{mn} $, $  B^{(q)}_{5m} =B^{*(-q)}_{5m} $. For
\begin{equation}
 B^{(q)}_{\alpha\beta}=\frac{1}{2}B^{(q)}_{mn}\sigma_{\alpha\beta}^{mn} \;,\;\;\;\;\;\;\; B^{(q)}_{\dot{\alpha}\dot{\beta}}=\frac{1}{2}B^{(q)}_{mn}\bar{\sigma}_{\dot{\alpha}\dot{\beta}}^{mn} \;,\;\;\;\;\;\;\;A^{(q)}_{\alpha\dot{\alpha}}=\sigma^{m}_{\alpha\dot{\alpha}} B^{(q)}_{5m}   \;,
\end{equation}
the action of the momentum operator $ P_{M} \equiv -iD_{M}$ on the 2-form is given by
\begin{eqnarray}
\label{814} && P_{0}B^{(q)}_{\alpha\beta}=-iH^{(q)}_{\alpha\beta}\;, \;\;\;\;\;\;\;\; P_{0}B^{(q)}_{\dot{\alpha}\dot{\beta}}=-iH^{(q)}_{\dot{\alpha}\dot{\beta}}\;, \;\;\;\;\;\;\;\;P_{0}A^{(q)}_{\alpha\dot{\alpha}}=-i\tilde{F}^{(q)}_{\alpha\dot{\alpha}}\;,\\ \label{814av} &&P_{\alpha\dot{\alpha}}B^{(q)}_{\beta\gamma}=i\epsilon_{\beta\alpha}F^{(q)}_{\gamma\dot{\alpha}}+i\epsilon_{\gamma\alpha}F^{(q)}_{\beta\dot{\alpha}}\;,\;\;\;\;\;\;P_{\alpha\dot{\alpha}}B^{(q)}_{\dot{\beta}\dot{\gamma}}=i\epsilon_{\dot{\beta}\dot{\alpha}}F^{(q)}_{\alpha\dot{\gamma}}+i\epsilon_{\dot{\gamma}\dot{\alpha}}F^{(q)}_{\alpha\dot{\beta}}\;,\\  &&\label{814avu} P_{5}B^{(q)}_{\alpha\beta}=i\tilde{H}^{(q)}_{\alpha\beta}\;, \;\;\;\;\;\;\;\; P_{5}B^{(q)}_{\dot{\alpha}\dot{\beta}}=-i\tilde{H}^{(q)}_{\dot{\alpha}\dot{\beta}}\;, \;\;\;\;\;\;\;\; P_{\beta\dot{\beta}}A^{(q)}_{\alpha\dot{\alpha}}=-i\epsilon_{\dot{\alpha}\dot{\beta}}\tilde{H}^{(q)}_{\alpha\beta}+i\epsilon_{\alpha\beta}\tilde{H}^{(q)}_{\dot{\alpha}\dot{\beta}}\;.
\end{eqnarray}
The self-duality condition (\ref{sd1}) indicates 
\begin{equation}\label{814e}
\tilde{H}^{(q)}_{\alpha\beta}=H^{(q)}_{\alpha\beta}\;, \;\;\;\;\;\;\;\; \tilde{H}^{(q)}_{\dot{\alpha}\dot{\beta}}=H^{(q)}_{\dot{\alpha}\dot{\beta}}  \;, \;\;\;\;\;\;\;\;\tilde{F}^{(q)}_{\alpha\dot{\alpha}}=F^{(q)}_{\alpha\dot{\alpha}}\;.
\end{equation}
$(P_{0}+P_{5})B^{(q)}_{\alpha\beta}=0  $. The field strengths obtained from $ A^{(q)}_{\alpha\dot{\alpha}} $ and $ (B^{(q)}_{\alpha\beta},B^{(q)}_{\dot{\alpha}\dot{\beta}}) $ are the same, just as that in (\ref{826t}), (\ref{826tt}) and (\ref{826ttt}).

Supersymmetry transformations in $ 5d $ also descend from the $ 6d $ theory \cite{1}. Starting from the $ 6d $ supersymmetry transformation 
\begin{equation}\label{de}
\delta B_{MN}=i\bar{\epsilon}\Gamma_{MN} \Psi
\end{equation}
of the 2-form tensor, the superalgebra gives the transformation rule of the whole multiplet together with the self-duality condition (\ref{self}). The $ 5+1 $ decomposition of (\ref{de}) gives
\begin{align}
 &Q^{A}_{\gamma} B^{(q)}_{\alpha\beta}=2(\epsilon_{\beta\gamma}\Psi^{A(q)}_{\alpha}+\epsilon_{\alpha\gamma}\Psi^{A(q)}_{\beta})
\;, &Q_{\dot{\gamma}}^{A} B^{(q)}_{\alpha\beta}&=0 \;,\\  &Q^{A}_{\gamma} B^{(q)}_{\dot{\alpha}\dot{\beta}}=0\;,&Q_{\dot{\gamma}}^{A} B^{(q)}_{\dot{\alpha}\dot{\beta}}&=2(\epsilon_{\dot{\beta}\dot{\gamma}} \Psi_{\dot{\alpha}}^{A(q)}+\epsilon_{\dot{\alpha}\dot{\gamma}} \Psi_{\dot{\beta}}^{A(q)}  ) \;,\\  &Q_{\beta}^{A} A^{(q)}_{\alpha\dot{\alpha}}=2\epsilon_{\beta\alpha}\Psi_{\dot{\alpha}}^{A(q)}\;,& Q_{\dot{\beta}}^{A} A^{(q)}_{\alpha\dot{\alpha}}&=2\epsilon_{\dot{\beta}\dot{\alpha}}\Psi^{A(q)}_{\alpha}\;.
\end{align}
Subsequent transformations follow from (\ref{kk4k})-(\ref{k4}).

Actions of central charges on $A^{(q)}_{\alpha\dot{\alpha}} $, $ B^{(q)}_{\alpha\beta} $ and $ B^{(q)}_{\dot{\alpha}\dot{\beta}} $ are given by 
\begin{equation}
Z^{AB}A^{(q)}_{\alpha\dot{\alpha}}
=iD_{\alpha\dot{\alpha}}\Phi^{AB(q)} \;,\;\;\;\;\;\;\;Z^{AB}_{\beta\dot{\beta}}A^{(q)}_{\alpha\dot{\alpha}}=  Z^{AB}_{\gamma\beta} A^{(q)}_{\alpha\dot{\alpha}}=Z^{AB}_{\dot{\gamma}\dot{\beta}} A^{(q)}_{\alpha\dot{\alpha}}=0  \;,
\end{equation}
\begin{eqnarray}
  && \label{834a}  Z^{AB}B^{(q)}_{\alpha\beta}
=  Z^{AB}_{\dot{\gamma}\dot{\beta}}B^{(q)}_{\beta\gamma}=0\;, \\  &&     Z^{AB}_{\alpha\dot{\alpha}}B^{(q)}_{\beta\gamma}=i(\epsilon_{\beta\alpha}D_{\gamma\dot{\alpha}}\Phi^{AB(q)}+\epsilon_{\gamma\alpha}D_{\beta\dot{\alpha}}\Phi^{AB(q)}   ) \;,    \\&&  \label{834aa}    Z^{AB}_{\gamma\beta}B^{(q)}_{\alpha\sigma}=2(\epsilon_{\alpha\gamma}\epsilon_{\beta\sigma}+\epsilon_{\sigma\gamma}\epsilon_{\beta\alpha})W^{AB(q)}
 \; ,   
\end{eqnarray}
and
\begin{eqnarray}
  &&   Z^{AB}B^{(q)}_{\dot{\alpha}\dot{\beta}}
=  Z^{AB}_{\gamma\beta}B^{(q)}_{\dot{\beta}\dot{\gamma}}=0\;,      \\  && Z^{AB}_{\alpha\dot{\alpha}}B^{(q)}_{\dot{\beta}\dot{\gamma}}=i(\epsilon_{\dot{\beta}\dot{\alpha}}D_{\alpha\dot{\gamma}}\Phi^{AB(q)}+\epsilon_{\dot{\gamma}\dot{\alpha}}D_{\alpha\dot{\beta}}\Phi^{AB(q)})\;,     \\ &&  Z^{AB}_{\dot{\gamma}\dot{\beta}}B^{(q)}_{\dot{\alpha}\dot{\sigma}}=-2(\epsilon_{\dot{\alpha}\dot{\gamma}}\epsilon_{\dot{\beta}\dot{\sigma}}+\epsilon_{\dot{\sigma}\dot{\gamma}}\epsilon_{\dot{\beta}\dot{\alpha}})W^{AB(q)}
\; ,
\end{eqnarray}
where $\Phi^{AB(q)} = -\Phi^{BA(q)}$ is the scalar field and $ W^{AB(q)}=W^{BA(q)} $ is an auxiliary field vanishing when $ \mathcal{N}=1 $. Actions of central charges on $ I(x) $ are listed in (\ref{k3})-(\ref{k33a}), which could be compared with (\ref{834a})-(\ref{834aa}).

In the abelian situation, the action of $ P_{5} $ on $ B^{(q)}_{mn} $ can be written explicitly,
\begin{equation}
P_{5}B^{(q)}_{mn}=-iH^{(q)}_{5mn}=-i(\partial_{5}B^{(q)}_{mn}+\partial_{n}B^{(q)}_{5m}+\partial_{m}B^{(q)}_{n5})=\frac{q}{R_{5}}B^{(q)}_{mn}+iF^{(q)}_{mn} \;, 
\end{equation}
where $F^{(q)}_{mn}\equiv\partial_{m}B^{(q)}_{5n}- \partial_{n}B^{(q)}_{5m} $, so 
\begin{eqnarray}
 && \label{sim}-iH^{(q)}_{\alpha\beta}=P_{0}B^{(q)}_{\alpha\beta}=-P_{5}B^{(q)}_{\alpha\beta}=-\frac{q}{R_{5}}B^{(q)}_{\alpha\beta}-iF^{(q)}_{\alpha\beta}\;,\\ && \label{simm}-iH^{(q)}_{\dot{\alpha}\dot{\beta}}=P_{0}B^{(q)}_{\dot{\alpha}\dot{\beta}}=P_{5}B^{(q)}_{\dot{\alpha}\dot{\beta}}=\frac{q}{R_{5}}B^{(q)}_{\dot{\alpha}\dot{\beta}}+iF^{(q)}_{\dot{\alpha}\dot{\beta}}\;.
\end{eqnarray}
For $ I(x) $, we have (\ref{k2}) which is similar to (\ref{sim}). Let
\begin{equation}
(\mathcal{I}_{\alpha\beta})^{b}_{a}(x)\equiv \epsilon_{\{\beta\underline{\gamma}}\mathcal{I}_{\alpha\},a}^{b,\gamma} (x)   = \int DU\; D_{\{\alpha}^{b}[u(x)]D_{\beta \}a}[u^{-1}(x)] UI(x)U^{-1}\;,
\end{equation}
then $ (\mathcal{I}_{\alpha\beta})^{a}_{a}=0 $. $(\mathcal{I}_{\alpha\beta})^{b}_{a}  $ is in the $ (1,0) $ representation of $ SU(2)_{L} \times SU(2)_{R} $ and the adjoint representation of $SU(N)$. From (\ref{k2}), 
\begin{equation}
[P_{0},(\mathcal{I}_{\alpha\beta})^{b}_{a}(x)] =-[P_{5},(\mathcal{I}_{\alpha\beta})^{b}_{a}(x)]=-\frac{1}{R_{5}} (\mathcal{I}_{\alpha\beta})^{b}_{a}(x)-i (\mathcal{F}_{\alpha\beta})^{b}_{a}(x)\;,
\end{equation}
where 
\begin{equation}
(\mathcal{F}_{\alpha\beta})^{b}_{a}(x)=-\frac{1}{4} \int DU\;D_{\{\alpha}^{b}[u(x)]D_{\beta\}a}[u^{-1}(x)] U(F^{\rho}_{\sigma})_{\rho\;0}^{\sigma}(x)I(x)U^{-1}\;.
\end{equation}
$ (\mathcal{I}_{\alpha\beta})^{b}_{a} $ behaves like $  (B^{(1)}_{\alpha\beta})^{b}_{a}  $. Similarly, from the anti-instanton operator $ I^{\dagger} (x)$, we may construct $ (\mathcal{I}^{\dagger}_{\dot{\alpha}\dot{\beta}})^{b}_{a}(x) $ corresponding to $   (B^{(-1)}_{\dot{\alpha}\dot{\beta}})^{b}_{a}  $ with
\begin{equation}
[P_{0},(\mathcal{I}^{\dagger}_{\dot{\alpha}\dot{\beta}})^{b}_{a}(x)] =[P_{5},(\mathcal{I}^{\dagger}_{\dot{\alpha}\dot{\beta}})^{b}_{a}(x)]=-\frac{1}{R_{5}} (\mathcal{I}^{\dagger}_{\dot{\alpha}\dot{\beta}})^{b}_{a}(x)+i (\mathcal{F}^{\dagger}_{\dot{\alpha}\dot{\beta}})^{b}_{a}(x)\;.
\end{equation}
When acting on the vacuum, 
\begin{equation}
P_{0}(\mathcal{I}_{\alpha\beta})^{b}_{a}(x) \vert\Omega\rangle=\frac{1}{R_{5}} (\mathcal{I}_{\alpha\beta})^{b}_{a}(x) \vert\Omega\rangle\;,\;\;\;\;\;\;\;\;P_{0}(\mathcal{I}^{\dagger}_{\dot{\alpha}\dot{\beta}})^{b}_{a}(x)  \vert\Omega\rangle=\frac{1}{R_{5}} (\mathcal{I}^{\dagger}_{\dot{\alpha}\dot{\beta}})^{b}_{a}(x) \vert\Omega\rangle\;.
\end{equation}
$ (\mathcal{I}_{\alpha\beta})^{b}_{a}(x) $ and $ (\mathcal{I}^{\dagger}_{\dot{\alpha}\dot{\beta}})^{b}_{a}(x) $ both create positive energy states. On the other hand, the chiral anti-instanton operator and the anti-chiral instanton operator related to $  (B^{(-1)}_{\alpha\beta})^{b}_{a}  $ and $ (B^{(1)}_{\dot{\alpha}\dot{\beta}})^{b}_{a}  $ cannot be constructed, and will create negative energy states. From (\ref{sim}) and (\ref{simm}), $  (B^{(-1)}_{\alpha\beta})^{b}_{a}  $ and $ (B^{(1)}_{\dot{\alpha}\dot{\beta}})^{b}_{a}  $ are conjugate momenta of $  (B^{(1)}_{\alpha\beta})^{b}_{a}  $ and $ (B^{(-1)}_{\dot{\alpha}\dot{\beta}})^{b}_{a}  $.

\subsection{Chiral multiplet from the instanton operator}

We have seen that $ I(x) $, and thus $\mathcal{I}^{b_{1}\cdots b_{m},p_{1}\cdots p_{n}}_{q_{1}\cdots q_{m},a_{1}\cdots a_{n}} $, has a lot in common with $ B_{ \alpha_{1}\alpha_{2} \cdots\alpha_{2J}       }$. As the operator realization of $ B_{ \alpha_{1}\alpha_{2} \cdots\alpha_{2J}       }$, $ \mathcal{I}^{b_{1}\cdots b_{m},p_{1}\cdots p_{n}}_{q_{1}\cdots q_{m},a_{1}\cdots a_{n}}  $ should also follow the supersymmetry transformation rule (\ref{rule1})-(\ref{8.10}) which is entirely based on (\ref{111a}) and (\ref{111b}). For $ \mathcal{I}^{b_{1}\cdots b_{m},p_{1}\cdots p_{n}}_{q_{1}\cdots q_{m},a_{1}\cdots a_{n}}  $, we have (\ref{Qw1}), but no counterpart for (\ref{111b}). In fact, (\ref{111a}) and (\ref{111b}) also indicate 
\begin{eqnarray}
&&Z^{AB}_{\{ \beta\dot{\beta}} B _{\alpha_{1}\alpha_{2} \cdots\alpha_{2J}       \}} =0\;,\\   &&\label{K}P_{\{ \beta\dot{\beta}} B _{\alpha_{1}\alpha_{2} \cdots\alpha_{2J}       \}} =0 \;,
\end{eqnarray}
and in particular, $P_{1\dot{\beta}} B _{1 \cdots 1    } =0  $. The situation is similar for the gauge field $ A_{\alpha\dot{\alpha}} $. From (\ref{333}), we have 
\begin{equation}\label{cov}
 P_{\{\beta\{\dot{\beta}}A_{\alpha\}\dot{\alpha}\}}=0 \;.
\end{equation}
Since $ P_{m}A_{n} =F_{mn}$, (\ref{cov}) is equivalent to $P_{\{m}A_{n\}} =0  $ which is automatically satisfied. Likewise, $ P_{\{ \beta\dot{\beta}} B _{\alpha_{1}      \alpha_{2}\}} =0 $ requires $P_{m}B_{nl} =P_{[m}B_{nl]} $, i.e. $ P_{m} $ is the covariant derivative of the 2-form. However, in the $ 5d $ gauge theory, $  P_{\beta\dot{\beta}} $ only acts as the covariant derivative of the 1-form.
\begin{equation}
i P_{\beta\dot{\beta}} \mathcal{I}_{\alpha_{1}      \alpha_{2}}=\partial_{\beta\dot{\beta}} \mathcal{I}_{\alpha_{1}      \alpha_{2}}-i[A_{\beta\dot{\beta}},\mathcal{I}_{\alpha_{1}      \alpha_{2}}]\;.
\end{equation}
$ P_{\beta\dot{\beta}} $ cannot recognize the spinor index of $ \mathcal{I}_{\alpha_{1}      \alpha_{2}} $. It is impossible to construct a local operator satisfying (\ref{K}) in the framework of the $ 5d $ gauge theory.

Of course, we may construct operators satisfying (\ref{111b}). For example, let $O_{\alpha\beta} = Q^{1}_{\{\alpha}Q^{2}_{\beta\}}I(x) $, then $ Q_{\{\gamma}^{A}O_{\alpha\beta\}} =0 $, but (\ref{111a}) is violated. To reconcile (\ref{111a}) and (\ref{111b}), let us recall the usual procedure to get the supermultiplet from a $ 1/2 $ BPS state. For a gauge invariant $ 1/2 $ BPS operator, the action of $ M $ broken supercharges (usually anti-commutative) gives a supermultiplet with $ 2^{M} $ states. On the other hand, for a $ 1/2 $ BPS state like monopoles, $ M $ broken supercharges can be decomposed into $ M/2 $ creation and $ M/2 $ annihilation operators, while the generated supermultiplet has $2^{M/2}  $ components. Starting from a $ 1/2 $ BPS state $ \vert b_{x}\rangle $ with $  Q_{\dot{\alpha}}^{A}\vert b_{x}\rangle=0 $, the broken supercharges are $  Q_{\alpha}^{A}$ which could be organized into the creation and the annihilation operators $ (Q_{2}^{1}  , Q_{2}^{2} ) $ and $(Q_{1}^{1}  , Q_{1}^{2} )$. $ \vert b_{x}\rangle $ is not annihilated by $ (Q_{1}^{1}  , Q_{1}^{2} ) $, since otherwise, it will be $ 3/4 $ BPS. Nevertheless, the superposition of $  \vert b_{x}\rangle  $ leads to $ \vert b(0)\rangle $ with $P_{\alpha\dot{\alpha}}\vert b(0)\rangle=0 $ and the action of $ Q_{1}^{A} $ could project it into $ \vert B\rangle_{11} =\frac{1}{2}Q_{1}^{1}   Q_{1}^{2} \vert b (0)\rangle $. $ \vert B\rangle_{11}  $ is the ground state with 
\begin{equation}\label{b11}
Q_{\dot{\alpha}}^{A} \vert B\rangle_{11} =0\;,\;\;\;\;\;\;\; Q_{1}^{A}\vert B\rangle_{11} =0\;.
\end{equation}
The action of $(Q_{2}^{1}  , Q_{2}^{2} )   $ on $ \vert B\rangle_{11}  $ results in a $ 4 $-component supermultiplet as required.

Now perform a boost $ K $ to send $ \vert b(0)\rangle  $ into $ \vert b (p_{1\dot{\alpha}})\rangle $ with the momentum $ p_{1\dot{\alpha}} $, i.e $ K \vert b(0)\rangle =\vert b (p_{1\dot{\alpha}})\rangle $. Such boost makes $ Q_{1}^{A} $ invariant, so $K\vert B\rangle_{11} =\frac{1}{2}Q_{1}^{1}   Q_{1}^{2} \vert b(p_{1\dot{\alpha}}) \rangle$. The action of $ Q_{\dot{\alpha}}^{A}  $ on $ K\vert B\rangle_{11} $ has an extra term:
\begin{equation}
Q_{\dot{\alpha}}^{A} K\vert B\rangle_{11} =-iP_{1\dot{\alpha}} Q_{1}^{A}\vert b(p_{1\dot{\alpha}})\rangle=-ip_{1\dot{\alpha}} Q_{1}^{A}\vert b(p_{1\dot{\alpha}})\rangle\;.
\end{equation}
If $ K\vert B\rangle_{11}$ is gauge equivalent to $ K \vert B\rangle_{11}-\frac{i}{2}p^{\dot{\gamma}}_{1}\vert a\rangle_{1\dot{\gamma}} $ for an arbitrary vector $ \vert a \rangle_{\gamma\dot{\gamma}}  $, then with $\vert a \rangle_{\gamma\dot{\gamma}}  $ taken to be a state with $Q_{\dot{\beta}}^{A}\vert a\rangle_{\gamma\dot{\gamma}}  =2\epsilon_{\dot{\beta}\dot{\gamma}}Q_{\gamma}^{A}  \vert b(p_{1\dot{\alpha}})\rangle$, we will have
\begin{equation}
Q_{\dot{\alpha}}^{A} \left( - \frac{i}{2}p^{\dot{\gamma}}_{1}\vert a\rangle_{1\dot{\gamma}} \right) =-ip_{1\dot{\alpha}} Q_{1}^{A}\vert b(p_{1\dot{\alpha}})\rangle\;.
\end{equation}
So the extra term can be gauged away if the gauge equivalence relation 
\begin{equation}
 B_{\alpha\beta} \sim  B_{\alpha\beta}-\frac{i}{2}P^{\dot{\gamma}}_{\{\alpha}a_{\beta\}\dot{\gamma}}
\end{equation}
arising from $ B_{mn}\sim B_{mn} +iP_{m}a_{n}$ is assumed. It is a common feature of the gauge field that the Lorentz transformation is accompanied by a gauge transformation \cite{sw}. Superposition of $K \vert B\rangle_{11}  $ for all of $ K $ produces a local $  \vert B\rangle_{11} $ satisfying (\ref{b11}). The definition of the momentum operator $ P_{\alpha\dot{\alpha}} $ is not modified, but the price paid is the introduction of an additional gauge symmetry.

More generally, we may expect the operator satisfying both (\ref{111a}) and (\ref{111b}) can be constructed if the gauge equivalence  
\begin{equation}
 B_{\alpha_{1}\cdots \alpha_{2J}} \sim  B_{\alpha_{1}\cdots \alpha_{2J}}-\frac{i}{2}P^{\dot{\gamma}}_{\{\alpha_{1}}a_{\alpha_{2}\cdots \alpha_{2J}\}\dot{\gamma}}
\end{equation}
is assumed, where the gauge transformation parameter $ a_{\alpha_{1}\cdots \alpha_{2J-1}\dot{\gamma}} $ is in the $ (J-\frac{1}{2}, \frac{1}{2}) $ representation of $ SU(2)_{L} \times SU(2)_{R}  $.

Similarly, consider a $ 4d $ $ \mathcal{N} =1$ supersymmetric theory for the chiral scalar multiplets. The superalgebra is $ \{Q_{\alpha},Q_{\dot{\alpha}}\}=2P_{\alpha\dot{\alpha}} $. There is no gauge field, so $ P_{\alpha\dot{\alpha}}  $ acts as $ \partial_{\alpha\dot{\alpha}}  $. To construct a vector multiplet with $ A_{\alpha\dot{\alpha}}  $ the highest spin state,  
\begin{equation}
Q_{\{\beta}A_{\alpha\}\dot{\alpha}}=Q_{\{\dot{\beta}}A_{\alpha\dot{\alpha}\}}=0
\end{equation}
must be satisfied, which gives the constraint (\ref{cov}) indicating $ P_{\alpha\dot{\alpha}}  $ acts as a covariant derivative on $  A_{\alpha\dot{\alpha}}  $. If $A_{\alpha\dot{\alpha}}   $ is taken to be $\frac{1}{2}Q_{\alpha}Q_{\dot{\alpha}} \vert a\rangle  $ for some nonperturbative state $ \vert a\rangle $, then $ Q_{\{\beta}A_{\alpha\}\dot{\alpha}}=0 $, but 
\begin{equation}
Q_{\{\dot{\beta}}A_{\alpha\dot{\alpha}\}}=P_{\alpha\{\dot{\beta}}Q_{\dot{\alpha}\}} \vert a\rangle\;.
\end{equation}
With the gauge equivalence $A_{\alpha\dot{\alpha}}\sim A_{\alpha\dot{\alpha}}-P_{\alpha\dot{\alpha}}\vert a \rangle  $ assumed, 
\begin{equation}
Q_{\{\dot{\beta}}A_{\alpha\dot{\alpha}\}}\sim Q_{\{\dot{\beta}}A_{\alpha\dot{\alpha}\}}-Q_{\{\dot{\beta}}P_{\alpha\dot{\alpha}\}}\vert a \rangle=P_{\alpha\{\dot{\beta}}Q_{\dot{\alpha}\}} \vert a\rangle-Q_{\{\dot{\beta}}P_{\alpha\dot{\alpha}\}}\vert a \rangle=0\;.
\end{equation}
So the realization of the higher form covariant derivative in a lower form theory requires the introduction of an additional gauge symmetry.

Now let us construct $ B_{\alpha_{1}\cdots \alpha_{2J}} $ from $  \mathcal{I}^{b_{1}\cdots b_{m},p_{1}\cdots p_{n}}_{q_{1}\cdots q_{m},a_{1}\cdots a_{n}}$ explicitly. With
\begin{equation}
(\mathcal{I}_{\alpha_{1}\cdots \alpha_{n},\beta_{1}\cdots \beta_{m}})^{b_{1}\cdots b_{m}}_{a_{1}\cdots a_{n}} \equiv \epsilon_{\alpha_{1}\gamma_{1}} \cdots \epsilon_{\alpha_{n}\gamma_{n}} \mathcal{I}^{b_{1}\cdots b_{m},\gamma_{1}\cdots \gamma_{n}}_{\beta_{1}\cdots \beta_{m},a_{1}\cdots a_{n}}  \;,
\end{equation}
$(\mathcal{I}_{\alpha_{1}\cdots \alpha_{n},\beta_{1}\cdots \beta_{m}})^{b_{1}\cdots b_{m}}_{a_{1}\cdots a_{n}} $ can be decomposed into the irreducible representations of $ SU(2)_{L} $ with the spin $ J $ ranging from $ 0 $ (for the even $ m+n $) or $ \frac{1}{2} $ (for the odd $ m+n $) to $ \frac{m+n}{2}  $. The gauge indices and the spinor indices are intertwined. From (\ref{6.25}), 
\begin{equation}
\delta^{a_{i}}_{b_{j}}(\mathcal{I}_{\alpha_{1}\cdots   \alpha_{n},\beta_{1}\cdots  \beta_{m}})^{b_{1}\cdots  b_{m}}_{a_{1}\cdots  a_{n}} =\epsilon_{\alpha_{i}\beta_{j}}(\mathcal{I}_{\alpha_{1}\cdots \alpha_{i-1}\alpha_{i+1}\cdots  \alpha_{n},\beta_{1}\cdots \beta_{j-1}\beta_{j+1}\cdots \beta_{m}})^{b_{1}\cdots b_{j-1}b_{j+1}\cdots b_{m}}_{a_{1}\cdots a_{i-1}a_{i+1}\cdots a_{n}} \;,
\end{equation}
the contraction of $ a_{i} $ and $ b_{j} $ also eliminates $ \alpha_{i} $ and $ \beta_{j} $. If $ \alpha_{i} $ and $ \beta_{j} $ are symmetric, $ \mathcal{I} $ will be traceless with respect to $ a_{i} $ and $ b_{j} $. When $ m=n=1 $, 
\begin{equation}\label{trace1}
\delta^{a}_{b}(\mathcal{I}_{\alpha,\beta})^{b}_{a} =\epsilon_{\alpha\beta}\mathcal{I} \;,
\end{equation}
the trace of $ (\mathcal{I}_{\alpha,\beta})^{b}_{a} $ is a scalar. In the following, we will ignore the gauge indices and simply use $ \mathcal{I}_{\alpha_{1}\cdots \alpha_{2J}} $ to represent operators with the spin $ J $. When the gauge group is $ SU(2) $, $ \mathcal{I}_{\alpha_{1}\cdots \alpha_{2J}}  $ is in the $ \mathbf{2}^{2J}$ symmetric representation of $ SU(2) $. But for $ SU(N) $ with $ N>2 $, the gauge representation of $ \mathcal{I}_{\alpha_{1}\cdots \alpha_{2J}} $ is not uniquely determined by $ J $.

Generic operators obtained through the action of $ Q_{\alpha}^{A} $ on $ \mathcal{I}_{\alpha_{1}\cdots \alpha_{2J}}  $ are $H^{n}\mathcal{I} $, $  Q H^{n}\mathcal{I} $, $ QQ H^{n}\mathcal{I}  $, $ QQ Q H^{n}\mathcal{I} $ and $ QQQQ H^{n}\mathcal{I}  $ with $ n=0,1,\cdots $. In the $ n=0 $ class, operators with the spin $ J $ can only be $Q_{\{\alpha_{2J}}^{1}Q_{\alpha_{2J-1}}^{2} \mathcal{I}_{\alpha_{1}\cdots \alpha_{2J-2}\}}   $ and $ Q_{\{\alpha_{1}}^{1}Q_{\alpha_{2}}^{2} Q^{1\rho}Q^{2\sigma}\mathcal{I}_{\rho\sigma\alpha_{3}\cdots \alpha_{2J}\}}  $. 
\begin{itemize}
\item[(1)] If 
\begin{equation}\label{866}
 B_{\alpha_{1}\cdots \alpha_{2J}} =\frac{1}{2}Q_{\{\alpha_{2J}}^{1}Q_{\alpha_{2J-1}}^{2} \mathcal{I}_{\alpha_{1}\cdots \alpha_{2J-2}\}}\;,
\end{equation}
then 
\begin{align}
 &Q_{\{\alpha}^{A} B_{\alpha_{1}\cdots \alpha_{2J}\}} =0\;,\\  &Q_{\dot{\gamma}}^{A}B_{\alpha_{1}\cdots\alpha_{2J} }=-iK_{\{\alpha_{1}\dot{\gamma}}Q_{\alpha_{2}}^{A} \mathcal{I}_{ \alpha_{3}\cdots\alpha_{2J}  \}  } \;.
\end{align}
With the gauge equivalence 
\begin{equation}\label{8.74}
 B_{\alpha_{1}\cdots \alpha_{2J}} \sim B_{\alpha_{1}\cdots \alpha_{2J}}+\delta B_{\alpha_{1}\cdots \alpha_{2J}} =  B_{\alpha_{1}\cdots \alpha_{2J}}-\frac{i}{2}K^{\dot{\alpha}}_{\{\alpha_{1}} a_{\alpha_{2}\cdots \alpha_{2J} \}\dot{\alpha}}
\end{equation}
assumed, if
\begin{eqnarray}
 \nonumber && Q_{\dot{\gamma}}^{A}(B_{\alpha_{1}\cdots\alpha_{2J} }+\delta B_{\alpha_{1}\cdots\alpha_{2J} })\\   \nonumber &=&-iK_{\{\alpha_{1}\dot{\gamma}}Q_{\alpha_{2}}^{A} \mathcal{I}_{ \alpha_{3}\cdots\alpha_{2J}  \}  }+\frac{i}{2}[Q_{\{\alpha_{1}}^{A},h ]a_{\alpha_{2}\cdots \alpha_{2J} \}\dot{\gamma}}-\frac{i}{2}K^{\dot{\alpha}}_{\{\alpha_{1}} Q_{\dot{\gamma}}^{A}a_{\alpha_{2}\cdots \alpha_{2J} \}\dot{\alpha}}\\ &=&0
\end{eqnarray}
for some $a_{\alpha_{2}\cdots \alpha_{2J} \dot{\alpha}}  $, $  Q_{\dot{\gamma}}^{A}B_{\alpha_{1}\cdots\alpha_{2J} }=0$ can be recovered through a gauge transformation. However, in this scenario, $ B_{\alpha_{1}\cdots \alpha_{2J}} $ is in the same gauge representation as $  \mathcal{I}_{\alpha_{1}\cdots \alpha_{2J-2}} $. In particular, when $G= SU(2) $, with the gauge indices added, (\ref{866}) can be explicitly written as 
\begin{equation}\label{876}
 B^{a_{1}\cdots a_{2J-2}}_{\alpha_{1}\cdots \alpha_{2J}} =\frac{1}{2}Q_{\{\alpha_{2J}}^{1}Q_{\alpha_{2J-1}}^{2} \mathcal{I}^{a_{1}\cdots a_{2J-2}}_{\alpha_{1}\cdots \alpha_{2J-2}\}}\;.
\end{equation}
When $ J $ takes the minimum value $ 1 $, the chiral 2-form is a $ SU(2) $ singlet $ B_{\alpha\beta} =\frac{1}{2}Q_{\{\alpha}^{1}Q_{\beta\}}^{2} \mathcal{I} $, with $B_{\alpha\beta} \sim   B_{\alpha\beta} -\frac{i}{2}K^{\dot{\alpha}}_{\{\alpha} a_{\beta \}\dot{\alpha}}$.
\item[(2)] Alternatively, from
\begin{equation}\label{8990}
 HB_{\alpha_{1}\cdots \alpha_{2J}} =\frac{i}{2} Q_{\{\alpha_{1}}^{1}Q_{\alpha_{2}}^{2} Q^{1\rho}Q^{2\sigma}\mathcal{I}_{\rho\sigma\alpha_{3}\cdots \alpha_{2J}\}} \;,
\end{equation}
we can also get $ B_{\alpha_{1}\cdots \alpha_{2J}} $ satisfying $ Q_{\{\alpha}^{A} B_{\alpha_{1}\cdots \alpha_{2J}\}} =0  $. $ Q_{\dot{\gamma}}^{A}B_{\alpha_{1}\cdots\alpha_{2J} }\neq 0 $, but a suitable gauge transformation $ B_{\alpha_{1}\cdots \alpha_{2J}} \rightarrow B_{\alpha_{1}\cdots \alpha_{2J}}+\delta B_{\alpha_{1}\cdots \alpha_{2J}}   $ could make $  Q_{\dot{\gamma}}^{A}(B_{\alpha_{1}\cdots\alpha_{2J} }+\delta B_{\alpha_{1}\cdots\alpha_{2J} })= 0$. A special situation is $ J=\frac{1}{2} $, for which
\begin{equation}\label{8991}
H B_{1}=\frac{i}{2} Q_{1}^{1}Q_{1}^{2}Q_{2}^{2} Q^{1\alpha}\mathcal{I}_{\alpha}  \;,\;\;\;\;\;\;
H B_{2}=\frac{i}{2} Q_{2}^{2}Q_{2}^{1}Q_{1}^{1} Q^{2\alpha}\mathcal{I}_{\alpha} \;.
\end{equation}
$ B_{\alpha_{1}\cdots \alpha_{2J}} $ defined in (\ref{8990}) and (\ref{8991}) carry the same gauge indices as $ \mathcal{I}_{\alpha_{1}\cdots \alpha_{2J}}  $. When $ J=1 $, from $ (\mathcal{I}_{\alpha\beta} )^{b}_{a} $, we may construct $ (B_{\alpha\beta})^{b}_{a} $ in the adjoint representation of $ SU(N) $ from 
\begin{equation}\label{8992}
 H(B_{\alpha\beta})^{b}_{a}=\frac{i}{2} Q_{\{\alpha}^{1}Q_{\beta\}}^{2}Q^{1\rho} Q^{2\sigma}(\mathcal{I}_{\rho\sigma} )^{b}_{a}\;.
\end{equation}
\end{itemize}

Starting from $ \mathcal{I}_{\alpha_{1}\cdots \alpha_{n},\beta_{1}\cdots \beta_{m}} $ directly, we may define $ B_{\alpha_{1}\cdots \alpha_{2J}}  $ with (1) and (2) combined together. For example, let
\begin{equation}
H B_{\alpha_{1}\cdots \alpha_{2J}} = \frac{i}{4}   Q_{\{\alpha_{1}}^{1}Q_{\alpha_{2}}^{2} \epsilon_{BA}Q^{A\rho}Q^{B\sigma}\mathcal{I}_{\underline{\rho\sigma}\alpha_{3}\cdots \alpha_{n},\alpha_{n+1}\cdots \alpha_{2J}\}}\;,
\end{equation}
where $ \mathcal{I}_{\rho\sigma\alpha_{3}\cdots \alpha_{n},\alpha_{n+1}\cdots \alpha_{2J}} $ can also be replaced by $  \mathcal{I}_{\rho\alpha_{3}\cdots \alpha_{n},\sigma\alpha_{n+1}\cdots \alpha_{2J}} $ or $ \mathcal{I}_{\alpha_{3}\cdots \alpha_{n},\rho\sigma\alpha_{n+1}\cdots \alpha_{2J}} $, then
\begin{eqnarray}
\nonumber && H B_{\alpha_{1}\cdots \alpha_{2J}} \\ \nonumber &=&\frac{i}{4}   Q_{\{\alpha_{1}}^{1}Q_{\alpha_{2}}^{2} \epsilon_{BA}Q^{A[\rho}Q^{B\sigma]}\mathcal{I}_{\underline{\rho\sigma}\alpha_{3}\cdots \alpha_{n},\alpha_{n+1}\cdots \alpha_{2J}\}}
+ \frac{i}{4}   Q_{\{\alpha_{1}}^{1}Q_{\alpha_{2}}^{2} \epsilon_{BA}Q^{A\{\rho}Q^{B\sigma\}}\mathcal{I}_{\underline{\rho\sigma}\alpha_{3}\cdots \alpha_{n},\alpha_{n+1}\cdots \alpha_{2J}\}}\\ \nonumber  &=&\frac{1}{2}Q_{\{\alpha_{1}}^{1}Q_{\alpha_{2}}^{2} \epsilon^{\rho\sigma}H\mathcal{I}_{\underline{\rho\sigma}\alpha_{3}\cdots \alpha_{n},\alpha_{n+1}\cdots \alpha_{2J}\}}+ \frac{i}{4}   Q_{\{\alpha_{1}}^{1}Q_{\alpha_{2}}^{2} \epsilon_{BA}Q^{A\{\rho}Q^{B\sigma\}}\mathcal{I}_{\underline{\rho\sigma}\alpha_{3}\cdots \alpha_{n},\alpha_{n+1}\cdots \alpha_{2J}\}}\\\label{883}&=&-\frac{1}{2}HQ_{\{\alpha_{1}}^{1}Q_{\alpha_{2}}^{2} \mathcal{I}_{\alpha_{3}\cdots  \alpha_{2J}\}}+\frac{i}{2}Q_{\{\alpha_{1}}^{1}Q_{\alpha_{2}}^{2} Q^{1\rho}Q^{2\sigma}\mathcal{I}_{\rho\sigma\alpha_{3}\cdots  \alpha_{2J}\}}\;, 
\end{eqnarray}
where 
\begin{equation}
\mathcal{I}_{\alpha_{3}\cdots  \alpha_{2J}}= \epsilon^{\sigma\rho}\mathcal{I}_{\rho\sigma\{\alpha_{3}\cdots \alpha_{n},\alpha_{n+1}\cdots \alpha_{2J}\}}\;,\;\;\;\;\;\;\;\;\mathcal{I}_{\rho\sigma\alpha_{3}\cdots  \alpha_{2J}}=\mathcal{I}_{\{\rho\sigma\alpha_{3}\cdots \alpha_{n},\alpha_{n+1}\cdots \alpha_{2J}\}}
\;. 
\end{equation}
In (\ref{883}), $  B_{\alpha_{1}\cdots \alpha_{2J}}  $ is the sum of two parts defined in the scenario (1) and (2) respectively. The first part can be taken as the trace whose gauge transformation law (\ref{8.74}) looks like the transformation of the abelian tensor field. In particular, when $ G=SU(2) $, $ m=n=1 $,
\begin{equation}
H (B_{\alpha\beta})_{a}^{b} = \frac{i}{4}   Q_{\{\alpha}^{1}Q_{\beta\}}^{2} \epsilon_{BA}Q^{A\rho}Q^{B\sigma}(\mathcal{I}_{\rho,\sigma})_{a}^{b}=-\frac{1}{2}\delta^{b}_{a}HQ_{\{\alpha}^{1}Q_{\beta\}}^{2} \mathcal{I}+\frac{i}{2} Q_{\{\alpha}^{1}Q_{\beta\}}^{2} Q^{1\rho}Q^{2\sigma}(\mathcal{I}_{\rho\sigma})_{a}^{b}\;.
\end{equation}
$ (B_{\alpha\beta})_{a}^{b} $ is in the $ \mathbf{N} \times \bar{\mathbf{N} } $ representation of $ SU(2) $ with the trace given by the first term. As is shown in (\ref{trace1}), the trace part of $(\mathcal{I}_{\rho,\sigma})_{a}^{b}  $ is a scalar $ \mathcal{I} $, which now gets the spin under the action of $Q_{\{\alpha}^{1}Q_{\beta\}}^{2}   $.

\section{Global symmetry enhancement in $ 5d $ $ \mathcal{N} =1$ $ SU(2) $ gauge theories}\label{Nf}

It is well known that for a $ 5d $ $ \mathcal{N} =1$ $ SU(2) $ gauge theory with $ N_{f} $ flavors, at the UV fixed point, the global symmetry will enhance from $ U(1)\times SO(2N_{f}) $ to $ E_{N_{f}+1} $, where $ U(1) $ is the instanton number and $ SO(2N_{f}) $ is the flavor group, $ N_{f} \leq 7$ \cite{L1,tw10a,GNO,tw10aa,tw10ac,1bb}. When the symmetry is enhanced to $ E_{N_{f}+1}  $, BPS states should also fall into the $ E_{N_{f}+1}  $ representation. Representations of the vector and hyper multiplets under the enhanced symmetry group is classified by \cite{Hoo1} from the geometric engineering in M-theory. In this section, for $ N_{f} =0,1,\cdots,4$, we will construct the instanton vector multiplets and hypermultiplets, which, together with the original multiplets, provide the complete representations of the $ E_{N_{f}+1} $ group in \cite{Hoo1}.

Consider the $ 5d $ $ \mathcal{N} =1$ gauge theory with the gauge group $ SU(2) $. Aside from a vector multiplet with the field content $ (\phi,\lambda^{A},A) $, there are $ N_{f} $ hypermultiplets $ (q_{s}^{A},\psi_{s}) $ in the fundamental representation of $ SU(2) $. $ s=1,\cdots,N_{f} $. On the $ 1 $-instanton background, $ \psi_{s} $ and $ \bar{\psi}_{s} $ contribute $ 2N_{f} $ fermionic zero modes $K_{s} $ and $\bar{K}_{s} $ with the conformal dimension $ 0 $. On the vacuum annihilated by $ \bar{K}_{s} $, we may construct instanton operators 
\begin{equation}\label{4.33}
\mathcal{I}^{a_{1}\cdots a_{m}}_{\alpha_{1}\cdots \alpha_{m};s_{1}\cdots s_{k}}(x)= \int DU\;D_{\alpha_{1}}^{a_{1}}[u(x)]\cdots D_{\alpha_{m}}^{a_{m}}[u(x)] UK_{s_{1}}    \cdots  K_{s_{k}} I(x)U^{-1}
\end{equation}
in the spinor representation of the flavor group $ SO(2N_{f}) $. $ \mathcal{I}^{a_{1}\cdots a_{m}}_{\alpha_{1}\cdots \alpha_{m};s_{1}\cdots s_{k}}=\mathcal{I}^{a_{1}\cdots a_{m}}_{\alpha_{1}\cdots \alpha_{m};[s_{1}\cdots s_{k}]} $, $ a_{i},\alpha_{i}=1,2$, $ s_{i}=1,\cdots,N_{f} $. $  \mathcal{I}^{a_{1}\cdots a_{m}}_{\alpha_{1}\cdots \alpha_{m};s_{1}\cdots s_{k}} $ carries the flavor charge $ k-\frac{N_{f}}{2} $.

Similar to the discussion around (\ref{4.3}), some $  \mathcal{I}^{a_{1}\cdots a_{m}}_{\alpha_{1}\cdots \alpha_{m};s_{1}\cdots s_{k}} $ in (\ref{4.33}) are $ 0 $. For the gauge transformation operator $ W $ with the transformation matrix $w=-1_{2}   $,
\begin{eqnarray}\label{9.2}
\nonumber  \mathcal{I}^{a_{1}\cdots a_{m}}_{\alpha_{1}\cdots \alpha_{m};s_{1}\cdots s_{k}}(x) &=&  \int DU\;D_{\alpha_{1}}^{a_{1}}[u(x)w]\cdots D_{\alpha_{m}}^{a_{m}}[u(x)w] UWK_{s_{1}}    \cdots  K_{s_{k}}W^{-1}W I(x)W^{-1}U^{-1}\\  &=&(-1)^{m+k}\mathcal{I}^{a_{1}\cdots a_{m}}_{\alpha_{1}\cdots \alpha_{m};s_{1}\cdots s_{k}}(x) \;, 
\end{eqnarray}
so $ \mathcal{I}^{a_{1}\cdots a_{m}}_{\alpha_{1}\cdots \alpha_{m};s_{1}\cdots s_{k}}(x)\neq 0 $ only for the even $ m+k $. In particular, the spin $ \frac{1}{2} $ operator in the fundamental representation of $ SU(2) $ can be constructed as 
\begin{equation}\label{9.33}
\mathcal{I}^{a}_{\alpha;s_{1}\cdots s_{k}}(x)= \int DU\;D_{\alpha}^{a}[u(x)] UK_{s_{1}}    \cdots  K_{s_{k}}   I(x)U^{-1}
\end{equation}
with the odd $ k $; while the spin $1$ operator in the adjoint (symmetric) representation of $ SU(2) $ can be constructed as 
\begin{equation}\label{9.34}
\mathcal{I}^{ab}_{\alpha \beta;s_{1}\cdots s_{k}}(x)= \int DU\;D_{\{\alpha}^{a}[u(x)] D_{\beta\}}^{b}[u(x)] UK_{s_{1}}    \cdots  K_{s_{k}}   I(x)U^{-1}
\end{equation}
with the even $ k $.

When $G= SU(N) $ and $ N\geq 3 $, for the $ I(x)\vert \Omega\rangle $ annihilated by $ \bar{K}_{s} $, in (\ref{6.25}), with $ I(x) $ replaced by $K_{s_{1}}    \cdots  K_{s_{k}}   I(x)  $, we will get $ \mathcal{I}^{b_{1}\cdots b_{m},p_{1}\cdots p_{n}}_{q_{1}\cdots q_{m},a_{1}\cdots a_{n};s_{1}\cdots s_{k}}(x) $ in the spinor representation of $ SO(2N_{f}) $. Since $ K_{s} $ carries the $ U(1) $ charge $ N-2 $, instead of (\ref{mnn}), $ m $ and $ n $ should satisfy
\begin{equation}
m-n-k=\frac{\Delta Q_{U(1)}}{N-2}\;.
\end{equation}
In particular, when and only when $ \Delta Q_{U(1)}=0 $, we will get $ \mathcal{I}_{q,a}^{b,p} $ and $\mathcal{I}_{q;s}^{b}  $, which could be mapped into the adjoint vector and the fundamental hyper of the theory. The rest $ \mathcal{I}^{b_{1}\cdots b_{n+k},p_{1}\cdots p_{n}}_{q_{1}\cdots q_{n+k},a_{1}\cdots a_{n};s_{1}\cdots s_{k}}(x) $ with $ k\geq 2 $ do not have the correspondence in the field theory. So, the symmetry enhancement occurs when $ \Delta Q_{U(1)}=0   $, and the enhancement pattern can only be $ U(1) \times SO(2N_{f}) \rightarrow SU(2) \times SO(2N_{f}) $, which is also the conclusion in \cite{1dc,Hoo3}.

The fermionic zero mode $K$ can be written as\footnote{In the $ 5d $ $ \mathcal{N}=2 $ theory, $\bar{\mu}^{A}_{a}=\int d^{4}x \; (\lambda^{A}_{\alpha})^{p}_{a}f_{p}^{\alpha }  $. Using (\ref{59a}), the action of $ Q_{\dot{\alpha}}^{B}  $ with $ \Omega^{BA} =0$ gives
\begin{eqnarray}\label{9.8i}
\nonumber Q_{\dot{\alpha}}^{B} \bar{\mu}^{A}_{a}I(x)\vert\Omega\rangle &=& \{Q_{\dot{\alpha}}^{B} ,\bar{\mu}^{A}_{a}\}I(x)\vert\Omega\rangle=\int d^{4}x \;\{Q_{\dot{\alpha}}^{B} ,  (\lambda^{A}_{\alpha})^{p}_{a}\}f_{p}^{\alpha }I(x)\vert\Omega\rangle=\int d^{4}x \;(D_{\alpha\dot{\alpha}}\phi^{AB} )_{a}^{p}f_{p}^{\alpha }I(x)\vert\Omega\rangle\\  &=&-\int d^{4}x \;(\phi^{AB} )_{a}^{p}D_{\alpha\dot{\alpha}}f_{p}^{\alpha }I(x)\vert\Omega\rangle=0 \;.
\end{eqnarray}
}
\begin{equation}
K=\int d^{4}x \; \psi^{p}_{\alpha}f_{p}^{\alpha }
\end{equation}
with $ f_{p}^{\alpha } $ satisfying the zero mode equation $  D_{\alpha\dot{\alpha}}f^{\alpha}_{p}=0$, where $ p=1,2 $, and $ A_{\alpha\dot{\alpha}} $ is taken to be (\ref{mn}). Under the action of $Q_{\dot{\alpha}}^{A}  $, from (\ref{from}), 
\begin{eqnarray}\label{9.10a}
\nonumber  Q_{\dot{\alpha}}^{A} KI(x)\vert\Omega\rangle &=&  \{Q_{\dot{\alpha}}^{A} ,K\}I(x)\vert\Omega\rangle=\int d^{4}x f_{p}^{\alpha }\{Q_{\dot{\alpha}}^{A} , \psi^{p}_{\alpha}\}I(x)\vert\Omega\rangle=\int d^{4}x\sqrt{2}if_{p}^{\alpha } D_{\alpha\dot{\alpha}}q^{Ap}I(x)\vert\Omega\rangle\\  &=&-\int d^{4}x\sqrt{2}i q^{Ap} D_{\alpha\dot{\alpha}}  f_{p}^{\alpha } I(x)\vert\Omega\rangle=0 \;, 
\end{eqnarray}
where the singular contribution to $  A_{\alpha\dot{\alpha}}  $ is $ a_{\alpha\dot{\alpha}} $ and the regular contribution gives zero after the integration. So 
\begin{equation}
Q_{\dot{\alpha}}^{A}  \mathcal{I}^{a}_{\alpha ;s_{1}\cdots s_{k}}(x) \vert\Omega\rangle=Q_{\dot{\alpha}}^{A}  \mathcal{I}^{ab}_{\alpha \beta;s_{1}\cdots s_{k}}(x)\vert\Omega\rangle =0\;.
\end{equation}
With $ \mathcal{I}_{\alpha} $ and $  \mathcal{I}_{\rho\sigma} $ in (\ref{8991}) and (\ref{8992}) replaced by (\ref{9.33}) and (\ref{9.34}), we can also obtain the corresponding $HB^{a}_{\alpha;s_{1}\cdots s_{k}}  $ and $ HB^{ab}_{\alpha \beta;s_{1}\cdots s_{k}} $ with 
\begin{equation}
Q_{\{\alpha}^{A}HB^{a}_{\beta\};s_{1}\cdots s_{k}}  =0\;,\;\;\;\;\;\;Q_{\{\gamma}^{A}HB^{ab}_{\alpha \beta\};s_{1}\cdots s_{k}} =0\;.
\end{equation}
Hence, with some additional gauge equivalence relation assumed, we may get the chiral operators $B^{a}_{\alpha;s_{1}\cdots s_{k}}    $ and $ B^{ab}_{\alpha \beta;s_{1}\cdots s_{k}}  $ satisfying the equations (\ref{ali22})-(\ref{ali33}) and (\ref{ali})-(\ref{alii}) just as $B^{a}_{\alpha}    $ and $ B^{ab}_{\alpha \beta}  $.

When the flavor symmetry is enhanced to $ E_{N_{f}+1}  $, the vector multiplets $ (F^{ab}_{\alpha\beta},\lambda_{\alpha}^{Aab} ,\phi^{ab}) $ and $(HB^{ab}_{\alpha\beta;s_{1}\cdots s_{k}},\Psi_{\alpha;s_{1}\cdots s_{k}}^{Aab} ,\Phi^{ab}_{s_{1}\cdots s_{k}})  $ may live in the same representation, while the hypermultiplets $ (\psi^{a}_{\alpha s},q_{s}^{Aa} ) $ and $ (HB^{a}_{\alpha;s_{1}\cdots s_{k}},\Psi_{s_{1}\cdots s_{k}}^{Aa} ) $ may live in the same representation. Supermultiplets related by a flavor symmetry transformation should follow the same supersymmetry transformation rule and the same equations of motion. For $B^{a}_{\alpha;s_{1}\cdots s_{k}}    $, we have
\begin{align}
 &\label{910}Q^{A}_{ \beta}(\sqrt{2} H B _{\alpha;s_{1}\cdots s_{k} } )=\sqrt{2}\epsilon_{\alpha\beta } H \Psi_{s_{1}\cdots s_{k}}^{A} \;,& Q^{B}_{ \beta} \Psi_{s_{1}\cdots s_{k}}^{A}&=\sqrt{2} i\epsilon^{BA}(\sqrt{2} H B _{\beta ;s_{1}\cdots s_{k}     })
\;,\\ & Q_{\dot{\beta}}^{A}(\sqrt{2}H B _{\alpha ;s_{1}\cdots s_{k}     })=-\sqrt{2} K_{\alpha\dot{\beta}}\Psi_{s_{1}\cdots s_{k}}^{A}
\;, &Q_{\dot{\beta}}^{B}\Psi_{s_{1}\cdots s_{k}}^{A} &=\sqrt{2}i\epsilon^{BA}(-\frac{\sqrt{2}}{2}K^{\beta}_{\dot{\beta}}B_{\beta;s_{1}\cdots s_{k} })
\end{align}
and 
\begin{equation}\label{912}
H(-\frac{\sqrt{2}}{2}K^{\beta}_{\dot{\gamma}}B_{\beta ;s_{1}\cdots s_{k}    })+K^{\beta}_{\dot{\gamma}}(\sqrt{2}HB_{ \beta ;s_{1}\cdots s_{k}   })-\frac{i}{\sqrt{2}} [Q_{\dot{\gamma}}^{A}, H] \Psi_{A ;s_{1}\cdots s_{k}   }=0
\end{equation}
as the fermionic equation of motion. The supersymmetry transformation of $ (\psi^{a}_{\alpha s},q_{s}^{Aa} ) $ is
\begin{align}
&\label{913}Q_{\beta}^{A} \psi_{\alpha s}  =
\sqrt{2}\epsilon_{\alpha\beta} Hq_{s}^{A}    \;,&
Q_{\beta}^{B}  q_{s}^{A}&= \sqrt{2}i \epsilon^{BA}\psi_{\beta s}
 \;,\\ &
Q_{\dot{\beta}}^{A}  \psi_{\alpha s} =-\sqrt{2}K_{\alpha\dot{\beta}}q_{s}^{A}
\;, &Q_{\dot{\beta}}^{B}  q_{s}^{A} &= \sqrt{2} i \epsilon^{BA} \psi_{\dot{\beta} s} \;,
\end{align}
from which, the fermionic equation of motion is obtained as 
\begin{equation} \label{915}
H\psi_{\dot{\gamma}s}+K^{\beta}_{\dot{\gamma}}\psi_{\beta s}-\frac{i}{\sqrt{2}} [Q_{\dot{\gamma}}^{A},H]q_{As}  =0\;.
\end{equation}
In (\ref{913})-(\ref{915}), $ H $ acts as $ P_{0}+Z $. When $ R_{5} \rightarrow \infty $, the action of $ H $ in (\ref{910})-(\ref{912}) also reduces to $ P_{0}+Z $. With (\ref{910})-(\ref{912}) compared with (\ref{913})-(\ref{915}), we find that the mapping
\begin{equation}
(\sqrt{2} H B _{\alpha;s_{1}\cdots s_{k} } ,\Psi_{s_{1}\cdots s_{k}}^{A},-\frac{\sqrt{2}}{2}K^{\beta}_{\dot{\gamma}}B_{\beta ;s_{1}\cdots s_{k}    })\Leftrightarrow ( \psi_{\alpha s} ,q_{s}^{A},\psi_{\dot{\gamma}s})
\end{equation}
is allowed under the flavor symmetry transformation. For $B^{ab}_{\alpha\beta;s_{1}\cdots s_{k}}    $, we have
\begin{eqnarray}
\label{910o} &&Q^{A}_{\gamma}( \frac{i}{2}H B _{\alpha\beta ;s_{1}\cdots s_{k}   } )=i\epsilon_{\alpha\gamma }  H\Psi^{A}_{\beta ;s_{1}\cdots s_{k}}+i\epsilon_{\beta\gamma }  H\Psi^{A}_{\alpha;s_{1}\cdots s_{k} }, \\ &&Q_{\dot{\gamma}}^{A} (\frac{i}{2}H B _{\alpha\beta ;s_{1}\cdots s_{k} } )=-i K_{\alpha\dot{\gamma}}\Psi ^{A}_{\beta;s_{1}\cdots s_{k}   }-i K_{\beta\dot{\gamma}}\Psi ^{A}_{\alpha ;s_{1}\cdots s_{k}   }  , \\ && Q^{B}_{ \beta} \Psi ^{A}_{\alpha ;s_{1}\cdots s_{k}     } =\epsilon^{BA}(\frac{i}{2}H B _{\alpha\beta ;s_{1}\cdots s_{k}    } + i\epsilon_{\alpha\beta  }H\Phi_{s_{1}\cdots s_{k}  }),\;Q_{\dot{\beta}}^{B}\Psi ^{A}_{\alpha ;s_{1}\cdots s_{k}        } =\frac{i}{3}\epsilon^{AB}K^{\beta}_{\dot{\beta}}B_{\alpha\beta ;s_{1}\cdots s_{k}  },
\\  && Q^{A}_{ \beta}H \Phi_{ s_{1}\cdots s_{k}} =-H\Psi_{\beta ;s_{1}\cdots s_{k}}^{A},\;\;\;\;Q_{\dot{\beta}}^{A} H\Phi_{ s_{1}\cdots s_{k}}=K^{\beta}_{\dot{\beta}} \Psi ^{A}_{\beta  ;s_{1}\cdots s_{k}      }.
\end{eqnarray}
With $ Q_{\dot{\beta}}^{A} \Phi_{s_{1}\cdots s_{k}   }\equiv-\Psi_{\dot{\beta};s_{1}\cdots s_{k}   }^{A}  $, the last equation gives the fermionic equation of motion
\begin{equation}\label{912o}
H\Psi_{\dot{\beta};s_{1}\cdots s_{k}   }^{A} +K^{\beta}_{\dot{\beta}} \Psi ^{A}_{\beta  ;s_{1}\cdots s_{k}       }+[H,Q_{\dot{\beta}}^{A}]\Phi_{s_{1}\cdots s_{k}   }=0.
\end{equation}
The supersymmetry transformation of $ (F^{ab}_{\alpha\beta},\lambda_{\alpha}^{Aab} ,\phi^{ab}) $ is
\begin{align}
 \label{913o}&Q^{A}_{\gamma} F_{\alpha\beta}=i \epsilon_{\alpha\gamma} H\lambda_{\beta}^{A}+i \epsilon_{\beta\gamma}H\lambda_{\alpha}^{A} , &Q_{\dot{\gamma}}^{A} F_{\alpha\beta} &=-i K_{\alpha\dot{\gamma}} \lambda_{\beta}^{A} -i K_{\beta\dot{\gamma}} \lambda_{\alpha}^{A} ,\\  &Q_{\beta}^{B} \lambda_{\alpha}^{A} =
\epsilon^{BA} (F_{\alpha\beta}+ i\epsilon_{\alpha\beta}  H\phi )\;,&Q_{\dot{\beta}}^{B} \lambda_{\alpha}^{A} &=i\epsilon^{AB} hA_{\alpha\dot{\beta}} \;,\\  &Q^{A}_{ \beta}H\phi=-H\lambda_{\beta}^{A}\;,&Q_{\dot{\beta}}^{A} H\phi&=K^{\beta}_{\dot{\beta}} \lambda^{A}_{\beta     }
\end{align}
with the fermionic equation of motion
\begin{equation}\label{915o}
H\lambda_{\dot{\beta}}^{A}+K^{\alpha}_{\dot{\beta}} \lambda^{A}_{\alpha       }+[ H,Q_{\dot{\beta}}^{A} ]\phi=0
\end{equation}
coming from the last equation. In (\ref{913o})-(\ref{915o}), $ H $ acts as $ P_{0}+Z $. When $ R_{5} \rightarrow \infty $, the action of $ H $ in (\ref{910o})-(\ref{912o}) also reduces to $ P_{0}+Z $. With (\ref{910o})-(\ref{912o}) compared with (\ref{913o})-(\ref{915o}), the mapping
\begin{equation}
(\frac{i}{2}H B _{\alpha\beta ;s_{1}\cdots s_{k}   }  ,\Psi^{A}_{\beta ;s_{1}\cdots s_{k}},\Phi_{ s_{1}\cdots s_{k}},\Psi_{\dot{\beta};s_{1}\cdots s_{k}   }^{A} )\Leftrightarrow ( F_{\alpha\beta},\lambda_{\beta}^{A},\phi,\lambda_{\dot{\beta}}^{A})
\end{equation}
is allowed under the flavor symmetry transformation.

Now we are ready to decompose $ 1+2^{N_{f}-1} $ vectors $\{ F_{\alpha\beta} , H B _{\alpha\beta ;s_{1}\cdots s_{k}   } \}$ and $ 2N_{f}+2^{N_{f}-1}  $ hypers $\{ \psi_{s} , \bar{\psi}_{s} , H B _{\alpha ;s_{1}\cdots s_{k}   } \}$ into the irreducible $ E_{N_{f}+1} $ representations in \cite{Hoo1}. We will only discuss the situation for $ N_{f} \leq 4$. When $ N_{f} \geq 4$, operators with more than $ 1 $ instanton numbers are required to fill the complete representation.

\subsection{$SU(2)$}

For the pure $ SU(2) $ gauge theory, the original global symmetry is $ U(1) $. The symmetry enhancement pattern is 
\begin{equation}
 U(1)  \rightarrow E_{1}= SU(2) \;.
\end{equation}
We use $ (n_{e}; I) $ to represent the electric charge and the instanton number. Then BPS states and the associated charges are given by \\

\noindent W boson: $ (1; 0) \sim (F_{\alpha\beta})^{ab} $\\
Instanton: $ (1; 1)     \sim H(B_{\alpha\beta})^{ab}  $\\

\noindent With the charge of the $ U(1) $ Carton subgroup of $ SU(2) $ defined as (see \cite{2} for the interpretation for such rearrangement of charges in $ E_{n} $ field theories)
\begin{equation}
q_{1}=\frac{1}{2}n_{e} -I\;,
\end{equation}
the vectors compose the $ 2 $ representation of $ SU(2) $:
\begin{equation}
 [ (F_{\alpha\beta})^{ab}    ,  H(B_{\alpha\beta})^{ab}  ]   \sim  [ (1; \frac{1}{2} )  ,   (1;-\frac{1}{2})   ]\;.
\end{equation}

\subsection{$SU(2)+1\mathbf{F}$}

For the $ SU(2) $ gauge theory with $ 1 $ flavor, the original global symmetry is $ U(1) \times SO(2)   $. The symmetry enhancement pattern is 
\begin{equation}
 U(1) \times SO(2)  \rightarrow  E_{2}=SU(2) \times U(1)\;. 
\end{equation}
We use $ (n_{e}; I,Q_{1}) $ to represent the electric charge, the instanton number and the flavor charge. Then BPS states and the associated charges are given by \\

\noindent W boson: $ (1; 0,0) \sim  (F_{\alpha\beta})^{ab} $\\
Quark: $ (\frac{1}{2}; 0,1)\sim (\psi_{\alpha} )^{a}$, $(\frac{1}{2}; 0,-1)  \sim (\bar{\psi}^{\alpha})^{a}$\\
Instanton: $ (1; 1,-\frac{1}{2}) \sim H(B_{\alpha\beta})^{ab}   \rightarrow  (\frac{1}{2}; 1,\frac{1}{2}) \sim  H(B_{\alpha;1})^{a}  $\\

\noindent With the charges of the $ U(1) \times U(1)$ Carton subgroup of $ SU(2) \times U(1) $ given by
\begin{equation}
q_{1}=\frac{1}{2}n_{e} -\frac{7}{8}I+\frac{1}{4}Q_{1},\;\;\;\;\;\;  q_{2}= \frac{1}{2}I+Q_{1},
  \end{equation}  
the vectors are in the $ 2_{0} $ representation
\begin{equation}
[(F_{\alpha\beta})^{ab},H(B_{\alpha\beta})^{ab} ]\sim   [ ( 1;\frac{1}{2},0)   ,  ( 1;-\frac{1}{2},0)   ],
\end{equation}
while the hypers are in the $ 2_{1}$ representation
\begin{equation}
[ (\psi_{\alpha} )^{a} ,  H(B_{\alpha;1})^{a}  ]\sim  [ ( \frac{1}{2};\frac{1}{2},1)  , ( \frac{1}{2};-\frac{1}{2},1)  ] 
\end{equation}
and the $ 1_{-1} $ representation
\begin{equation}
(\bar{\psi}^{\alpha})^{a} \sim    ( \frac{1}{2};0,-1)  \;.
\end{equation}

\subsection{$SU(2)+2 \mathbf{F}$}

For the $ SU(2) $ gauge theory with $ 2 $ flavors, the original global symmetry is $ U(1) \times SO(4)   $. The symmetry enhancement pattern is 
\begin{equation}
 U(1) \times SO(4)  \rightarrow  E_{3}=SU(3) \times SU(2)\;.
\end{equation}
We use $ (n_{e}; I,Q_{1},Q_{2}) $ to represent the electric charge, the instanton number and the flavor charges. Then BPS states and the associated charges are given by \\

\noindent W boson: $ (1; 0,0,0) \sim ( F_{\alpha\beta})^{ab}$\\
Quark: $ (\frac{1}{2}; 0,1,0) , (\frac{1}{2}; 0,0,1)  , (\frac{1}{2}; 0,-1,0)  , (\frac{1}{2}; 0,0,-1) \sim (\psi_{\alpha;s} )^{a},(\bar{\psi}^{\alpha}_{s} )^{a}$\\
Instanton: 
\begin{eqnarray}
\nonumber && (1; 1,-\frac{1}{2},-\frac{1}{2}) \sim H(B_{\alpha\beta})^{ab}   \rightarrow[  (\frac{1}{2}; 1,\frac{1}{2},-\frac{1}{2}), (\frac{1}{2}; 1,-\frac{1}{2},\frac{1}{2}) ]  \sim H(B_{\alpha;s})^{a} \\ \nonumber &\rightarrow& (1; 1,\frac{1}{2},\frac{1}{2}) \sim H(B_{\alpha\beta;12})^{ab} 
\end{eqnarray}
$ s=1,2 $. With the charges of the $ U(1) \times U(1) \times U(1)$ Carton subgroup of $ SU(3) \times SU(2)$ given by 
\begin{equation}
 q_{1}=Q_{1}+Q_{2},\;\;\;\;\;  q_{2}= n_{e}-\frac{3}{2}I-\frac{1}{2}Q_{1}-\frac{1}{2}Q_{2},\;\;\;\;\;  q_{3}=\frac{1}{2}Q_{1}-\frac{1}{2}Q_{2},
\end{equation}
the vectors are in the $ (\bar{3},1)$ representation with
\begin{equation}
[  ( F_{\alpha\beta})^{ab} ,H(B_{\alpha\beta;12})^{ab} , H(B_{\alpha\beta})^{ab}  ]\sim [    (1; 0,1,0),(1; 1,-1,0) , (1; -1,0,0)   ],
\end{equation}
while the hypers are in the $ (3,2) $ representation with 
\begin{eqnarray}
 &&    [ (\psi_{\alpha;1} )^{a}  , (\bar{\psi}^{\alpha}_{2} )^{a}  ,   H(B_{\alpha;1})^{a}  ]\sim [ ( \frac{1}{2};1,0,\frac{1}{2})  ,(\frac{1}{2};  -1,1,\frac{1}{2})  ,  ( \frac{1}{2}; 0,-1,\frac{1}{2}) ] \\  & & [(\psi_{\alpha;2} )^{a} ,  (\bar{\psi}^{\alpha}_{1} )^{a}  ,  H(B_{\alpha;2})^{a}  ] \sim [ (\frac{1}{2}; 1,0,-\frac{1}{2})  , (\frac{1}{2}; -1,1,-\frac{1}{2})  ,  ( \frac{1}{2};0,-1,-\frac{1}{2}) ].
\end{eqnarray}

\subsection{$SU(2)+3 \mathbf{F}$}

For the $ SU(2) $ gauge theory with $ 3 $ flavors, the original global symmetry is $ U(1) \times SO(6)   $. The symmetry enhancement pattern is 
\begin{equation}
 U(1) \times SO(6)  \rightarrow E_{4}= SU(5) \;.
\end{equation}
We use $  (n_{e}; I,Q_{1}, Q_{2}, Q_{3}) $ to represent the electric charge, the instanton number and the flavor charges. Then BPS states and the associated charges are given by
\\

\noindent W boson: $ (1; 0,0,0,0) \sim (F_{\alpha\beta})^{ab}$\\
Quark: 
\begin{eqnarray}
\nonumber &&  (\frac{1}{2}; 0,1,0,0)  , (\frac{1}{2}; 0,0,1,0)  , (\frac{1}{2}; 0,0,0,1), (\frac{1}{2}; 0,-1,0,0)   , (\frac{1}{2}; 0,0,-1,0)   , (\frac{1}{2}; 0,0,0,-1)  \\\nonumber  &\sim & (\psi_{\alpha;s} )^{a},(\bar{\psi}^{\alpha}_{s} )^{a}
\end{eqnarray}
Instanton: 
\begin{eqnarray}
\nonumber && (1; 1,-\frac{1}{2},-\frac{1}{2},-\frac{1}{2}) \sim H(B_{\alpha\beta})^{ab} \\\nonumber   &\rightarrow & [(\frac{1}{2}; 1,\frac{1}{2},-\frac{1}{2},-\frac{1}{2})   , (\frac{1}{2}; 1,-\frac{1}{2},\frac{1}{2},-\frac{1}{2})   , (\frac{1}{2}; 1,-\frac{1}{2},-\frac{1}{2},\frac{1}{2})  ] \sim H(B_{\alpha;s})^{a}       \\\nonumber   &\rightarrow & [(1; 1,-\frac{1}{2},\frac{1}{2},\frac{1}{2})   , (1; 1,\frac{1}{2},-\frac{1}{2},\frac{1}{2}) , (1; 1,\frac{1}{2},\frac{1}{2},-\frac{1}{2})]  \sim H(B_{\alpha\beta;s_{1}s_{2}})^{ab}     \\\nonumber  &\rightarrow & [ (\frac{1}{2}; 1,\frac{1}{2},\frac{1}{2},\frac{1}{2})  ] \sim H(B_{\alpha;123})^{a}
\end{eqnarray}
$ s,s_{1},s_{2}=1,2 $. With the charges of the $ U(1) \times U(1) \times U(1)\times U(1)$ Carton subgroup of $ SU(5) $ given by 
\begin{eqnarray}
\nonumber  q_{1}&=&Q_{2}+Q_{3}\;,\;\;\;\;\; q_{2}= Q_{1}-Q_{2}\;,\;\;\;\;\;  q_{3}=Q_{2}-Q_{3}\;,\\  q_{4}&=&n_{e}-\frac{5}{4}I-\frac{1}{2}Q_{1}-\frac{1}{2}Q_{2}+\frac{1}{2}Q_{3}\;,
\end{eqnarray}
the vectors are in the $ \bar{5}$ representation with 
\begin{eqnarray}
\nonumber &&[ (F_{\alpha\beta})^{ab}),  H(B_{\alpha\beta})^{ab}  , H(B_{\alpha\beta;23})^{ab}  , H(B_{\alpha\beta;13})^{ab} , H(B_{\alpha\beta;12})^{ab} ] \\  &\sim & [ ( 1;0,0,0,1),   (1;-1, 0,0,0) ,( 1;1,-1,0,0) , (1; 0,1,-1,0) , ( 1;0,0,1,-1)  ]\;,
\end{eqnarray}
while the hypers are in $ 10 $ representation with
\begin{eqnarray}
\nonumber && [ (\psi_{\alpha;1} )^{a} , (\bar{\psi}^{\alpha}_{1} )^{a}  ,(\psi_{\alpha;2} )^{a} , (\bar{\psi}^{\alpha}_{2} )^{a} , (\psi_{\alpha;3} )^{a} , (\bar{\psi}^{\alpha}_{3} )^{a},   H(B_{\alpha\beta;1})^{a}   , H(B_{\alpha\beta;2})^{a}   , H(B_{\alpha\beta;3})^{a} , H(B_{\alpha\beta;123})^{a}  ]\\\nonumber   &\sim &  [ (\frac{1}{2}; 0,1,0,0)  , ( \frac{1}{2};0,-1,0,1)  , (\frac{1}{2}; 1,-1,1,0)  ,  ( \frac{1}{2};-1,1,-1,1)  ,  (\frac{1}{2}; 1,0,-1,1), (\frac{1}{2};-1,0,1,0),\\& & ( \frac{1}{2};-1,1,0,-1)  , ( \frac{1}{2};0,-1,1,-1)  , (\frac{1}{2}; 0,0,-1,0)  ,  ( \frac{1}{2};1,0,0,-1)  ]\;.
\end{eqnarray}

\subsection{$SU(2)+4 \mathbf{F}$}

For the $ SU(2) $ gauge theory with $ 4 $ flavors, the original global symmetry is $ U(1) \times SO(8)   $. The symmetry enhancement pattern is 
\begin{equation}
 U(1) \times SO(8)  \rightarrow  E_{5}=Spin(10) \;.
\end{equation}
We use $ (n_{e}; I,Q_{1}, Q_{2}, Q_{3},Q_{4}) $ to represent the electric charge, the instanton number and the flavor charges. Then BPS states and the associated charges are given by
\\

\noindent W boson: $  (1; 0,0,0,0,0)\sim (F_{\alpha\beta})^{ab} $\\
Quark: 
\begin{eqnarray}
\nonumber &&(\frac{1}{2}; 0,1,0,0,0)  , (\frac{1}{2}; 0,0,1,0,0)  , (\frac{1}{2}; 0,0,0,1,0)  , (\frac{1}{2}; 0,0,0,0,1)  ,\\ \nonumber && (\frac{1}{2}; 0,-1,0,0,0)  , (\frac{1}{2}; 0,0,-1,0,0)  , (\frac{1}{2}; 0,0,0,-1,0)  , (\frac{1}{2}; 0,0,0,0,-1) \sim (\psi_{\alpha;s} )^{a},(\bar{\psi}^{\alpha}_{s} )^{a}
\end{eqnarray}
Instanton: 
\begin{eqnarray}
\nonumber && (1; 1,-\frac{1}{2},-\frac{1}{2},-\frac{1}{2},-\frac{1}{2}) \sim H(B_{\alpha\beta})^{ab}\\ \nonumber & \rightarrow &  (\frac{1}{2}; 1,\frac{1}{2},-\frac{1}{2},-\frac{1}{2},-\frac{1}{2})   , (\frac{1}{2}; 1,-\frac{1}{2},\frac{1}{2},-\frac{1}{2},-\frac{1}{2})   , \\ \nonumber && (\frac{1}{2}; 1,-\frac{1}{2},-\frac{1}{2},\frac{1}{2},-\frac{1}{2})   , (\frac{1}{2}; 1,-\frac{1}{2},-\frac{1}{2},-\frac{1}{2},\frac{1}{2})  \sim H(B_{\alpha;s})^{a}    \\\nonumber  &\rightarrow & 
(1; 1,-\frac{1}{2},\frac{1}{2},\frac{1}{2},-\frac{1}{2})   , (1; 1,\frac{1}{2},-\frac{1}{2},\frac{1}{2},-\frac{1}{2}) , (1; 1,\frac{1}{2},-\frac{1}{2},-\frac{1}{2},\frac{1}{2})  , \\ \nonumber && 
(1; 1,-\frac{1}{2},-\frac{1}{2},\frac{1}{2},\frac{1}{2})   , (1; 1,-\frac{1}{2},\frac{1}{2},-\frac{1}{2},\frac{1}{2}) , (1; 1,\frac{1}{2},\frac{1}{2},-\frac{1}{2},-\frac{1}{2})\sim H(B_{\alpha\beta;s_{1}s_{2}})^{ab}   \\\nonumber & \rightarrow & 
 (\frac{1}{2}; 1,-\frac{1}{2},\frac{1}{2},\frac{1}{2},\frac{1}{2}) ,(\frac{1}{2}; 1,\frac{1}{2},-\frac{1}{2},\frac{1}{2},\frac{1}{2}),(\frac{1}{2}; 1,\frac{1}{2},\frac{1}{2},-\frac{1}{2},\frac{1}{2}) ,(\frac{1}{2}; 1,\frac{1}{2},\frac{1}{2},\frac{1}{2},-\frac{1}{2}) \sim H(B_{\alpha;s_{1}s_{2}s_{3}})^{a} \\\nonumber &\rightarrow &   (1; 1,\frac{1}{2},\frac{1}{2},\frac{1}{2},\frac{1}{2}) \sim H(B_{\alpha\beta;1234})^{ab}  
\end{eqnarray}
$ s,s_{1},s_{2},s_{3}=1,2 $. With the charges of the $ U(1) \times U(1) \times U(1)\times U(1)\times U(1)$ Carton subgroup of $ Spin(10) $ defined as  
\begin{eqnarray}
\nonumber  q_{1}&=& -\frac{1}{2}Q_{1}-\frac{1}{2} Q_{2}-\frac{1}{2} Q_{3}-\frac{1}{2}Q_{4}\;, \;\;\;\;\; q_{2}= -\frac{1}{2}Q_{1}+\frac{1}{2} Q_{2}+\frac{1}{2} Q_{3}-\frac{1}{2}Q_{4} \;,  \\ \nonumber  q_{3}  & = &\frac{1}{2}Q_{1}-\frac{1}{2} Q_{2}+\frac{1}{2} Q_{3}-\frac{1}{2}Q_{4}\; , \;\;\;\;\;  q_{4}=-\frac{1}{2}Q_{1}-\frac{1}{2} Q_{2}+\frac{1}{2} Q_{3}+\frac{1}{2}Q_{4} \;, \;\;\;\;\;  q_{5}= n_{e}- I\;,\\
\end{eqnarray}
the vectors are in the $ 10 $ representation with
\begin{eqnarray}\label{9.49}
\nonumber && [ (F_{\alpha\beta})^{ab} , (1; 2,0,0,0,0),  H(B_{\alpha\beta})^{ab} ,H(B_{\alpha\beta;23})^{ab}, H(B_{\alpha\beta;13})^{ab}, H(B_{\alpha\beta;34})^{ab} ,  \\ \nonumber &  & H(B_{\alpha\beta;1234})^{ab}, H(B_{\alpha\beta;14})^{ab},H(B_{\alpha\beta;24})^{ab}  ,  H(B_{\alpha\beta;12})^{ab} ] \\ \nonumber &   &\sim \\ \nonumber &   & [ (1; 0,0,0,0,1), (1; 0,0,0,0,-1), \\ \nonumber &  &  (1;1, 0,0,0,0) ,(1;0,1,0,0,0) , ( 1;0,0,1,0,0), ( 1;0,0,0,1,0) ,\\  &  & (1;-1, 0,0,0,0) ,(1;0,-1,0,0,0) , ( 1;0,0,-1,0,0), (1; 0,0,0,-1,0)  ]\;,
\end{eqnarray}
while the hypers are in the $ 16 $ representation with
\begin{eqnarray}
\nonumber && 
[(\psi_{\alpha;1} )^{a}  , (\psi_{\alpha;2} )^{a}  , (\psi_{\alpha;3} )^{a}  ,(\psi_{\alpha;4} )^{a}  ,(\bar{\psi}^{\alpha}_{1} )^{a}  , (\bar{\psi}^{\alpha}_{2} )^{a}  , (\bar{\psi}^{\alpha}_{3} )^{a} , (\bar{\psi}^{\alpha}_{4} )^{a},
  \\ \nonumber &  & H(B_{\alpha;1})^{a}   , H(B_{\alpha;2})^{a}      , H(B_{\alpha;3})^{a}      , H(B_{\alpha;4})^{a}    ,
H(B_{\alpha;234})^{a}   ,H(B_{\alpha;134})^{a}  ,H(B_{\alpha;124})^{a}   ,H(B_{\alpha;123})^{a}  ]\\ \nonumber &   &\sim
 \\ \nonumber 
&   & [ (\frac{1}{2}; -\frac{1}{2}, -\frac{1}{2},\frac{1}{2},-\frac{1}{2},\frac{1}{2}), (\frac{1}{2};- \frac{1}{2}, \frac{1}{2},-\frac{1}{2},-\frac{1}{2},\frac{1}{2}), (\frac{1}{2}; -\frac{1}{2}, \frac{1}{2},\frac{1}{2},\frac{1}{2},\frac{1}{2}), (\frac{1}{2};- \frac{1}{2}, -\frac{1}{2},-\frac{1}{2},\frac{1}{2},\frac{1}{2}), 
\\ \nonumber &  &  (\frac{1}{2}; \frac{1}{2},\frac{1}{2},-\frac{1}{2},\frac{1}{2},\frac{1}{2}), (\frac{1}{2}; \frac{1}{2},-\frac{1}{2},\frac{1}{2},\frac{1}{2},\frac{1}{2}), ( \frac{1}{2};\frac{1}{2},-\frac{1}{2},-\frac{1}{2},-\frac{1}{2},\frac{1}{2}), (\frac{1}{2}; \frac{1}{2},\frac{1}{2},\frac{1}{2},-\frac{1}{2},\frac{1}{2}),
\\ \nonumber &  &   ( \frac{1}{2};\frac{1}{2},-\frac{1}{2},\frac{1}{2},-\frac{1}{2},-\frac{1}{2}), (\frac{1}{2}; \frac{1}{2},\frac{1}{2},-\frac{1}{2},-\frac{1}{2},-\frac{1}{2}), (\frac{1}{2}; \frac{1}{2},\frac{1}{2},\frac{1}{2},\frac{1}{2},-\frac{1}{2}), (\frac{1}{2}; \frac{1}{2},-\frac{1}{2},-\frac{1}{2},\frac{1}{2},-\frac{1}{2}),
\\ \nonumber &  &   (\frac{1}{2}; -\frac{1}{2},\frac{1}{2},-\frac{1}{2},\frac{1}{2},-\frac{1}{2}), (\frac{1}{2}; -\frac{1}{2},-\frac{1}{2},\frac{1}{2},\frac{1}{2},-\frac{1}{2}), (\frac{1}{2};- \frac{1}{2},-\frac{1}{2},-\frac{1}{2},-\frac{1}{2},-\frac{1}{2}), (\frac{1}{2}; -\frac{1}{2},\frac{1}{2},\frac{1}{2},-\frac{1}{2},-\frac{1}{2})  ]\;.\\
\end{eqnarray}
Note that in (\ref{9.49}),  a vector multiplet with the instanton number $ 2 $ and the electric charge $ 1 $ is required to complete the $ 10 $ representation.

\section{Conclusion and discussion}\label{con1}

The goal of this paper is to find the BPS spectrum of the instanton Hilbert space $ \mathcal{H}_{q} $. When $ q=1 $, we construct $ I(x)\vert\Omega\rangle $ with $Q_{\dot{\alpha}}^{A}  I(x)\vert\Omega\rangle=0 $ and then $ (P_{0} +P_{5})  I(x)\vert\Omega\rangle=0$. The action of the gauge transformation $ U $ gives degenerate states $UI(x)\vert \Omega \rangle $ which could be organized into $  \mathcal{I}_{\alpha_{1}\cdots \alpha_{2J}} (x) \vert \Omega \rangle $ in the $ (J,0) $ representation of the $ SO(4)\cong SU(2)_{L} \times SU(2)_{R} $ rotation group and the definite representation of the gauge group. For the $ 5d $ $ \mathcal{N}=1 $ gauge theory, the spectrum of BPS instanton states matches quite well with the spectrum of string webs in the brane webs realization. The constructed instanton states have $ \rho=0 $. The question is whether there are more BPS states in $ \mathcal{H}_{1} $ coming from the $  \rho >0$ sectors. We cannot exclude such possibility, but the superconformal index in $ 5d $ gauge theories only receives the contribution from the point-like instantons (anti-instantons) \cite{5a}, and in the radial quantization formalism, states created by instanton operators in $ S^{4} \times \mathbb{R}_{+}$ are supersymmetric only when $ \rho=0 $ \cite{tw10}. When $ q >1$, $ G=SU(N) $, BPS instanton states with $ \rho_{i}=0 $ can be built from 
\begin{equation}
\left\lbrace U_{1}I(x_{1}) U_{2}I(x_{2}) \cdots U_{q}I(x_{q})  \vert \Omega \rangle|\;\forall \; x_{i},\forall \; U_{i} \in \mathcal{G}[SU(N)]\right\rbrace 
\end{equation}
labeled by $ q( 4N-1 ) $ parameters. The spectrum with the definite spin and the electric charges are more complicated.

In this paper, we mainly focus on $ 5d $ $ \mathcal{N}=1 $ gauge theories. For the $ 5d $ $ \mathcal{N}=2 $ $ SU(N) $ gauge theory with $ N\geq 3 $, based on the fermionic zero mode analysis, $ \mathcal{I}^{b_{1}\cdots b_{m},p_{1}\cdots p_{n}}_{q_{1} \cdots q_{m},a_{1}\cdots a_{n}} \vert \Omega \rangle$ constructed from $ I(x) \vert \Omega \rangle $ with $\mu^{A a}  I(x) \vert \Omega \rangle =0 $ for $ A=1,\cdots,4 $ and $ a=3,\cdots,N $ should satisfy $m-n=-2N  $. So the spin $ 1 $ state $ (\mathcal{I}_{\alpha\beta} )_{a}^{b} \vert \Omega \rangle$ in the adjoint representation of $ SU(N) $ does not exist, while for $ \mathcal{N}=1 $ gauge theories, such state exists only when $ \pm \kappa =N-\frac{1}{2} N_{f}    $. For the generic $ q $, the allowed gauge representations will be $ q $-dependent. Instanton states in the $ 5d $ $ \mathcal{N}=2 $ theory are KK modes of the $6d$ $(2,0)  $ theory. It seems that states with the different KK momentum do not share the same gauge structure.

$   (\mathcal{I}_{\alpha\beta} )_{a}^{b}$, although has a lot in common with the selfdual tensor $ (B_{\alpha\beta} )_{a}^{b} $, cannot be directly identified with it. The momentum operator $  P_{\alpha\dot{\alpha}}  $ acts as a 2-form covariant derivative on $ B_{\alpha\beta} $ with $ P_{\{\alpha\dot{\alpha}} B_{\beta\gamma\}} =0$. In the framework of the $ 5d $ gauge theory, we cannot construct a local operator satisfying this relation without modifying the definition of $ P_{\alpha\dot{\alpha}}  $. Instead, an additional gauge equivalence relation, which, in the abelian case, is just $ B_{\alpha\beta} \sim  B_{\alpha\beta}-\frac{i}{2}P^{\dot{\gamma}}_{\{\alpha}a_{\beta\}\dot{\gamma}} $ is introduced to make $ B_{\alpha\beta} $ from $ \mathcal{I}_{\alpha\beta}  $.

 %%%%%%%%%%%%%%%%%%%%%%%%%%%%%%%%%%
\section*{Acknowledgments}

The work is supported in part by NSFC under the Grant No. 11605049.

%%%%%%%%%%%%%%%%%%%%%%%%%%%%%%%%%%

\begin{appendix}

\section{Spinor algebra}\label{A}

In $ 5d $ Lorentzian flat spacetime, a suitable $4\times 4  $ matrix representation of the Dirac matrices is given by
\begin{equation}       
\gamma^{0}=i\left[                 
  \begin{array}{cc}   
   -1_{2} & 0_{2} \\  
   0_{2} & 1_{2} \\  
  \end{array}
\right] \;, \;\;\;\;  \gamma^{m}=\left[                 
  \begin{array}{cc}   
   0_{2} &  -i \sigma_{\alpha\dot{\alpha}}^{m } \\  
i  \bar{\sigma}^{m\dot{\alpha}\alpha}& 0_{2} \\  
  \end{array}
\right]  \;, \;\;\;\;  \begin{array}{c}   
\sigma^{m} =(i \vec{\tau}, 1_{2})  \\  
  \;\;\;   \;\; \bar{\sigma}^{m} =( -i\vec{\tau},1_{2})  \\  
  \end{array}    \;,         
\end{equation}
where $ m=1,2,3,4 $, $ \alpha,\dot{\alpha}=1,2 $, $ \vec{\tau} $ are Pauli matrices. $ \gamma^{\mu} $ for $ \mu=0,1,\ldots,4 $ satisfy $\{\gamma^{\mu},\gamma^{\nu}\}  =2\eta^{\mu\nu}$. $ \gamma^{0} \gamma^{1}  \gamma^{2}  \gamma^{3}  \gamma^{4} =i1_{4} $. The matrix $ \gamma^{5} $ is diagonal
\begin{equation}       
\gamma^{5}\equiv -\gamma^{1}\gamma^{2}\gamma^{3}\gamma^{4}=\left[                 
  \begin{array}{cc}   
   1_{2} & 0_{2} \\  
  0_{2} &- 1_{2} \\  
  \end{array}
\right]  \;.
\end{equation}
Projections of a non-chiral four-component spinor with $\frac{1}{2} (1\pm \gamma^{5}) $ yield the two-component chiral spinors. $  \gamma^{mn}=\frac{1}{2}\epsilon^{mnkl}\gamma_{kl}\gamma^{5} $, where 
\begin{equation}       
\gamma^{mn}= \frac{1}{2}(\gamma^{m}\gamma^{n}-\gamma^{n}\gamma^{m})
=\left[                
  \begin{array}{cc}   
\sigma^{mn}& 0_{2} \\  
  0_{2} &  \bar{\sigma}^{mn}  \\  
  \end{array}
\right] 
\end{equation}
with
\begin{equation}
( \sigma^{mn} )_{\alpha}^{\beta}=\frac{1}{2}(\sigma^{m}\bar{\sigma}^{n}-\sigma^{n}\bar{\sigma}^{m})_{\alpha}^{\beta}\;,\;\;\;\;\;\;\;\;\; ( \bar{\sigma}^{mn} )^{\dot{\alpha}}_{\dot{\beta}}=\frac{1}{2}(\bar{\sigma}^{m}\sigma^{n}-\bar{\sigma}^{n}\sigma^{m})^{\dot{\alpha}}_{\dot{\beta}}\;.
\end{equation}
$ \sigma^{mn} $ and $\bar{\sigma}^{mn}  $ are selfdual and anti-selfdual:
 \begin{equation}
\sigma^{mn}=\frac{1}{2}\epsilon^{mnkl}\sigma_{kl}\;,\;\;\;\;\;\;\;\;\;\;  \bar{\sigma}^{mn}=-\frac{1}{2}\epsilon^{mnkl}\bar{\sigma}_{kl}\;.
 \end{equation}
The contraction relations are 
\begin{eqnarray}
\nonumber &&\sigma^{m}_{\alpha\dot{\alpha}}\bar{\sigma}^{\dot{\beta}\beta}_{m}=2\delta^{\beta}_{\alpha}\delta^{\dot{\beta}}_{\dot{\alpha}}\;,\\ \nonumber &&  \sigma^{n}_{\alpha\dot{\alpha}}(\sigma_{nm})_{\beta\gamma}=\epsilon_{\alpha\beta}\sigma_{m\gamma\dot{\alpha}}+\epsilon_{\alpha\gamma}\sigma_{m\beta\dot{\alpha}}\;,\;\;\;\;\;\;\;\;\; 
\sigma^{n}_{\alpha\dot{\alpha}}(\bar{\sigma}_{nm})_{\dot{\beta}\dot{\gamma}}=\epsilon_{\dot{\beta}\dot{\alpha}}\sigma_{m\alpha\dot{\gamma}}+\epsilon_{\dot{\gamma}\dot{\alpha}}\sigma_{m\alpha\dot{\beta}}\;,\\ &&  (\sigma^{mn})^{\beta}_{\alpha}(\sigma_{mn})^{\tau}_{\gamma}=4(\delta^{\beta}_{\alpha}\delta^{\tau}_{\gamma}-2\delta^{\tau}_{\alpha}\delta^{\beta}_{\gamma})\;,\;\;\;\;\;\;\;\;\; 
(\bar{\sigma}^{mn})_{\dot{\beta}}^{\dot{\alpha}}(\bar{\sigma}_{mn})_{\dot{\tau}}^{\dot{\gamma}}=4(\delta^{\dot{\alpha}}_{\dot{\beta}}\delta^{\dot{\gamma}}_{\dot{\tau}}-2\delta^{\dot{\alpha}}_{\dot{\tau}}\delta^{\dot{\gamma}}_{\dot{\beta}}) \;,
\end{eqnarray}
where $ \epsilon_{\alpha\beta} =-\epsilon^{\alpha\beta}  $, $\epsilon_{\dot{\alpha}\dot{\beta}}=-\epsilon^{\dot{\alpha}\dot{\beta}}  $, $ \epsilon_{\beta\alpha}\epsilon^{\alpha\gamma} =\delta^{\gamma}_{\beta}$, $ \epsilon_{\dot{\beta}\dot{\alpha}}\epsilon^{\dot{\alpha}\dot{\gamma}}=\delta^{\dot{\gamma}}_{\dot{\beta}} $. $\epsilon^{12}=1  $. $ \sigma_{\alpha\dot{\alpha}}^{m}=-\epsilon_{\dot{\alpha}\dot{\beta}}\bar{\sigma}^{m\dot{\beta}\beta}\epsilon_{\beta\alpha} $. $\epsilon^{\beta\alpha}\xi_{\alpha}=\xi^{\beta}$, $\epsilon_{\dot{\beta}\dot{\alpha}}\xi^{\dot{\alpha}}=\xi_{\dot{\beta}}$, $\xi\chi=\xi_{\alpha}\epsilon^{\alpha\beta}\chi_{\beta}$, $\bar{\xi}\bar{\chi}=\bar{\xi}^{\dot{\alpha}}\epsilon_{\dot{\alpha}\dot{\beta}}
\bar{\chi}^{\dot{\beta}}$. The charge conjugation matrix $ C $ is 
\begin{equation}\label{C}       
C=\left[                
  \begin{array}{cc}   
  \epsilon^{\alpha\beta} & 0_{2} \\  
  0_{2} &   \epsilon_{\dot{\alpha}\dot{\beta}} \\  
  \end{array}
\right] \;, \;\;\;\;\;\;\;C^{-1}=\left[                
  \begin{array}{cc}   
  \epsilon_{\alpha\beta} & 0_{2} \\  
  0_{2} &   \epsilon^{\dot{\alpha}\dot{\beta}} \\  
  \end{array}
\right] \;.
\end{equation}
$ C \gamma^{\mu}C^{-1}=\gamma^{\mu T} $.

\section{$ 5d $ $\mathcal{N}=2  $ chiral multiplet }\label{B}

For the $ 5d $ $ \mathcal{N}=2 $ supersymmetry, starting from $ B_{1\cdots 1} $ with
\begin{equation}\label{AB}
Q^{A}_{\dot{\beta}} B _{1\cdots 1}   =0\;,\;\;\;\;\;\;\;\;Q^{A}_{1} B _{1\cdots 1} =0\;,\;\;\;\;\;\;\;\;A=1,2,3 ,4,
\end{equation}
successive actions of supercharges generate
\begin{equation}\label{b.2}
 B _{1\cdots 1} \rightarrow  Q_{2}^{A} B _{1\cdots 1}  \rightarrow Q_{2}^{A}Q_{2}^{B} B _{1\cdots 1} \rightarrow Q_{2}^{A}Q_{2}^{B}Q_{2}^{C} B _{1\cdots 1}\rightarrow Q_{2}^{1}Q_{2}^{2}Q_{2}^{3}Q_{2}^{4}  B _{1\cdots 1}\;.
\end{equation}
The $ SU(2)_{L} $ transformation of (\ref{AB}) gives 
\begin{equation}
Q^{A}_{\dot{\beta}} B _{ \alpha_{1}\alpha_{2} \cdots\alpha_{2J}       }   =0\;,\;\;\;\;\;\;\;\; Q^{A}_{\{ \beta} B _{\alpha_{1}\alpha_{2} \cdots\alpha_{2J}       \}} =0 \;.
\end{equation}
When $ J\geq 2 $, the action of supercharges on $ B _{ \alpha_{1}\alpha_{2} \cdots\alpha_{2J}       } $ produces a supermultiplet consisting of $ 5 $ component fields with the spins ranging from $ J$ to $ J-2 $.

When $ J=1 $, we get a tensor multiplet with fields $ (B_{\alpha\beta},\Psi_{\alpha}^{A} ,\Phi^{AB}) $. In (\ref{b.2}), $ Q_{2}^{A}B_{11}\sim \Psi_{1}^{A} $, $Q_{2}^{A}Q_{2}^{B}B_{11}\sim \Phi^{AB}   $, but 
\begin{equation}
Q_{2}^{A}Q_{2}^{B} Q_{2}^{C}B _{1 1}\sim Q_{2}^{A}Q_{2}^{B} Q_{1}^{C}B _{1 2} \sim   Q_{1}^{C}Q_{2}^{A}Q_{2}^{B} B _{1 2}
\sim Q_{1}^{C}Q_{2}^{A}Q_{1}^{B} B _{2 2} \sim -Q_{1}^{C}Q_{1}^{B} Q_{2}^{A}B _{2 2}     =0\;,
\end{equation}
where terms involving anti-commutators are dropped. The multiplet terminates at $  \Phi^{AB} $.

When $ J\geq 2 $, we have
\begin{eqnarray}
\nonumber &&Q_{2}^{A}Q_{2}^{B} Q_{2}^{C}B _{1 1 1\cdots 1}\sim Q_{2}^{A}Q_{2}^{B} Q_{1}^{C}B _{1 2 1\cdots 1}\sim  Q_{1}^{C}Q_{2}^{A}Q_{2}^{B} B _{1 2 1\cdots 1}\\   &\sim &Q_{1}^{C}Q_{2}^{A}Q_{1}^{B} B _{2 2 1\cdots 1} \sim -Q_{1}^{C}Q_{1}^{B} Q_{2}^{A}B _{2 2 1\cdots 1}\sim -Q_{1}^{C}Q_{1}^{B}Q_{1}^{A}  B _{2 22\cdots 1} \;,
\end{eqnarray}
\begin{equation}
Q_{2}^{A}Q_{2}^{B} Q_{2}^{C}Q_{2}^{D}B _{1 1 1\cdots 1}\sim Q_{2}^{D}Q_{1}^{C}Q_{1}^{B}Q_{1}^{A}  B _{2 22\cdots 1}\sim -Q_{1}^{C}Q_{1}^{B}Q_{1}^{A} Q_{2}^{D} B _{2 22\cdots 1}\sim -Q_{1}^{C}Q_{1}^{B}Q_{1}^{A} Q_{1}^{D} B _{2 222\cdots 1}
\end{equation}
The multiplet is composed by $ (B _{ \alpha_{1} \cdots\alpha_{2J}       },\Psi^{A} _{ \alpha_{1} \cdots\alpha_{2J-1}       }, \Phi^{AB} _{ \alpha_{1} \cdots\alpha_{2J-2}       } ,\tilde{\Psi}^{A} _{ \alpha_{1} \cdots\alpha_{2J-3}       } ,\tilde{B} _{ \alpha_{1} \cdots\alpha_{2J-4}       } )$.

\end{appendix}

\end{document}